\documentclass[11pt]{article}
\usepackage{graphicx} 
\usepackage{amsmath,amsthm,amssymb}
\usepackage[margin=1in]{geometry}

\newtheorem{theorem}{Theorem}
\newtheorem{lemma}{Lemma}
\newtheorem{corollary}{Corollary}
\newtheorem{proposition}{Proposition}
\newtheorem{definition}{Definition}

\newtheorem{claim}{Claim}

\usepackage{cite}
\usepackage{xcolor}
\usepackage[
    hypertexnames=false,
    colorlinks=true,
    linkcolor=blue,
    citecolor=blue,
    urlcolor=blue
]{hyperref}
\usepackage{enumitem}

\usepackage{color-edits}
\addauthor[Sid]{sb}{teal}
\addauthor[Kamesh]{km}{magenta}
\addauthor[Tianle]{tj}{blue}

\usepackage{tikz}
\usetikzlibrary{calc}
\usepackage{pgfplots}
\usepackage{placeins}

\newcommand{\leanverified}[1]{%
  \par\smallskip\noindent
  \fcolorbox{black!35}{black!3}{%
    \parbox{\dimexpr\linewidth-2\fboxsep-2\fboxrule\relax}{%
      \small\textsf{\textbf{Lean-verified:}} #1}}%
  \par\smallskip}

\DeclareMathOperator{\E}{\mathbb{E}}
\newcommand{\OPT}{\mbox{OPT}}
\newcommand{\REV}{\mbox{REV}}
\newcommand{\SPM}{\mbox{SPM}}

\newcommand{\TSlack}{\operatorname{T\text{-}Slack}}
\newcommand{\PSlack}{\operatorname{P\text{-}Slack}}

\makeatletter
\def\@fnsymbol#1{\ensuremath{\ifcase#1\or \dagger\or \ddagger\or
   \mathsection\or \mathparagraph\or \|\or **\or \dagger\dagger
   \or \ddagger\ddagger \else\@ctrerr\fi}}
\makeatother
\newcommand*\samethanks[1][\value{footnote}]{\footnotemark[#1]}

\title{Beyond Worst-Case Online Algorithms via Input Subsampling}
\author{Siddhartha Banerjee\thanks{School of Operations Research and Information Engineering, Cornell University, Ithaca, NY 14853. Email: \textsf{sbanerjee@cornell.edu}. Supported by AFOSR grant FA9550-23-1-0068.}  \and Tianle Jiang\thanks{Department of Computer Science, Duke University, Durham, NC 27708-0129. Emails: \textsf{kamesh@cs.duke.edu}, \textsf{tianle.jiang@duke.edu}. Supported by NSF grant IIS-2402823.}  \and Kamesh Munagala\samethanks[2]}
\date{}

\begin{document}

\maketitle
\begin{abstract}
We study whether classical competitive ratios survive input subsampling, when both the online algorithm and the offline optimum are evaluated on the same subsampled input. We consider the canonical problems of prophet inequalities and ski-rental. In these settings, for any single subsampling rate, the classical worst-case ratios remain unchanged. We ask a stronger question: can one adversarial input remain worst-case at two distinct rates $p>q$?

For prophet inequalities, in the model where the policy knows the sampling rate and learns item survival online, every instance has competitive ratio strictly below $2$ at one of the two rates, and we determine the tight simultaneous guarantee. For ski rental, in the fluid model where sampling scales the horizon, the answer depends on $q/p$. Against adversarial distributions over horizons, one unrestricted randomized algorithm achieves a strict simultaneous improvement over $e/(e-1)$ for every distribution exactly when $q/p$ is below approximately $0.34$. Requiring the algorithm to buy by the breakeven time lowers this threshold to approximately $0.31$, showing that unlike classical ski-rental algorithms, buying later can provably help in this setting.

Our results therefore reveal a new form of beyond-worst-case fragility under input subsampling: instances that are worst-case at one rate need not remain hard at two rates simultaneously.
\end{abstract}

\section{Introduction}
\label{sec:introduction}

Competitive analysis evaluates an online algorithm by comparing its performance with an offline optimum on a worst-case input. When an instance has a natural \emph{scale} -- the number of arrivals or the duration of its horizon -- the adversary may choose a different hard input at every scale. Consequently, the classical competitive ratio can hide an important structural question: \emph{is a worst-case instance intrinsically hard, or is its hardness finely tuned to the scale at which it is evaluated?}

We study this question through \emph{independent subsampling}. Starting from a single base `adversarial' input, we retain each item independently with probability \(p\), and evaluate both the online algorithm and the offline optimum on the retained input. This is a natural way to vary the effective scale of an online instance while changing it as little as possible: every item is treated symmetrically, and surviving items retain their values/relative order. 
At the same time, subsampling is milder than other perturbations that are known to improve online guarantees: unlike random-order models, it preserves relative order~\cite{esfandiari2017prophet,azar2018prophet}, and unlike targeted removal, it does not selectively delete particular distributions or realizations~\cite{liu2021variable,correa2025residual}. Moreover, its different scales are naturally coupled: for \(q<p\), a \(q\)-sample can be obtained by further subsampling a \(p\)-sample at rate \(q/p\).

The effect of changing scale can vary dramatically across online problems. At one extreme, some hard inputs are essentially self-similar under subsampling. Paging~\cite{borodin1998online} gives a simple example. A standard hard distribution for randomized paging draws each request independently and uniformly from a set of \(K+1\) pages, where \(K\) is the cache size. Independent subsampling at any positive rate leaves the retained requests independent and uniform. Hence the same hard distribution yields the classical \(H_K\) lower bound simultaneously at every sampling rate.\footnote{More precisely, for any fixed finite collection of rates, a standard concentration argument yields a single sequence approaching the \(H_K\) lower bound simultaneously for the collection.}

At the opposite extreme, hardness may rely on fine-scale regularity that random thinning disrupts. In binary prediction, the regret of follow-the-leader is exactly proportional to the number of lead changes~\cite{banerjee2024smart}, so highly regular sequences such as \(0101\ldots\) are worst-case; random deletion quickly destroys this structure. Relatedly, Bernoulli sampling has been used to regularize fictitious play and leader-based dynamics~\cite{li2018sampled}.

Between these extremes lies the phenomenon studied in this paper. The worst-case competitive ratio may remain unchanged at every individual scale: once the sampling rate is fixed, the adversary can simply retune its instance and recover the classical lower bound. Nevertheless, the instances attaining this bound may themselves be scale-sensitive: \emph{no single base instance may remain nearly worst-case at two distinct sampling rates}. In other words, classical competitive analysis can be scale-invariant even when its worst-case instances are not.

We investigate this phenomenon in two foundational online stopping problems whose classical hard instances depend delicately on scale:

\begin{itemize}
\item \textbf{Prophet inequalities.}
An adversary chooses a sequence of independent nonnegative random variables. Under \(p\)-subsampling, each variable is independently retained with probability \(p\), and both the online algorithm and the prophet are evaluated on the retained sequence. Here subsampling changes the effective number of opportunities while leaving the distributions of the surviving variables unchanged.

\item \textbf{Ski rental.}
An adversary chooses a base horizon \(B\). We use the continuous-time formulation, in which viewing the instance at rate \(p\) scales its effective horizon to \(pB\). Thus the same underlying instance is evaluated at two different length scales.
\end{itemize}

For either problem, looking at one rate at a time reveals nothing new at the level of the classical competitive ratio: after the rate is fixed, the adversary can tune an instance to that rate and force the factor \(2\) for prophet inequalities~\cite{krengel1977semiamarts,samuelcahn1984threshold} or \(e/(e-1)\) for randomized ski rental~\cite{karlin1994nonuniform,borodin1998online}. We instead fix two rates \(p>q\) and require the adversary to choose one base instance that is evaluated at both. Our central question is:
\emph{Can a single instance be simultaneously worst-case at distinct scales?}

Our results show that the landscape is subtler than either of the extreme examples above. For prophet inequalities, worst-case instances are fragile for every pair of distinct rates: at least one of the two induced instances must have competitive ratio strictly smaller than \(2\). For ski rental, by contrast, fragility and robustness are separated by a sharp threshold in the scale ratio \(\rho=q/p\).

Our viewpoint is closely related to smoothed analysis, which asks whether worst-case behavior survives a random perturbation of an adversarial input~\cite{spielman2004smoothed}; related perturbation-based viewpoints have also been studied in online learning and online algorithms~\cite{rakhlin2011smoothed,BecchettiEtAl2006MultilevelFeedback,coester2025smoothed}. The distinction is that smoothed analysis typically fixes a perturbation level and asks for the worst performance after perturbation. Our question concerns the \emph{stability of the worst-case instances themselves as the perturbation level changes}. Indeed, at every fixed subsampling rate the worst-case competitive ratio in our two problems is exactly the classical one; the phenomenon appears only when we ask whether the nearly extremal instances at two different rates overlap. Thus subsampling serves not merely as a source of randomness, but as a way to expose scale sensitivity that the usual worst-case ratio at any single perturbation level necessarily discards.

\subsection{Our results}
We give a tight characterization of how worst-case instances behave across two scales for prophet inequalities and ski rental. In both problems, we fix two sampling rates $0<q<p\leq 1$ and denote $\rho=q/p$. The adversary must commit to one underlying base instance -- or distribution over base instances -- whose induced instances are evaluated at the two scales.

\medskip
\noindent\textbf{Prophet inequalities.}
We consider two input models. In the stronger \emph{revealed-subsequence} model, the algorithm is given the surviving sequence of distributions before observing their values and runs the optimal online policy for that sequence; it does not need to know the sampling rate. In the weaker \emph{revealed-rate} model, the algorithm knows the original distributions and the sampling rate, but does not know which variables survive the sampling. Our proofs focus on the weaker model, where we analyze the optimal dynamic program adapted to input subsampling, and the positive results we obtain directly extend to the stronger model, since revealing the surviving sequence only gives the stronger model more information.

For a base instance $\mathcal I$, let $CR_{\mathcal I}(r)$ denote the ratio between the prophet's value and the value of the optimal rate-$r$ dynamic program. For any single sampling rate $r$, we show that the classical factor $2$ remains tight: there is an instance $\mathcal{I}_r$ with $CR_{\mathcal{I}_r}(r)\to 2$. We augment the standard two-item hard instance: now it consists of many deterministic unit values followed by a rare high-value, with parameters tuned so that the optimal algorithm is nearly indifferent between accepting a surviving unit value and waiting.

In contrast, when there are two different sampling rates $p,q$, we prove
\[
\sup_{\mathcal I}
\min\{CR_{\mathcal I}(p),CR_{\mathcal I}(q)\}
\le
2-\frac{\rho(1-\rho)}{2(3-\rho)}
<2.
\]
Thus, although the adversary can choose a different tight instance at each individual rate, no single base instance can be nearly tight at both rates simultaneously. More generally, we characterize the exact Pareto frontier between $CR_{\mathcal I}(p)$ and $CR_{\mathcal I}(q)$ for the weaker model, and give asymptotically matching instances. 

\medskip
\noindent\textbf{Randomized ski rental.}
For classical ski rental (with unit buying cost, and unit per period rental cost), the optimal randomized competitive ratio is
$e/(e-1)\approx 1.58$, achieved by the truncated-exponential equalizer distribution~\cite{karlin1994nonuniform,borodin1998online}. This continues to hold for any single subsampling rate.

In contrast, we evaluate one buy-time distribution at the two scaled horizons $pB$ and $qB$, where $\rho=q/p<1$. The algorithm knows the scale ratio $\rho=q/p$ and chooses a single buy-time distribution that is evaluated at both horizons $pB$ and $qB$. This allows us to ask whether the same adversarial input remains hard at both scales even when the strategy is chosen for that comparison. Already for a deterministic adversary choosing a single horizon $B$, simultaneous hardness is fragile: by slightly perturbing the classical exponential buying distribution, the algorithm can concentrate its nearly worst-case performance into a narrow range of horizons. Since $pB$ and $qB$ differ by the factor $\rho$, they cannot both lie in this range. Consequently, for every fixed $\bar\rho<1$, there is an algorithm that obtains a uniform strict improvement over $e/(e-1)$ at one of the two scales whenever $\rho\le \bar\rho$.

Our technically and conceptually more interesting results are for a {\em distributional adversary}, which chooses a probability distribution over base horizons. 

In this setting, whether simultaneous improvement is possible depends sharply on $\rho$. We obtain an exact threshold for each of two algorithm classes. Below the threshold, one algorithm achieves a uniform strict improvement: for every adversarial distribution, at least one of the two scales has competitive ratio smaller than $e/(e-1)$ by a fixed amount. At or above the threshold, every algorithm admits an adversarial distribution that prevents any strict simultaneous improvement. For algorithms constrained to buy by the breakeven time, this threshold is $\rho_\star\approx0.311$; for arbitrary algorithms, it is
$\rho_{\mathrm{all}}\approx0.338$.
{\em The existence of these thresholds and the gap between them are surprising} and show that in this model, unlike classical ski rental, buying by the breakeven time is not without loss for distributional adversaries.

Finally, for distributional adversaries, we show that when the algorithm knows only an upper bound on $\rho$, a single algorithm cannot obtain a uniform strict improvement over $e/(e-1)$.

\paragraph{Technical highlights.}
In addition to the conceptual contribution, our paper involves several new technical ideas. At a high level, at each scale separately, classical competitive analysis gives exactly the same worst-case ratio, so the scalar ratio itself contains no indication of the simultaneous improvement. Our proofs therefore retain one additional layer of information -- the signed slack in the classical tight guarantee -- and study how this slack can trade off across two linked scales.

For prophet inequalities, the relevant slack is the difference between the online value and one half of the prophet's value. We revisit the standard dynamic-programming proof and identify two losses that are normally discarded: replacing the changing continuation values by a common threshold, and upper-bounding the prophet's maximum by a baseline plus the sum of excesses above it. An instance can make both losses small at one sampling rate, but not at two distinct rates simultaneously. A simple version of the argument uses two cutoffs to divide all instances into three cases and argue that in each case, some loss is large for $p$ or $q$. This certifies an improvement below $2$ that depends on $\rho$.  
For the exact frontier, we refine this argument using the two continuation thresholds; independence reduces the remaining optimization to one over products of failure probabilities, yielding both the Pareto frontier and matching instances.

For ski rental, we study the residual curve: the difference between an algorithm's expected cost and the target multiple of the offline optimum~\cite{karlin2003tcp}, viewed as a function of the horizon. The problem then reduces to two sets of conditions: conditions governing when a residual curve can be realized by a ski-rental algorithm, and the two-scale inequalities governing simultaneous improvement over $\frac{e}{e-1}$. We derive a necessary weighted-integral condition for realizability and a partial converse under the regularity conditions used in our constructions.
For the impossibility results, we combine the two-scale inequalities with nonnegative weights and show that they contradict the necessary realizability condition. For the positive results, we construct residual curves that satisfy the two-scale inequalities and the sufficient realization conditions, recursively balancing positive residual at one scale with negative residual at a smaller scale. Strikingly, the same recurrence governs both the positive construction and the nonnegative weights in the impossibility proof, allowing the two arguments to meet at an exact threshold.

\subsection{Related work}
\label{subsec:related-work}

\noindent\textbf{Random perturbations and beyond worst case.} 
In addition to the smoothed analysis work cited above, another line of beyond-worst-case work augments the online algorithm with predictions, samples, or distributional advice. For example, Diakonikolas et al.~\cite{diakonikolas2021distributional} study online algorithms with distributional advice for both prophet inequalities and ski rental. These models, like ours, introduce information or structure beyond the realized online input, but the role of that information is different: our objective is to understand how one adversarial base instance behaves when viewed at two scales.

\medskip
\noindent\textbf{Prophet inequalities.}
The classical prophet inequality of Krengel, Sucheston, and Garling~\cite{krengel1977semiamarts} shows that a gambler observing independent nonnegative random variables can obtain at least half of the expected maximum. Samuel-Cahn later showed that this guarantee can be achieved by a single-threshold rule~\cite{samuelcahn1984threshold}. Prophet inequalities have since been extended to richer feasibility constraints, including matroids and related mechanism-design settings~\cite{kleinberg2019matroid}, and are closely connected to simple auction mechanisms~\cite{chawla2010sequential,lucier2017prophet}.

Several recent variants study settings in which part of the input is removed or reordered before the online decision is made. Liu et al.~\cite{liu2021variable} introduce a variable-decomposition framework and show that, for non-identical distributions, one can recover the optimal i.i.d.-prophet constant $\beta\approx0.745$ after removing a small constant number of distributions chosen by the algorithm. Correa et al.~\cite{correa2025residual} study residual prophet inequalities, where the top $k$ realized values are removed before the remaining sequence is observed online. A different source of randomness appears in the prophet-secretary model, which keeps every variable but presents them in uniformly random order~\cite{esfandiari2017prophet,azar2018prophet}. Random order permits guarantees strictly better than the classical factor-$2$ bound.

These models are closely related philosophically to ours, but perturb the input in different ways. Targeted removal may select particular distributions or realizations, while random-order models alter the arrival order. Independent subsampling instead removes every variable symmetrically and preserves the relative order of the survivors. More importantly, our objective is not merely to improve the guarantee after perturbation, but to ask whether one base instance can remain nearly tight at two different subsampling rates.

The i.i.d. prophet setting provides another important benchmark~\cite{hill1982comparisons,abolhassani2017beating,correa2021posted}. Work on i.i.d. prophets has also studied comparisons finer than the competitive ratio by characterizing the possible gaps between the gambler and the prophet~\cite{hill1981additive,hill1982comparisons,kertz1986complete}. Our use of the signed gap from the classical prophet guarantee is similar in spirit, though our goal is to control how this gap behaves simultaneously at two linked sampling rates.

\medskip
\noindent\textbf{Ski rental.}
Ski rental is one of the canonical examples in competitive analysis. The deterministic optimum is $2$, and the optimal randomized competitive ratio is $e/(e-1)$. The optimal randomized algorithm uses the classical equalizer distribution for the buy time. This and related rent-or-buy problems were studied in~\cite{karlin1994nonuniform,karlin2003tcp}; see also the textbook treatment of~\cite{borodin1998online}.

Ski rental has also been a central test problem for learning-augmented online algorithms. Purohit et al.~\cite{purohit2018predictions} introduced ski-rental algorithms with machine-learned predictions and consistency--robustness tradeoffs. Gollapudi and Panigrahi~\cite{gollapudi2019experts} studied rent-or-buy algorithms with multiple experts, and subsequent work considered multi-shop and other variants with learned advice~\cite{wang2020multishop}. Related stochastic models include semi-stochastic ski rental~\cite{madry2011semistochastic}, distributional predictions~\cite{cui2026distributional,kim2026distributional}, and prior-independent finite-horizon ski rental~\cite{hartline2026prior}. These works provide the algorithm with a predicted horizon, expert advice, samples, or information about the horizon distribution, whereas our model asks how one adversarial base horizon behaves under two linked scalings.

\medskip
\noindent\textbf{Alternative competitive objectives.}
Dinitz et al.~\cite{dinitz2024controlling_tail_risk_ski_rental} optimize expected competitive ratio subject to bounds on the tail of the realized competitive ratio. Their characterization produces non-monotone, discontinuous buying distributions on recursively scaled time intervals. This superficially resembles our alternating constructions, where the inequality at $y$ is coupled to the inequality at~$\rho y$. The objectives are nevertheless different: their setting admits buying by the breakeven time without loss, while in our distributional two-scale model allowing later purchases strictly enlarges the range of $\rho$ for which simultaneous improvement is possible.

Alternative performance measures for prophet inequalities and ski rental compare the expectation of the realized ratio (EoR) with the classical ratio of expectations (RoE)~\cite{ezra2025prophet_expected_competitive_ratio,chen2026new_metric_ski_rental}. In particular, Chen and Zhang~\cite{chen2026new_metric_ski_rental} optimize the minimum of EoR and RoE for ski rental. Their two coordinates correspond to different performance measures applied to one horizon distribution, whereas ours apply the same competitive measure to one base instance at two different scales.

\section{The Prophet Inequality}
\label{sec:prophet}

In the classical prophet inequality, a gambler sequentially observes
independent, nonnegative random variables
$
    X_1,X_2,\ldots,X_n
$
drawn from known distributions, with finite expectations. Upon observing the realized value of $X_i$,
the gambler must immediately and irrevocably decide whether to accept $X_i$
and stop, or reject it and continue. The classical theorem of~\cite{krengel1977semiamarts,samuelcahn1984threshold} states that the expected offline optimum is at most
twice the expected reward of an appropriate online threshold policy, and this factor is tight, as we discuss below. 

\subsection{Subsampling model}
We now study the same problem under independent subsampling. For
$r\in(0,1]$, let $S_r\subseteq[n]$ be the random subset obtained by retaining
each item independently with probability $r$. Define the subsampled prophet
benchmark
\[
    \OPT(r)
    =
    \mathbb E\left[\max_{i\in S_r}X_i\right].
\]
The expectation is over both the realized set $S_r$ and the values of the variables $X_i$. We set the maximum over the empty set equal to $0$.

We consider two input models: In the stronger \emph{revealed-subsequence} model, the online algorithm is told the realized subsequence $S_r$, and then runs the optimal online policy for this surviving sequence of distributions; we use $\widetilde v(r)$ to denote its expected reward.
For our proof, it is more convenient to analyze a weaker \emph{revealed-rate} model where the online algorithm knows the subsampling rate $r$ but not the realized subsequence $S_r$. We use $v_i(r)$ to denote the optimal DP value-function from the suffix
$i,i+1,\ldots,n$ in this unrevealed-subsampling process, with
$v_{n+1}(r)=0$, and satisfying the following recurrence:
\begin{equation}
\begin{aligned}
    v_i(r)=
    r\,\mathbb E\bigl[\max\{X_i,v_{i+1}(r)\}\bigr]
    +(1-r)v_{i+1}(r).
\end{aligned}
\label{eq:prophet-dp-model}
\end{equation}
The expected reward of the DP is $v_1(r)$. Clearly
$
    \widetilde v(r)\ge v_1(r),
$
because revealing $S_r$ only gives the online algorithm more information. Hence any bound of the form
$
    \frac{\OPT(r)}{v_1(r)}\le c
$
also holds, a fortiori, for the more natural \emph{revealed-subsequence} model:
$
    \frac{\OPT(r)}{\widetilde v(r)}\le c.
$
We therefore focus on $v_1(r)$.

We restrict attention to nontrivial instances, meaning that at least one
$X_i$ is positive with positive probability. For every $r>0$, such an
instance has $v_1(r)>0$. We measure performance by the competitive ratio
\[
    CR(r)=\frac{\OPT(r)}{v_1(r)}.
\]

The classical prophet inequality says $CR(1)\le2$, and the factor $2$ is tight; to see this, consider a hard instance where $X_1=1$ and $X_2=(1/\epsilon)\cdot\text{Bernoulli}(\epsilon)$ for some sufficiently small $\epsilon>0$. As we show below, the factor remains tight for every single subsampling probability $r$.

\paragraph{Our objective.} Our goal is to understand whether this single-rate tightness can occur simultaneously at two different subsampling probabilities. Fix
$
    0<q<p\le1
$, with scale ratio
$
    \rho=\frac qp\in(0,1).
$
For a given instance, we evaluate the two competitive ratios
\[
    CR(q)=\frac{\OPT(q)}{v_1(q)},
    \qquad
    CR(p)=\frac{\OPT(p)}{v_1(p)}.
\]
The simultaneous  ratio of the instance is
$
    \min\{CR(q),CR(p)\},
$
because the algorithm is considered successful if it beats the classical bound at least one of the two rates.
Thus the question is whether the simultaneous competitive ratio
\[
    \sup_{\text{instances}}\min\{CR(q),CR(p)\}
\]
is still equal to $2$, as it is for each rate separately, or whether it is
strictly smaller than $2$ once the same instance must be evaluated at both
rates. We prove the latter, and in fact characterize the tight value.
Note that we allow $p=1$; in other words, qualitatively, our result says that for any instance, and any random subset of that instance, cannot both simultaneously attain the worst-case competitive ratio.

\subsection{Single-rate tightness}

We first prove that $CR(r)\leq2$ for every fixed $r$ using the DP formulation. The argument is analogous to the classical prophet-inequality proof for $r=1$ (i.e., without subsampling).
We give the full argument because its two sources of slack will be useful later.

\begin{proposition}
\label{prop:single_rate_competitive_ratio}
For every fixed $r\in(0,1]$, $CR(r)\leq 2$.
\end{proposition}

\begin{proof}
Fix a subsampling probability $r\in(0,1]$. Let $v_i(r)$ denote the optimal expected reward from the suffix $X_i,\ldots,X_n$ when each item survives independently with probability $r$, with $v_{n+1}(r)=0$. The optimal threshold at item $i$ is the continuation value $\tau_i(r)=v_{i+1}(r)$, and the DP recurrence is
\begin{align}
    v_i(r)&=r\,\mathbb E[\max\{X_i,\tau_i(r)\}]+(1-r)\tau_i(r) \notag\\
    &=\tau_i(r)+r\,\mathbb E[(X_i-\tau_i(r))^+]. \label{eq:prophet_dp_recurrence}
\end{align}
In particular, $v_i(r)\ge v_{i+1}(r)$, so every continuation threshold satisfies $\tau_i(r)\le v_1(r)$. Telescoping the recurrence gives
\begin{equation} 
    v_1(r)=r\sum_{i=1}^n\mathbb E[(X_i-\tau_i(r))^+]. 
\label{eq:prophet_dp_telescoping} \end{equation}
To upper bound the value of the prophet, we use the following standard single-threshold relaxation.
\begin{claim}
\label{claim:prophet_relaxation}
    For every $r\in(0,1]$ and every $t\ge0$, if $M_r=\max_{i\in S_r}X_i$, then
\begin{equation} 
    \OPT(r)=\mathbb E[M_r]\le t+r\sum_{i=1}^n\mathbb E[(X_i-t)^+].  
\label{eq:prophet-relaxation-bound}
\end{equation}
\end{claim}
\begin{proof}
    Indeed, pointwise,
    \[
    M_r\le t+(M_r-t)^+\le t+\sum_{i\in S_r}(X_i-t)^+,
    \]
    and taking expectations gives the claim.
\end{proof}

Combining the telescoping identity \eqref{eq:prophet_dp_telescoping} with
the relaxation \eqref{eq:prophet-relaxation-bound} at $t=v_1(r)$ gives
\begin{align*}
    \OPT(r)&\le v_1(r)+r\sum_{i=1}^n\mathbb E[(X_i-v_1(r))^+]
    \le v_1(r)+r\sum_{i=1}^n\mathbb E[(X_i-\tau_i(r))^+]=2v_1(r).
\end{align*}
Thus $CR(r)=\OPT(r)/v_1(r)\le2$ for every fixed subsampling probability.
\end{proof}

We next show that this bound is again tight for every fixed rate.

\begin{proposition}[Single-rate tightness]
\label{prop:single_rate_tightness}
For every fixed $r\in(0,1]$, there exists a sequence of instances such that $CR(r)\to2$.
\end{proposition}

\begin{proof}
Fix $r\in(0,1]$ and choose $0<\varepsilon<1$. Consider an instance with two stages. The first stage consists of $k$ deterministic variables $X_1=X_2=\ldots=X_k=1$; the second stage consists of one rare large variable $Z$, where
\[
    Z=\begin{cases}
        \frac{1}{\varepsilon r}, & \text{with probability }\varepsilon,\\
        0, & \text{otherwise}.
    \end{cases}
\]
Under $r$-subsampling, the expected value of the final variable from the algorithm's perspective is
\[
    r\mathbb E[Z]=r\cdot \varepsilon\cdot \frac{1}{\varepsilon r}=1.
\]
Therefore, while processing the deterministic $1$'s, the optimal DP is indifferent between accepting a realized $1$ and continuing. More explicitly, the continuation value at the final stage is $1$. At each preceding stage, the current deterministic variable either fails to survive, in which case the continuation value remains $1$, or survives and offers a value equal to the continuation value. Working backward through all $k$ deterministic variables therefore gives $v_1(r)=1$.

We now calculate the prophet's value. The event that $Z$ both takes its large value and survives subsampling has probability $r\varepsilon$. On this event, the prophet receives $1/(r\varepsilon)$. On the complementary event, the prophet receives $1$ exactly when at least one of the $k$ deterministic variables survives. These events are independent, and the latter event has probability $1-(1-r)^k$. Hence
\[
    \OPT(r)=r\varepsilon\frac{1}{r\varepsilon}+(1-r\varepsilon)\bigl(1-(1-r)^k\bigr)=1+(1-r\varepsilon)\bigl(1-(1-r)^k\bigr).
\]
Because $r>0$, we have $(1-r)^k\to0$ as $k\to\infty$. Thus, for each fixed $\varepsilon>0$,
\[
    \OPT(r)\longrightarrow 2-r\varepsilon
    \qquad\text{as }k\to\infty.
\]
Sending $\varepsilon\to0$ gives $\OPT(r)\to2$. Since $v_1(r)=1$ throughout, it follows that $CR(r)\to2$.
\end{proof}

The construction places the online policy at an indifference point: a safe value $1$ and the rare final value have the same continuation value. Changing the sampling rate breaks this balance. The next subsection makes this observation precise.

\subsection{Exploiting slacks of the DP analysis}
\label{sec:prophet-dp-slacks}

\paragraph{Decomposition of the slack.} Write $v_r=v_1(r)$ for convenience. The last step of the proof of Proposition~\ref{prop:single_rate_competitive_ratio} has two possible sources of looseness: we use the relaxed upper bound for $\OPT(r)$, and we replace each $\mathbb E[(X_i-\tau_i(r))^+]$ by the smaller quantity $\mathbb E[(X_i-v_r)^+]$. Therefore, we can decompose the gap between $\OPT(r)$ and $2v_r$ into a threshold slack plus a prophet slack as follows:
\begin{align}
    2v_r-\OPT(r)={}&\underbrace{r\sum_{i=1}^n\left(\mathbb E[(X_i-\tau_i(r))^+]-\mathbb E[(X_i-v_r)^+]\right)}_{\operatorname{T\text{-}Slack}(r)} \notag\\
    &+\underbrace{\left(v_r+r\sum_{i=1}^n\mathbb E[(X_i-v_r)^+]-\OPT(r)\right)}_{\operatorname{P\text{-}Slack}(r)}. \label{eq:two_slack_decomposition}
\end{align}
Both slack terms are nonnegative. Since $\tau_i(r)\le v_r$, we have
$\mathbb E[(X_i-\tau_i(r))^+]\ge\mathbb E[(X_i-v_r)^+]$ for every $i$; hence each summand in the threshold slack is nonnegative. This slack measures the loss from replacing the DP thresholds $\tau_i(r)$ by the single global threshold $v_r$. The prophet slack is nonnegative by Claim~\ref{claim:prophet_relaxation} and measures the loss from applying that relaxation. Thus, whenever $v_r>0$, the competitive ratio is
\begin{equation} CR(r)=\frac{\OPT(r)}{v_r}=2-\frac{\operatorname{T\text{-}Slack}(r)+\operatorname{P\text{-}Slack}(r)}{v_r}. \label{eq:two_slack_ratio} \end{equation}
Thus a strict improvement over $2$ follows whenever at least one of the two slacks is at least a constant multiple of $v_r$.

\paragraph{The two slacks under subsampling.} The decomposition also explains why the tight instance from Proposition~\ref{prop:single_rate_tightness} works. The deterministic block has flat thresholds equal to $1$, so the threshold slack on this block is zero. The rare final item contributes only $r\varepsilon$ to the threshold slack. Meanwhile,
\[
    \OPT(r)=1+(1-r\varepsilon)(1-(1-r)^k),
\]
while
\[
    v_1(r)+r\sum_i\mathbb E[(X_i-v_1(r))^+]=2-r\varepsilon.
\]
Therefore the prophet slack is $(1-r\varepsilon)(1-r)^k$, which vanishes as $k\to\infty$. After sending $k\to\infty$ and then $\varepsilon\to0$, both slacks vanish and $CR(r)\to2$.

The same construction provides intuition of why two different rates cannot be simultaneously tight. First consider the tight instance at $r=p$, but evaluate it at the lower rate $q$, and write $\rho=q/p$. At rate $q$, the last-item continuation value is $\rho$. Moving backward across the deterministic block raises the DP value from roughly $\rho$ to $1$, while $v_1(q)\to1$ and $\OPT(q)\to1+\rho$. Hence $CR(q)\to1+\rho$. In the decomposition, the prophet slack is asymptotically negligible, but the thresholds are no longer flat: in particular, $v_1(q)>\tau_k(q)\approx\rho$. The resulting threshold slack satisfies $\operatorname{T\text{-}Slack}(q)\to1-\rho$.

Conversely, consider the tight instance at $q$, but evaluate it at $p$. Now the last-item continuation value is $1/\rho>1$, so the online algorithm rejects all deterministic $1$'s and $v_1(p)=1/\rho$. The thresholds before the last item are flat, so the threshold slack is asymptotically zero. However, the static-threshold upper bound uses the baseline $v_1(p)=1/\rho$, while the prophet often receives only a deterministic value $1$ below this baseline. This creates prophet slack, and in normalized form $\operatorname{P\text{-}Slack}(p)/v_1(p)\to1-\rho$.

\subsection{A simple two-rate improvement}
\label{sec:prophet-simple-bound}

In this section, we turn the intuition about the two slacks into a quantitative bound and show that, for two rates, the better competitive ratio is strictly smaller than $2$. The proof is intended to be simple and intuitive; the constants are not optimized.

\begin{theorem}
\label{thm:simple_two_rate_prophet}
For every finite instance and every $0<q<p\le1$, with $\rho=q/p$,
\[
    \min\{CR(q),CR(p)\}\le 2-\frac{\rho^2(1-\rho)}{27}<2.
\]
\end{theorem}

\begin{proof}
Write $v_r=v_1(r)$ and $\tau_i(r)=v_{i+1}(r)$, as above. Fix a
parameter $\delta$ with $0<\delta<(1-\rho)/2$; we choose its value at the
end of the proof. The nontriviality assumption gives $v_p,v_q>0$, so all
ratios below are well defined.

At a high level, we distinguish between two cases. If $v_q$ is close to $v_p$, then the
single-threshold relaxation at the lower rate $q$ using $v_p$ already improves the factor $2$. If $v_q$ is separated
from $v_p$, we place a threshold $T$ midway between them. When there is
little expected mass above $T$ at rate $p$, the prophet often receives a
value below $T$, and this creates prophet slack at rate $p$. When there is
substantial mass above $T$, the same instance creates threshold slack at rate
$q$. Thus either the two DP values are close and the lower rate improves, or
they are separated and one of the two kinds of slack is large.

We now quantify these three possibilities.

\paragraph{Case 1: $v_q\ge (1-\delta)v_p$.}
Apply the single-threshold relaxation \eqref{eq:prophet-relaxation-bound} at rate $q$
with $t=v_p$. This gives
\begin{align}
    \OPT(q)
    &\le v_p+q\sum_i\mathbb E[(X_i-v_p)^+] \notag\\
    &\le v_p+\rho p\sum_i\mathbb E[(X_i-\tau_i(p))^+] \notag\\
    &=v_p(1+\rho) \notag\\
    &\le \frac{v_q}{1-\delta}(1+\rho).
\label{eq:simple-close-rates-opt}
\end{align}
For the second line, use $q=\rho p$ and $\tau_i(p)\le v_p$, which follows
from the monotonicity of the DP values in
\eqref{eq:prophet_dp_recurrence}. Lowering the threshold can only increase
the positive part. The equality in the third line is the telescoping identity
\eqref{eq:prophet_dp_telescoping}, and the last line is the assumption of
Case~1. Dividing \eqref{eq:simple-close-rates-opt} by $v_q$ gives
\begin{equation}
    CR(q)\le \frac{1+\rho}{1-\delta}
    =2-\frac{1-\rho-2\delta}{1-\delta}.
\label{eq:simple_case_close}
\end{equation}
Our choice $\delta<(1-\rho)/2$ makes the improvement in
\eqref{eq:simple_case_close} positive.

\paragraph{Case 2: $v_q<(1-\delta)v_p$.}
Let $\Delta=v_p-v_q$ and set $T=(v_p+v_q)/2$. The case assumption gives
\begin{equation}
    v_p-T=T-v_q=\frac{\Delta}{2}>\frac{\delta v_p}{2}.
\label{eq:simple-midpoint-gap}
\end{equation}
For each item, write
$\pi_i=\Pr[X_i>T]$, and define
$m_r(T)=r\sum_i\pi_i$. This is the expected number of retained values above
$T$ at rate $r$, and in particular
\begin{equation}
    m_q(T)=\rho m_p(T).
\label{eq:simple-cross-rate-mass}
\end{equation}

\paragraph{Prophet slack.} If $m_p(T)$ is small, we can lower bound the prophet slack at rate $p$. Let
$M_p=\max_{i\in S_p}X_i$. The event $M_p>T$ requires at least one retained
item with value above $T$, so the union bound gives
\begin{equation}
    \Pr[M_p\le T]\ge 1-p\sum_i\Pr[X_i>T]=1-m_p(T).
\label{eq:simple-union-bound}
\end{equation}
Moreover, the sum of the excesses of all retained values dominates the excess
of their maximum. Taking expectations yields
\begin{equation}
    p\sum_i\mathbb E[(X_i-v_p)^+]
    =\mathbb E\!\left[\sum_{i\in S_p}(X_i-v_p)^+\right]
    \ge\mathbb E[(M_p-v_p)^+].
\label{eq:simple-max-excess}
\end{equation}
Using the definition of $\PSlack(p)$ in
\eqref{eq:two_slack_decomposition}, followed by
\eqref{eq:simple-max-excess}, we obtain
\begin{align}
    \PSlack(p)
    &\ge v_p+\mathbb E[(M_p-v_p)^+]-\mathbb E[M_p] \notag\\
    &=\mathbb E[(v_p-M_p)^+] \notag\\
    &\ge (v_p-T)\Pr[M_p\le T] \notag\\
    &\ge (v_p-T)(1-m_p(T)).
\label{eq:simple-prophet slack-lower}
\end{align}
The equality uses the pointwise identity
$v_p+(x-v_p)^+-x=(v_p-x)^+$. The next line restricts the expectation to
the event $M_p\le T$, where $(v_p-M_p)^+\ge v_p-T$, and the final line uses
\eqref{eq:simple-union-bound}. Consequently, whenever $m_p(T)\le1$,
\eqref{eq:two_slack_ratio}, \eqref{eq:simple-midpoint-gap}, and
\eqref{eq:simple-prophet slack-lower} give
\begin{equation}
CR(p)\le 2-\frac{\PSlack(p)}{v_p}
\le 2-\frac{\delta}{2}(1-m_p(T)).
\label{eq:simple_prophet_slack_case}
\end{equation}

\paragraph{Threshold slack.} Otherwise, if $m_p(T)$ is large, we can lower bound the threshold slack at rate $q$. Since
$T>v_q\ge\tau_i(q)$, the difference between the two positive parts is
nonnegative for every value of $X_i$, and on the event $X_i>T$ it equals
$v_q-\tau_i(q)$. Therefore
\begin{equation}
    \mathbb E[(X_i-\tau_i(q))^+]-\mathbb E[(X_i-v_q)^+]
    \ge \pi_i\bigl(v_q-\tau_i(q)\bigr).
\label{eq:simple-threshold-event}
\end{equation}
The gap $v_q-\tau_i(q)$ can itself be bounded by telescoping the DP
increments:
\begin{align}
    v_q-\tau_i(q)
    &=\sum_{j=1}^i\bigl(v_j(q)-v_{j+1}(q)\bigr) 
    =q\sum_{j=1}^i\mathbb E[(X_j-\tau_j(q))^+] \notag\\
    &\ge q\sum_{j=1}^i(T-\tau_j(q))\pi_j \notag\\
    &\ge q(T-v_q)\sum_{j=1}^i\pi_j.
\label{eq:simple-threshold-gap}
\end{align}
The second line uses the DP increment
\eqref{eq:prophet_dp_recurrence}. For the third line, restrict each
expectation to $X_j>T$; on this event,
$(X_j-\tau_j(q))^+\ge T-\tau_j(q)$. The final line follows from
$\tau_j(q)\le v_q$.

Combining the definition of $\TSlack(q)$ in
\eqref{eq:two_slack_decomposition} with
\eqref{eq:simple-threshold-event}, and \eqref{eq:simple-threshold-gap} gives
\begin{align}
    \TSlack(q)
    &\ge q^2(T-v_q)
    \sum_{i=1}^n\sum_{j=1}^i\pi_i\pi_j \notag\\
    &\ge \frac{q^2(T-v_q)}{2}\left(\sum_{i=1}^n\pi_i\right)^2 \notag\\
    &=\frac{T-v_q}{2}m_q(T)^2
    =\frac{\rho^2(T-v_q)}{2}m_p(T)^2.
\label{eq:simple-threshold slack-lower}
\end{align}
The first inequality uses the elementary identity
\begin{align*}
    \sum_{i=1}^n\sum_{j=1}^i\pi_i\pi_j
    =\frac12\left(\left(\sum_{i=1}^n\pi_i\right)^2
    +\sum_{i=1}^n\pi_i^2\right)
    \ge\frac12\left(\sum_{i=1}^n\pi_i\right)^2.
\end{align*}
The last equality uses \eqref{eq:simple-cross-rate-mass}. Since
$v_p>v_q>0$, \eqref{eq:simple-midpoint-gap} implies
$(T-v_q)/v_q>\delta/2$, equations
\eqref{eq:two_slack_ratio} and \eqref{eq:simple-threshold slack-lower} yield
\begin{equation}
    CR(q)\le 2-\frac{\TSlack(q)}{v_q}
    <2-\frac{\rho^2\delta}{4}m_p(T)^2.
\label{eq:simple_threshold_slack_case}
\end{equation}

\paragraph{Completing the proof.} We now combine the bounds. If $m_p(T)\le2/3$, then
\eqref{eq:simple_prophet_slack_case} gives
$CR(p)\le2-\delta/6$. If $m_p(T)>2/3$, then
\eqref{eq:simple_threshold_slack_case} gives
$CR(q)<2-\rho^2\delta/9$. Together with the close-values bound
\eqref{eq:simple_case_close}, these two alternatives show that
\begin{equation}
    \min\{CR(q),CR(p)\}
    \le 2-\min\left\{
    \frac{1-\rho-2\delta}{1-\delta},
    \frac{\delta}{6},
    \frac{\rho^2\delta}{9}
    \right\}.
\label{eq:simple-three-bounds}
\end{equation}
Finally, choose $\delta=(1-\rho)/3$. The three improvements in
\eqref{eq:simple-three-bounds} become
\begin{equation}
    \frac{1-\rho}{2+\rho}\geq \frac{1-\rho}{3},
    \qquad
    \frac{1-\rho}{18},
    \qquad
    \frac{\rho^2(1-\rho)}{27}.
\label{eq:simple-three-specialized-bounds}
\end{equation}
Because $0<\rho<1$,  it is easy to see that the third quantity is the smallest.
Therefore
\[
    \min\{CR(q),CR(p)\}\le 2-\frac{\rho^2(1-\rho)}{27}.
    \qedhere
\]
\end{proof}

\subsection{The tight two-rate result}
\label{sec:prophet-tight-result}

Theorem~\ref{thm:simple_two_rate_prophet} used one global cutoff $T$ and
only the total mass above it. That coarse summary establishes a
gap below $2$, while leaving information about the positions of the DP
thresholds unused. The sharp analysis retains that information.

Specifically, we use the following survival-conversion function to
compare the same collection of independent events at the two sampling rates.

\begin{definition}[Survival conversion function]
\label{def:Gamma}
For $0<q<p\le1$, define $\Gamma_{q,p}:[0,1]\to[0,1]$ by
\[
    \Gamma_{q,p}(\phi)
    =
    \inf
    \left\{
        \prod_{i=1}^n(1-p\theta_i):
        n\ge1,\ \theta_i\in[0,1],\
        \prod_{i=1}^n(1-q\theta_i)\ge \phi
    \right\}.
\]
We also set $\Gamma_{q,p}(0)=0$.
\end{definition}

For $\phi\in[0,1]$, define
\[
    f(\phi)=\frac {1+\rho-(1-\rho)\phi^2}{1-(1-\rho)\phi},
    \qquad
    g(\phi)=2-(1-\rho)\phi\Gamma_{q,p}(\phi).
\]
The tight simultaneous competitive ratio is
\begin{equation}
    \zeta(q,p)
    =
    \max_{\phi\in[0,1]}
    \min\left\{
        f(\phi),
        g(\phi)
    \right\}.
    \label{eq:prophet-zeta-definition}
\end{equation}

\begin{theorem}[Tight two-rate subsampled-prophet theorem]
\label{thm:tight_two_rate_prophet}
For every finite instance and every $0<q<p\le1$,
\[
    \min\{CR(q),CR(p)\}\le\zeta(q,p)<2.
\]
Moreover, for every $0<q<p\le1$, there is a sequence of finite instances for
which the left-hand side converges to $\zeta(q,p)$.
\end{theorem}

The supporting lemmas and the full proof appear in
Appendix~\ref{app:prophet-tight-proof}.

\paragraph{Proof roadmap.}
Write $a=v_1(q)$ and $b=v_1(p)$ for the algorithm's values at the two
sampling rates. We first show that
$\rho b\le a\le b$. We describe the position of $a$ in this interval by
$
    \phi=\frac{b-a}{(1-\rho)b}\in[0,1].
$
Thus $\phi=0$ when the two algorithmic values coincide, while $\phi=1$
when they have the largest separation allowed by the scaling inequality.
The proof reduces every instance to this single parameter.

To bound the prophet's value, we split each item relative to the continuation
thresholds at rates $q$ and $p$. The contribution above the higher threshold
is controlled directly by the dynamic program. The remaining contribution
lies between the two thresholds and is bounded by $b-a$. Replacing each such
contribution by an independent variable that is either $0$ or $b-a$ can only
increase their expected maximum. After this replacement, the relevant
quantity is the probability that all these variables are zero.

At rate $q$, the dynamic-programming recurrence makes the resulting product
telescope and shows that this probability is at least $\phi$. The function
$\Gamma_{q,p}$ converts this statement into a lower bound on the corresponding
probability at rate $p$. Substitution gives $CR(q)\le f(\phi)$ and
$CR(p)\le g(\phi)$. The first bound increases from $1+\rho$ to $2$ as $\phi$
runs from $0$ to $1$, while the second decreases from $2$ to $1+\rho$.
Their unique crossing therefore bounds at least one of the two competitive
ratios strictly below $2$.

For tightness, the construction in Appendix~\ref{app:prophet-tight-proof}
chooses middle-range events that attain the conversion bound
$\Gamma_{q,p}$, together with baseline and rare high-value items that make
the preceding inequalities approach equalities. Hence the crossing value
describes the exact frontier.

\begin{figure}[t]
\centering
\begin{tikzpicture}
\begin{axis}[
    width=0.5\textwidth,
    height=0.5\textwidth,
    axis equal image,
    xmin=1.77, xmax=2.02,
    ymin=1.77, ymax=2.02,
    axis lines=left,
    xlabel={$CR(q)$},
    ylabel={$CR(p)$},
    xtick={9/5,1.95265,2},
    xticklabels={$\frac95$,$c^\star\approx1.95265$,$2$},
    ytick={9/5,1.95265,2},
    yticklabels={$\frac95$,$c^\star$,$2$},
    samples=100,
    clip=false
]

\pgfplotsinvokeforeach{0,...,16}{
    \addplot[
        thick,
        blue,
        domain={0.6^(#1+1)}:{0.6^#1}
    ]
    (
        {(1.8 - 0.2*x^2)/(1 - 0.2*x)},
        {2 - 0.2*x*(0.5^#1)*(1 - 0.5*(1 - x/(0.6^#1))/0.4)}
    );
}

\addplot[dashed, gray, domain=1.77:2.02] {x};

\addplot[only marks, mark=*, mark size=2pt, blue]
coordinates {(9/5,2) (2,9/5)};

\addplot[only marks, mark=*, mark size=2.2pt, red]
coordinates {(1.95265,1.95265)};

\addplot[only marks, mark=*, mark size=2pt]
coordinates {(2,2)};

\node[anchor=south] at (axis cs:9/5,2)
{$\left(\frac95,2\right)$};

\node[anchor=west] at (axis cs:2,9/5)
{$\left(2,\frac95\right)$};

\node[anchor=east, red] at (axis cs:1.95265,1.95265)
{$(c^\star,c^\star)$};

\node[anchor=south east] at (axis cs:2,2)
{$(2,2)$};

{frontier};

\node[anchor=north west] at (axis cs:1.97,1.97)
{$y=x$};

\end{axis}
\end{tikzpicture}
\caption{Frontier of $(CR(q),CR(p))$ for $(q,p)=(0.4,0.5)$ and its intersection with $y=x$.}
\label{fig:left-hull-diagonal-p04-q05}
\end{figure}
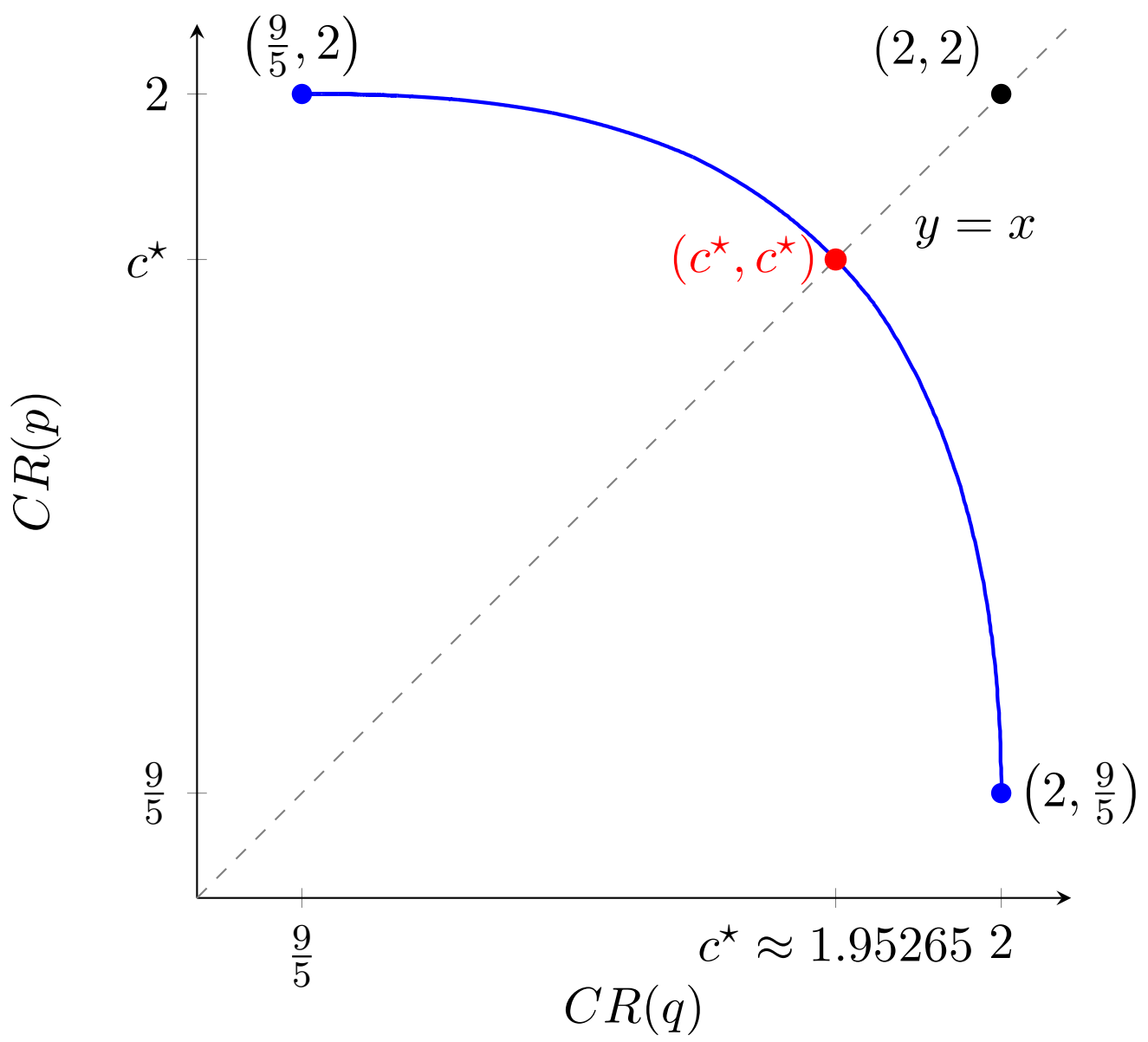

Figure~\ref{fig:left-hull-diagonal-p04-q05} illustrates the curve $\left(f(\phi),g(\phi)\right)$ parameterized by $\phi$ when $q=0.4,p=0.5$. It can be interpreted as follows: for each point $(x,y)$ on the curve, any instance such that $CR(q)=x$ must satisfy $CR(p)\leq y$.

Thus the feasible region of $(CR(q),CR(p))$ lies below the curve, which is strictly separated from the worst-case point $(2,2)$. The intersection of the curve with the line $y=x$ gives the global worst case for $\min\{CR(q),CR(p)\}$. In this case, we have $\min\{CR(q),CR(p)\}\leq 1.95265$.

\paragraph{A closed-form corollary.}
The exact value $\zeta(q,p)$ is given by the one-dimensional frontier above. The next corollary gives a simple closed-form upper bound depending only on the ratio $\rho=q/p$. It is not always tight, but it is explicit and strictly improves the classical factor $2$ for every $q<p$.

\begin{corollary}[Closed-form simultaneous bound]
\label{cor:closed_form_prophet}
For every finite instance and every $0<q<p\le1$, with $\rho=q/p$, we have
\[
\min\{CR(q),CR(p)\}\le B(\rho):=\frac{\rho^2-5\rho+12}{2(3-\rho)}=2-\frac{\rho(1-\rho)}{2(3-\rho)}<2.
\]
\end{corollary}

The proof of Corollary~\ref{cor:closed_form_prophet} also appears in Appendix~\ref{app:prophet-tight-proof}.

\subsection{Application to sequential posted pricing}
\label{sec:applications}

The prophet formulation also has a standard interpretation in single-item auction design. Consider a seller with one item and $n$ asymmetric bidders. Bidder $i$ has value $V_i$ drawn independently from a known absolutely continuous regular distribution $F_i$ with positive density $f_i$, finite optimal revenue, and a strictly increasing virtual-value function. Let $\varphi_i(v)=v-\frac{1-F_i(v)}{f_i(v)}$ denote bidder $i$'s virtual value. By Myerson's lemma, the expected revenue of any truthful mechanism equals its expected virtual surplus. Thus the revenue of the optimal auction is the expected maximum nonnegative virtual value:
\[
\mathbb E\left[\max_i \varphi_i(V_i)^+\right], \qquad \varphi_i(V_i)^+=\max\{\varphi_i(V_i),0\}.
\]

Sequential posted pricing mechanisms provide the corresponding online analogue. In a sequential posted pricing mechanism, bidders are approached in a fixed order and each bidder is offered a take-it-or-leave-it price. Under regularity, choosing a price is equivalent to choosing a threshold in virtual-value space. Thus, if we define $X_i=\varphi_i(V_i)^+$, then the optimal sequential posted pricing policy is the dynamic program for the prophet problem on the variables $X_i$; the continuation value $v_{i+1}$ determines the virtual threshold, and the posted price is $\varphi_i^{-1}(v_{i+1})$. This is the standard reduction between sequential posted pricing and prophet inequalities; see, e.g.,~\cite{chawla2010sequential}.

We now allow bidder participation to fluctuate. Under $r$-participation, each bidder independently participates with probability $r$. The mechanism knows $r$, and each bidder's participation is revealed when that bidder is reached. Let $\REV(r)$ denote the expected revenue of the Myerson auction run on the realized participating subset, and let $\SPM(r)$ denote the expected revenue of the optimal sequential posted pricing mechanism tuned to participation rate $r$. In virtual-value notation,
\[
    \REV(r)=\OPT(r), \qquad \SPM(r)=v_1(r).
\]
Therefore the results above apply directly to the competitive ratio $\REV(r)/\SPM(r)$.

\begin{corollary}[Two participation rates]
\label{cor:spm_two_rates}
Fix two participation rates $0<q<p\le1$, and let $\rho=q/p$. For every collection of bidder distributions satisfying the preceding assumptions,
\[
    \min\left\{\frac{\REV(q)}{\SPM(q)},\frac{\REV(p)}{\SPM(p)}\right\}\le \zeta(q,p)<2,
\]
where $\zeta(q,p)$ is the tight two-rate competitive ratio from Theorem~\ref{thm:tight_two_rate_prophet}. 
\end{corollary}

\begin{proof}
Apply Theorem~\ref{thm:tight_two_rate_prophet} to the nonnegative virtual-value variables $X_i=\varphi_i(V_i)^+$. For each participation rate $r$, the subsampled prophet benchmark $\OPT(r)$ is exactly the expected Myerson revenue on the participating subset, while the DP value $v_1(r)$ is exactly the revenue of the optimal sequential posted pricing mechanism tuned to rate $r$. Thus
\[
    \frac{\REV(r)}{\SPM(r)}=\frac{\OPT(r)}{v_1(r)}=CR(r).
\]
The stated inequalities are therefore immediate from Theorem~\ref{thm:tight_two_rate_prophet} and Corollary~\ref{cor:closed_form_prophet}.
\end{proof}

\section{Simultaneous Competitive Ratio for Ski Rental}
\label{sec:ski-rental}

This section studies simultaneous competitive ratio for the ski-rental problem. We first state the model and the objectives. We then move to proving our results for adversaries that choose a deterministic horizon, as well as those that choose a distribution over horizons.

Our approach starts with the classical view of the optimal randomized ski-rental algorithm as an equalizer: it achieves the same competitive ratio for all horizons. Our question asks whether this balance can persist when the same base horizon is viewed at two related lengths, $y$ and $\rho y$. To study this, we compare the cost of an algorithm at each horizon to the target competitive ratio and capture the gap via a residual curve. Against a deterministic adversary, it is enough for the residual to be nonpositive at one of $y$ and $\rho y$ for every $y$. A distributional adversary can mix horizons before the two ratios are compared, so this pointwise approach no longer suffices. The resulting condition couples the curve at $y$ and $\rho y$ in a single inequality. This relation drives both our constructions and our impossibility proofs, first through simple bounds and then through the sharp thresholds.

\subsection{Model and objectives}
\label{sec:ski-model-objective}

We first recall the ski-rental problem. A skier needs skis for an unknown horizon $y\geq 0$. Renting costs $r>0$ per unit time, while buying costs $b>0$ and permits free use thereafter. The skier learns the horizon only when it ends. Thus a deterministic online algorithm is specified by a buy time $x\in[0,\infty]$: it rents until time $x$ and, if $y \ge x$, buys at time $x$. Here $x=\infty$ means that the algorithm never buys. Its cost at horizon $y$ is
\[
c^{b,r}_x(y)=
\begin{cases}
ry, & y<x,\\
rx+b, & y\geq x,
\end{cases}
\qquad
c^{b,r}_\infty(y)=ry.
\]
The offline optimum knows the horizon in advance and either rents throughout or buys immediately. Its cost is therefore
\[
\OPT_{b,r}(y)=\min\{ry,b\}.
\]

The breakeven time is $b/r$. Scaling costs by $b$ and time by $b/r$, define the normalized horizon and buy time by
$\widehat y=\frac{ry}{b}$
and $\widehat x=\frac{rx}{b}$.
Under this normalization, renting costs one per unit time and buying costs one. We henceforth drop the hats and write
\[
c_x(y)=
\begin{cases}
y, & y<x,\\
x+1, & y\geq x,
\end{cases}
\qquad
c_\infty(y)=y,
\]
and
\[
\OPT(y)=\min\{y,1\}.
\]

A randomized ski-rental algorithm is specified by a distribution over a buy time $X\in[0,\infty]$. If $F$ is the CDF of $X$, its expected cost at horizon
$y$ is
$
C_F(y)=\mathbb E[c_X(y)].
$
For $y>0$, its pointwise competitive ratio is
\[
CR_F(y)=\frac{C_F(y)}{\OPT(y)}.
\]
We write
$R_{\mathrm{ski}}=\frac{e}{e-1}$
for the classical optimal randomized competitive ratio.

\paragraph{Two-rate model.} In our model, the adversary now chooses a base horizon $B\ge0$. We use the fluid approximation in which the random retained duration at rate $p$ is replaced by its law-of-large-numbers limit $pB$. Clearly, if there is only one subsampling rate $p$, it is impossible to do better than the classical competitive ratio of $\frac{e}{e-1}$, because the adversary can simply scale the input horizon by a factor of $1/p$.

Therefore, we consider two different subsampling rates $1 \ge p>q>0$, so the two realized horizons are $pB$ and $qB$. It is convenient to normalize the larger realized horizon by setting $y=pB$. Throughout this section we write $\rho=q/p\in(0,1)$, and use the equivalent setting where the adversary chooses $y$ and the two realized horizons are $y$ and $\rho y$. 

For a deterministic adversary, the goal is to find one algorithm $F_\rho$ and one constant $\eta>0$ such that, for every $y>0$,
\[
\min\{CR_{F_\rho}(y),CR_{F_\rho}(\rho y)\}\le R_{\mathrm{ski}}-\eta.
\]
For a distributional adversary, the adversary chooses a probability distribution $\mu$ over base horizons. The objective measures the ratio of expected costs
\[
CR_{F}(\mu)
=
\frac{\mathbb E_{y\sim\mu}[C_{F}(y)]}
     {\mathbb E_{y\sim\mu}[\OPT(y)]}.
\]
For $y\sim\mu$, let $\rho\mu$ denote the distribution of $\rho y$. The goal is to find $F_\rho$ and $\eta>0$ such that every nondegenerate\footnote{A distribution is nondegenerate if it is not concentrated
at $y=0$.} $\mu$ satisfies
\[
\min\{CR_{F_\rho}(\mu),CR_{F_\rho}(\rho\mu)\}
\le R_{\mathrm{ski}}-\eta.
\]

Several positive results below use algorithms whose buy times are supported on $[0,1]$. The impossibility results state their support and assumptions explicitly, and as we show later, unlike classic ski rental, a support on $[0,1]$ is not without loss of generality in our case.

\subsection{Residual curves and realization}
\label{sec:ski-residual-realization}

Our analysis starts from the classic randomized ski-rental algorithm. It chooses a buy time in $[0,1]$ according to the density $f^\star(x)=e^x/(e-1)$, with CDF $F^\star(x)=(e^x-1)/(e-1)$ on $[0,1]$. It is an equalizer: for every realized horizon $y$, its expected cost is $R_{\mathrm{ski}}\OPT(y)$~\cite{karlin1994nonuniform,karlin2003tcp,borodin1998online}. 

Since our goal is to beat $R_{\mathrm{ski}}$ in the two-rate model, we need to deviate from the equalizer. It is convenient to measure cost relative to an improved target. Given an algorithm $F$ and a target improvement $\eta>0$, define its residual at ratio $R_{\mathrm{ski}}-\eta$ by
\[
A_{F,\eta}(y):=C_F(y)-(R_{\mathrm{ski}}-\eta)\OPT(y).
\]
Then at every realized horizon $y>0$,
\[
CR_F(y)=R_{\mathrm{ski}}-\eta+\frac{A_{F,\eta}(y)}{\OPT(y)}.
\]
Thus the target ratio is met at horizon $y$ exactly when $A_{F,\eta}(y)\leq0$. Since the equalizer is optimal in the classical setting, we cannot hope to achieve $A_{F,\eta}(y)\leq 0$ at every horizon $y$. Fortunately, this is also not required in the two-rate model: an algorithm only needs to achieve the target ratio for an appropriate subset of horizons. As we will show later, for deterministic adversaries, for each $y>0$ the algorithm only needs to ensure that at least one of $A_{F,\eta}(y)\leq 0$ and $A_{F,\eta}(\rho y)\leq 0$ holds; for distributional adversaries, a stronger condition is needed.

In this section, we reduce designing ski-rental algorithms to designing the residual curve. First, we leverage the exponential distribution, which was used to prove the lower bound for classic ski-rental, to derive a necessary condition (Lemma~\ref{lem:positive-compensation-mass}) for what residual curves can be realized by a ski-rental algorithm. Next, we prove a weaker reverse direction: under mild additional conditions, every such curve can be converted back to a ski-rental algorithm that preserves a constant fraction of the residual at every horizon. Subsequent sections thus focus on showing possibility or impossibility of constructing a residual curve that simultaneously satisfies the condition in Lemma~\ref{lem:positive-compensation-mass} and the additional conditions for ensuring a simultaneous competitive ratio better than $R_{\mathrm{ski}}$.

\paragraph{Necessary condition of residual curves.} For any measurable $A$ satisfying $\int_0^\infty |A(y)|e^{-y}dy<\infty$, define its exponentially weighted mass by
\[
\mathcal L(A):=e\int_0^\infty A(y)e^{-y}dy.
\]

\begin{lemma}[Positive weighted mass]
\label{lem:positive-compensation-mass}
Let $F$ be a randomized ski-rental algorithm, let $R<R_{\mathrm{ski}}$, and let $A(y)=C_F(y)-R\cdot \OPT(y)$. Then $\mathcal L(A)>0$.
\end{lemma}

\begin{proof}
Suppose the horizon $y$ is drawn from the distribution on $[0,\infty)$ with density $e^{-y}$. Then
\[
\int_0^{\infty}\OPT(y)e^{-y}dy=\int_0^1ye^{-y}dy+\int_1^{\infty}e^{-y}dy=1-\frac1e.
\]
For any finite deterministic buy time $x\geq0$,
\[
\int_0^{\infty}c_x(y)e^{-y}dy=\int_0^xye^{-y}dy+\int_x^{\infty}(x+1)e^{-y}dy=1.
\]
The same identity holds for $x=\infty$, since then $c_\infty(y)=y$. Therefore, for any randomized algorithm $F$, including one with arbitrary buying times, Tonelli's theorem gives
\[
\int_0^{\infty}C_F(y)e^{-y}dy=\mathbb E_{X\sim F}\left[\int_0^{\infty}c_X(y)e^{-y}dy\right]=1.
\]
It follows that
\begin{align*}
\mathcal L(A) &= e\int_0^{\infty}C_F(y)e^{-y}dy - R\cdot e\int_0^{\infty}\OPT(y)e^{-y}dy\\
&=e-R(e-1)=(R_{\mathrm{ski}}-R)(e-1)>0. \qedhere
\end{align*}
\end{proof}

In particular, for the residual at target ratio $R=R_{\mathrm{ski}}-\eta$, $\mathcal L(A_{F,\eta})=(e-1)\eta$. Thus a candidate algorithm must leave a positive amount of residual somewhere. 

\paragraph{Realizing a residual curve.} We next give a partial converse to Lemma~\ref{lem:positive-compensation-mass}. For a candidate residual curve $A$ with positive weighted mass, define
\[
\eta_A:=\frac{\mathcal L(A)}{e-1},
\]
and
\[
\Phi_A(y):=A(y)-\eta_A\OPT(y)+e^y\int_0^y\bigl(A(x)-\eta_A\OPT(x)\bigr)e^{-x}dx.
\]
The function $\Phi_A$ is the perturbation of the classical buying CDF corresponding to $A$: under mild regularity conditions, $F^\star+\varepsilon\Phi_A$ is a valid buying distribution whose residual is $\varepsilon A$.

\begin{lemma}[Realizing a residual curve]
\label{lem:certificate-realization}
Let $A:[0,\infty)\to\mathbb R$ be right-continuous and piecewise continuously differentiable with finitely many pieces and jumps. Suppose that $A(0)=0$, $\mathcal L(A)>0$, and $A$ has a finite limit as $y\to\infty$. Suppose every jump of $A$ at or below time $1$ is upward. Suppose also
that there is a finite $M\ge0$ such that
$
    \Phi_A'(y)\ge -M
$
at every point $y\in(0,1)$ where $\Phi_A$ is differentiable, and that
$\Phi_A$ is nondecreasing on $[1,\infty)$. Then, for every sufficiently small $\varepsilon>0$, there is a randomized ski-rental algorithm whose residual curve at target ratio $R_{\mathrm{ski}}-\varepsilon\eta_A$ is $\varepsilon A$. If $A$ is constant on $[1,\infty)$, the resulting algorithm is supported on $[0,1]$.
\end{lemma}

The conditions in Lemma~\ref{lem:certificate-realization} have a direct interpretation in terms of the buying distribution. Below the breakeven time, the classical algorithm has strictly positive density, so sufficiently small bounded perturbations remain feasible; upward jumps of $A$ become positive buying atoms because $\Phi_A$ has the same jumps. Above the breakeven time, the classical CDF is already flat at $1$, so the perturbation $\Phi_A$ itself must be nondecreasing. The proof and a few convenient admissibility checks are given in Appendix~\ref{sec:curve-realization-appendix}.

\subsection{Deterministic adversaries}
\label{sec:ski-deterministic}

We show that for deterministic adversaries, the simultaneous competitive ratio beats $e/(e-1)$, even if the algorithm only knows an upper bound $\alpha$ on $\rho=q/p$, assuming $\alpha<1$. This serves as a warm-up for our main results on distributional adversaries.

We first note that against deterministic adversaries, buying by the breakeven time is without loss of generality. The proof appears in Appendix~\ref{app:extra}.

\begin{lemma}
\label{lem:det-wlog}
When the adversary chooses a deterministic horizon, restricting attention to randomized algorithms that buy by breakeven time is without loss of generality.
\end{lemma}

In the two-rate model, after the adversary chooses horizon $y$, an algorithm is evaluated at both $y$ and $\rho y$ and it only needs to beat $R_{\mathrm{ski}}$ at one of them. Therefore, a residual curve at target ratio $R_{\mathrm{ski}}-\eta$ succeeds against a deterministic adversary whenever
\[
\min\{A_{F,\eta}(y),A_{F,\eta}(\rho y)\}\leq0
\qquad\text{for every }y>0.
\]
This suggests a simple construction that places all positive residual inside a single interval that is narrow enough so that for any $y>0$, the interval cannot contain both $y$ and $\rho y$. Elsewhere the residual is simply set to zero.

\begin{theorem}[Deterministic simultaneous improvement]
\label{thm:deterministic-simultaneous-improvement}
Fix $\alpha\in(0,1)$. There exists a randomized ski-rental algorithm $F$ supported on $[0,1]$, and a constant $\varepsilon>0$ such that for every $0<\rho\leq\alpha$ and every base horizon $y>0$,
\[
\min\{CR_F(y),CR_F(\rho y)\}\leq\frac{e}{e-1}-\varepsilon.
\]
\end{theorem}

\begin{proof}
Set $\ell=\alpha/2$ and $u=1/2$, so that $\ell=\alpha u$. Let
\[
A(y)=
\begin{cases}
\kappa (y-\ell)^2(u-y)^2, & \ell<y<u,\\
0, & \text{otherwise},
\end{cases}
\]
where $\kappa>0$ is chosen so that $\mathcal L(A)=e-1$. Then $\eta_A=1$.

For any $y>0$ and $0<\rho\leq\alpha$, the two horizons $y$ and $\rho y$ cannot both lie in $(\ell,u)$: otherwise $\rho y>\ell=\alpha u>\alpha y$, contradicting $\rho\leq\alpha$. Hence at least one of $A(y)$ and $A(\rho y)$ is zero.

We now check the conditions of Lemma~\ref{lem:certificate-realization}.
The curve $A$ is continuously differentiable, satisfies $A(0)=0$, has no jumps, and converges to zero. 
Since $\eta_A=1$, the derivative $\Phi_A'$ is bounded below on $(0,1)$ because $A$ and $A'$ are bounded there. Moreover,
\[
\int_0^\infty\bigl(A(x)-\OPT(x)\bigr)e^{-x}dx=0,
\]
because $\mathcal L(A)=e-1=\mathcal L(\OPT)$. Since $A(x)=0$ and $\OPT(x)=1$ for $x\ge1$, for every $y\ge1$ we have
\[
\Phi_A(y)
=-1-e^y\int_y^\infty(-1)e^{-x}dx
=0.
\]
Thus $\Phi_A$ is nondecreasing on $[1,\infty)$, and all the conditions of Lemma~\ref{lem:certificate-realization} hold. 
It follows that there is a
constant $\varepsilon>0$, possibly depending on $\alpha$, such that
$\varepsilon A$ is the residual curve of an algorithm supported on
$[0,1]$ at target ratio $R_{\mathrm{ski}}-\varepsilon$. Since at least one
of $A(y)$ and $A(\rho y)$ is zero, this algorithm achieves the claimed
simultaneous competitive ratio.
\end{proof}

\subsection{Distributional adversaries: Preliminaries}
\label{sec:ski-dist-prelim}

We now move to our main results about distributional adversaries, where we show a sharp threshold of $\rho$ between possibility and impossibility of improving the competitive ratio. Further, unlike the deterministic case, buying by breakeven time is no longer optimal.

The distributional model uses the same residual $A_{F,\eta}$ as above, but the adversary may mix several horizons before the two rates are compared. For a deterministic horizon $y$, the target ratio is achieved whenever $A_{F,\eta}$ is nonpositive at either $y$ or $\rho y$. For a distribution over horizons, the analogous condition concerns the expected residual at the two scales. This distinction is the source of the additional difficulty.

\subsubsection{Why distributional adversaries are harder}

A distributional adversary can mix horizons that are bad in different ways. Recall that $\mu$ is the distribution of the base horizon $y$ chosen by the adversary. An algorithm beats $R_{\mathrm{ski}}$ by $\eta>0$ if and only if
\begin{equation}
\label{eq:two-scale-begin}
\min\left\{\mathbb E_{y\sim\mu}[A_{F,\eta}(y)],\mathbb E_{y\sim\mu}[A_{F,\eta}(\rho y)]\right\}\leq0
\qquad\text{for every }\mu.
\end{equation}
For a deterministic adversary, the residual curve only needs at least one of $A_{F,\eta}(y)$ and $A_{F,\eta}(\rho y)$ to be nonpositive at every $y$. The following example shows why the distributional condition~\eqref{eq:two-scale-begin} is stronger.

Suppose $\rho=1/2$, and for some target improvement $\eta$, the curve $A_{F,\eta}$ has values
\[
A_{F,\eta}(1/4)=-0.1,\qquad A_{F,\eta}(1/2)=0.2,\qquad A_{F,\eta}(1)=-0.1.
\]
Then the deterministic condition holds at both $y=1/2$ and $y=1$. However, a distributional adversary can put probability $1/2$ on $y=1/2$ and probability $1/2$ on $y=1$, giving
\[
\mathbb E_y[A_{F,\eta}(y)]=\frac12A_{F,\eta}(1/2)+\frac12A_{F,\eta}(1)=0.05>0,
\]
and
\[
\mathbb E_y[A_{F,\eta}(\rho y)]=\frac12A_{F,\eta}(1/4)+\frac12A_{F,\eta}(1/2)=0.05>0.
\]
Thus the algorithm violates~\eqref{eq:two-scale-begin} and fails to beat $R_{\mathrm{ski}}$ against this distributional adversary.

\subsubsection{A stronger condition from Farkas' lemma}

The above counterexample indicates that we need a stronger condition on the residual curve. We use the following special case of Farkas' lemma.

\begin{lemma}
\label{lem:two-dimensional-farkas}
Let $V\subseteq\mathbb R^2$. Exactly one of the following statements holds.
\begin{enumerate}
\item The convex hull of $V$ intersects $\mathbb R_{>0}^2$.
\item There exists a nonzero vector $\lambda=(\lambda_1,\lambda_2)\in\mathbb R_{\geq0}^2$ such that
\[
\lambda\cdot v\leq0
\qquad\text{for every }v\in V.
\]
\end{enumerate}
In words, either $conv(V)$ intersects $\mathbb R_{>0}^2$, or there is a hyperplane passing through $(0,0)$ that separates it from $\mathbb R_{>0}^2$.
\end{lemma}

\begin{lemma}
\label{lem:two-scale-farkas-certificate}
A residual curve $A=A_{F,\eta}$ satisfies \eqref{eq:two-scale-begin} if and only if there exists $\gamma>0$ such that
\begin{equation}
A(y)+\gamma A(\rho y)\leq0
\qquad\text{for every }y\geq0.
\label{eq:two-scale-A}
\end{equation}
\end{lemma}
\begin{proof}
For brevity, write $A:=A_{F,\eta}$ and
$R:=R_{\mathrm{ski}}-\eta$.

First suppose that $A$ satisfies \eqref{eq:two-scale-begin}.  We begin
with the endpoint $y=0$.  Let $p:=\Pr[X=0]$.  For every $0<y\le1$, the
event $X=0$ incurs cost $1$, and hence
\[
    A(y)=C_F(y)-Ry\ge p-Ry.
\]
If $p>0$, then both $A(y)$ and $A(\rho y)$ are positive for all
sufficiently small $y>0$.  Taking $\mu$ to be the point mass at such a
$y$ contradicts \eqref{eq:two-scale-begin}.  Therefore $p=0$, which
implies $C_F(0)=A(0)=0$.

Now set
\[
    V
    :=
    \left\{
        \bigl(A(y),A(\rho y)\bigr):y>0
    \right\}.
\]
Every finitely supported probability distribution on $(0,\infty)$ is
nondegenerate.  Its expectation vector is a finite convex combination
of points in $V$, and every point in $\operatorname{conv}(V)$ arises in
this way.  Hence \eqref{eq:two-scale-begin} implies that
$\operatorname{conv}(V)$ does not intersect $(0,\infty)^2$.
Lemma~\ref{lem:two-dimensional-farkas} therefore gives
$\lambda_1,\lambda_2\ge0$, not both zero, such that
\[
    \lambda_1A(y)+\lambda_2A(\rho y)\le0
    \qquad\text{for every }y>0.
\]

If $A$ is nonpositive on $(0,\infty)$, then any $\gamma>0$ gives the two-scale inequality~\eqref{eq:two-scale-A} directly.  Otherwise both separating coefficients are positive.  Indeed, if $\lambda_1=0$, then $A(\rho y)\le0$ for every $y>0$.  Since $\rho y$ ranges over all of $(0,\infty)$, this would make $A$ nonpositive everywhere on that interval.  The case $\lambda_2=0$ is similar.  We may therefore divide by $\lambda_1$ and set $\gamma:=\lambda_2/\lambda_1>0$, obtaining
\[
    A(y)+\gamma A(\rho y)\le0
    \qquad\text{for every }y>0.
\]
Since $A(0)=0$, the same inequality also holds at $y=0$.

Conversely, suppose that such a $\gamma>0$ exists that makes the two-scale inequality~\eqref{eq:two-scale-A} true.  We first verify that $A$ is bounded.  For $0\le y\le1$, every deterministic buying time has cost at most $y+1\le2$, so $C_F(y)\le2$ and $A$ is bounded on $[0,1]$. For $t\ge1$, applying \eqref{eq:two-scale-A} at $t/\rho$ and using $C_F(t/\rho)\ge C_F(t)$ gives $0\ge A(t/\rho)+\gamma A(t)\ge(1+\gamma)A(t)$. Thus $-R\le A(t)\le0$ on $[1,\infty)$, so $A$ is bounded there as well. Consequently, all expectations below are finite for every probability
distribution $\mu$.  

Taking expectations in the two-scale inequality~\eqref{eq:two-scale-A}
gives
\[
    \mathbb E_{y\sim\mu}[A(y)]
    +
    \gamma\mathbb E_{y\sim\mu}[A(\rho y)]
    \le0.
\]
The two expectations therefore cannot both be positive, proving
\eqref{eq:two-scale-begin}.
\end{proof}

Combining Lemma~\ref{lem:positive-compensation-mass} and Lemma~\ref{lem:two-scale-farkas-certificate} reduces the rest of our analysis to examining residual curves under the following two conditions:
\[
\mathcal L(A)>0,
\qquad
A(y)+\gamma A(\rho y)\leq0
\quad(y\geq0).
\]
We refer to the second condition as the \emph{two-scale inequalities}. Positive results construct admissible residual curves satisfying both conditions, while negative results show that the two-scale inequalities force $\mathcal L(A)\leq0$. The following additional consequence is used repeatedly in the impossibility proofs.

\begin{lemma}[Tail of a residual curve]
\label{lem:compensation-tail}
Let $F$ be a randomized ski-rental algorithm with simultaneous distributional competitive ratio at most $R<R_{\mathrm{ski}}$ at the two scales, and set $A(y)=C_F(y)-R\OPT(y)$. Then $A(t)\leq0$ for every $t\geq1$. Consequently, $C_F$ is bounded, the algorithm buys eventually almost surely, and
\[
\lim_{t\to\infty}C_F(t)=\mathbb E[X+1]<\infty.
\]
\end{lemma}
\begin{proof}
Set $\eta=R_{\mathrm{ski}}-R>0$, so that $A=A_{F,\eta}$. By
Lemma~\ref{lem:two-scale-farkas-certificate}, there exists $\gamma>0$
for which $A$ satisfies \eqref{eq:two-scale-A}.
Fix $t\geq1$ and apply the two-scale inequality~\eqref{eq:two-scale-A} at $y=t/\rho$. Both offline optima equal $1$, and the cost curve $C_F$ is nondecreasing, so
\[
A(t/\rho)=C_F(t/\rho)-R\geq C_F(t)-R=A(t).
\]
Therefore
\[
0\geq A(t/\rho)+\gamma A(t)\geq(1+\gamma)A(t),
\]
which proves $A(t)\leq0$.

It follows that $C_F(t)\leq R$ for $t\geq1$, so the cost curve is bounded. For every buy time $X$, the deterministic cost $c_X(t)$ increases to $X+1$ as $t\to\infty$, with the value $+\infty$ when $X=\infty$. Monotone convergence now gives $\lim_{t\to\infty}C_F(t)=\mathbb E[X+1]<\infty$, which also shows that the algorithm buys eventually almost surely.
\end{proof}

\subsection{Distributional adversaries: Main theorems}
\label{sec:ski-randomized-results}

Our main result shows that the distributional model has a sharp phase transition in the scale ratio $\rho$. There is a threshold value such that below it, one can beat the classical competitive ratio simultaneously, while above it every algorithm is stuck at the classical ratio. For algorithms that buy by the breakeven time, the threshold is $\rho_\star\approx0.311$. For arbitrary algorithms, the threshold is $\rho_{\mathrm{all}}\approx0.338$. 

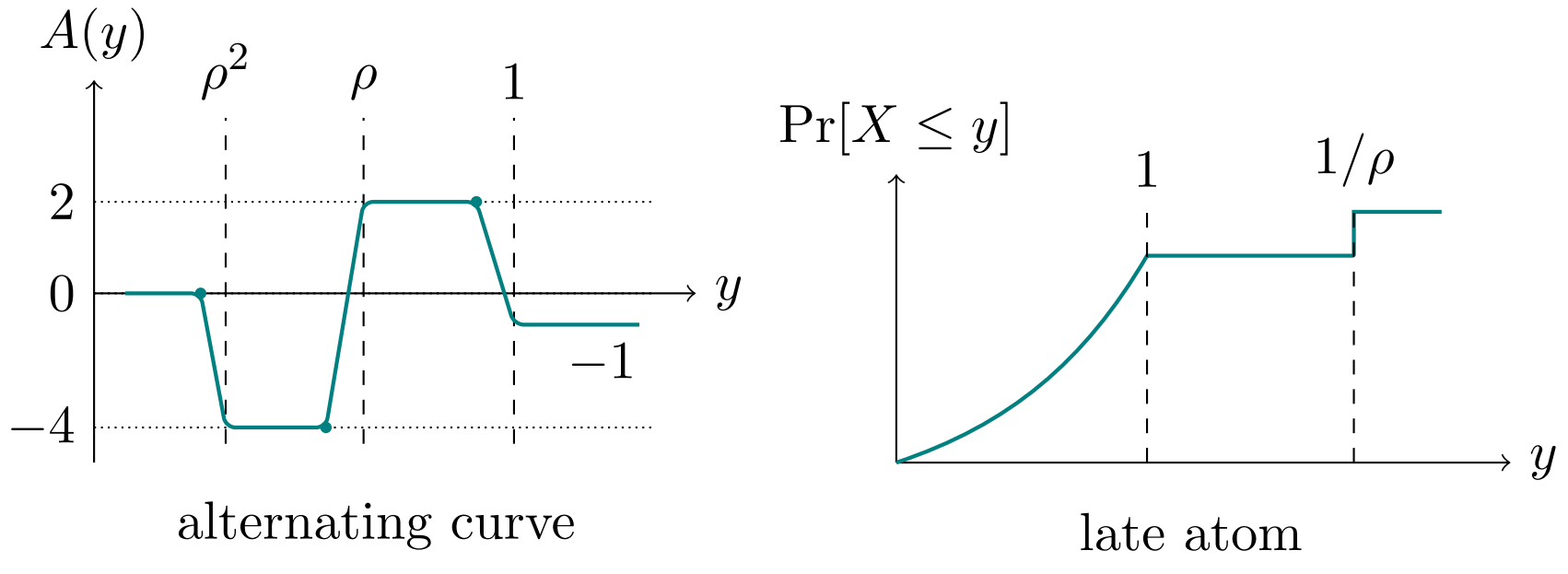
\begin{figure}[htbp]
\centering
\begin{tikzpicture}[scale=0.9]
\begin{scope}
\draw[->] (0,1.35) -- (4.8,1.35) node[right] {$y$};
\draw[->] (0,0) -- (0,3.05) node[above] {$A(y)$};
\draw[densely dotted] (0,2.08) -- (4.45,2.08);
\draw[densely dotted] (0,1.35) -- (4.45,1.35);
\draw[densely dotted] (0,0.28) -- (4.45,0.28);
\node[left] at (0,2.08) {$2$};
\node[left] at (0,1.35) {$0$};
\node[left] at (0,0.28) {$-4$};
\draw[dashed] (1.05,0.15) -- (1.05,2.75) node[above] {$\rho^2$};
\draw[dashed] (2.15,0.15) -- (2.15,2.75) node[above] {$\rho$};
\draw[dashed] (3.35,0.15) -- (3.35,2.75) node[above] {$1$};
\draw[thick, teal, rounded corners=2pt]
    (0.25,1.35) -- (0.85,1.35)
    -- (1.05,0.28)
    -- (1.85,0.28)
    -- (2.15,2.08)
    -- (3.05,2.08)
    -- (3.35,1.10)
    -- (4.35,1.10);
\fill[teal] (0.85,1.35) circle (1.3pt);
\fill[teal] (1.85,0.28) circle (1.3pt);
\fill[teal] (3.05,2.08) circle (1.3pt);
\node[below] at (4.05,1.10) {$-1$};
\node at (2.25,-0.5) {alternating curve};
\end{scope}
\begin{scope}[xshift=6.4cm]
\draw[->] (0,0) -- (4.9,0) node[right] {$y$};
\draw[->] (0,0) -- (0,2.3) node[above] {$\Pr[X\le y]$};
\draw[thick, teal] (0,0)
    .. controls (0.9,0.3) and (1.5,0.8) .. (2.0,1.65)
    -- (3.65,1.65);
\draw[thick, teal] (3.65,1.65) -- (3.65,2.0) -- (4.35,2.0);
\draw[dashed] (2.0,0) -- (2.0,2.05) node[above] {$1$};
\draw[dashed] (3.65,0) -- (3.65,2.05) node[above] {$1/\rho$};
\node at (2.35,-0.55) {late atom};
\end{scope}
\end{tikzpicture}
\caption{Two shapes used in the distributional analysis. The left panel
shows the smoothed version of the four-level residual curve
behind Theorem~\ref{thm:warmup-positive}. The right panel shows the
late-atom buy-time CDF shape available to arbitrary algorithms, which is
why their sharp threshold is larger.}
\label{fig:ski-buy-shapes}
\end{figure}

Two aspects make our results interesting. First, feasibility turns out to have a sharp one-dimensional boundary in $\rho$, even though the problems at different values of $\rho$ are not naturally nested: changing $\rho$ changes the two scales coupled by the two-scale inequality~\eqref{eq:two-scale-A}, and a construction for one ratio does not directly yield a construction for another. The threshold instead emerges from the common primal--dual recurrence underlying our proofs. Second, and more interestingly, there is a strict gap between these constants: buying by the breakeven time is not without loss of generality for distributional adversaries. The right panel of Figure~\ref{fig:ski-buy-shapes} previews the extra freedom: an arbitrary algorithm can buy at a single time $T>1$ with small probability.

Specifically, we show the following two theorems. The first theorem is proved in Appendix~\ref{app:tight-buy-by-one}, and the second in Appendix~\ref{app:tight-general}.

\begin{theorem}[Known ratio, buy by the breakeven time]
\label{thm:tight-buy-by-one}
For a fixed $\rho$, there exists a randomized ski-rental algorithm supported on $[0,1]$ whose simultaneous distributional competitive ratio is strictly below $e/(e-1)$ if and only if
$
    \rho<\rho_\star \approx0.311.
$
\end{theorem}

\begin{theorem}[Known ratio, arbitrary algorithms]
\label{thm:tight-general}
For a fixed $\rho$, there exists a randomized ski-rental algorithm with arbitrary buying times whose simultaneous distributional competitive ratio is strictly below $e/(e-1)$ if and only if
$
    \rho<\rho_{\mathrm{all}} \approx0.338.
$
\end{theorem}

To build towards these results, as a warm up, in Sections~\ref{sec:1/10} and~\ref{sec:3/5} we present simple versions of the positive and impossibility constructions and prove two suboptimal bounds. We then show in Section~\ref{sec:strict-late-buying-separation} that at $\rho = 1/3$, there is a separation between the two algorithm classes, and finally show in Section~\ref{sec:imposs_unknown} that knowing $\rho $ is necessary to derive any improvement over $e/(e-1)$.

\paragraph{Proof idea.}
\label{sec:ski-warmup}
At a high level, the positive and negative arguments are two sides of the same coin. As mentioned before, Lemma~\ref{lem:two-scale-farkas-certificate} shows that a successful residual curve satisfies \eqref{eq:two-scale-A} for some $\gamma>0$, while Lemma~\ref{lem:positive-compensation-mass} shows that it must also satisfy $\mathcal L(A)>0$.  For a positive result, we construct $A$ by making the two-scale inequalities~\eqref{eq:two-scale-A} tight wherever possible: positive residual at one scale is canceled by negative residual at the next smaller scale, producing the oscillating pattern seen in Figure~\ref{fig:ski-buy-shapes}. 
For a negative result, we use the same inequalities in the opposite direction: we combine the two-scale inequalities~\eqref{eq:two-scale-A} with nonnegative dual weights so that their sum controls the weighted mass of $A$ and make $\mathcal{L}(A)>0$ impossible. 

Technically, the sharp thresholds are derived via refinements of the ideas presented for the warm-up settings in Sections~\ref{sec:1/10},~\ref{sec:3/5}, and~\ref{sec:strict-late-buying-separation}. In Lemma~\ref{lem:general-weighted-tail}, we extend the dual-weight framework and upper bound $\mathcal{L}(A)$ by a weighted integral that only depends on the tail of the residual $A$ after time $1$. In Lemma~\ref{lem:general-recursive-extension} we present a general construction of the oscillating residual curve, which starts from a prescribed tail after time $1$ and extends it recursively to earlier scales by making the two-scale inequalities tight. As the number of oscillations tends to infinity, the weighted mass of the resulting curves converges to exactly the same expression as the upper bound in Lemma~\ref{lem:general-weighted-tail}. Thus the negative and positive arguments match symbolically: $\mathcal{L}(A)$ can be made positive if and only if there is a residual tail that makes the expression positive. The remaining work in obtaining the sharp thresholds is therefore to determine for which regime of $\rho$ is this possible, and it turns out that there is a single threshold separating possibility and impossibility. The proofs use specific analytic properties of the dual weights that we verify using Lean.

This perspective also explains why buying by the breakeven time can be suboptimal in the two-rate model. If the algorithm buys by time $1$, its residual is constant after time $1$ and Lemma~\ref{lem:general-weighted-tail} would imply a stronger upper bound on $\mathcal{L}(A)$, while a general algorithm can induce a non-constant residual tail. We give a concrete illustration of the strict benefit of buying after the breakeven time for $\rho=1/3$, where we show that constant tails cannot give positive weighted mass, while a linear tail corresponding to buying at time $3$ does.

\subsubsection{Simple possibility result for $\rho \le 1/10$}
\label{sec:1/10}
We first present an algorithm whose simultaneous competitive ratio beats $e/(e-1)$ when $\rho \le 1/10$. This serves to illustrate the first idea of using an oscillating residual curve, and the setting $\rho \le 1/10$ keeps the number of oscillations small and the math simple.

\begin{theorem}
\label{thm:warmup-positive}
Suppose  $0<\rho\le 1/10$. Then there is a randomized ski-rental algorithm, supported on $[0,1]$, whose simultaneous competitive ratio is strictly smaller than $R_{\mathrm{ski}}$.
\end{theorem}

\begin{proof}
We construct a residual curve with positive weighted mass satisfying the two-scale inequalities~\eqref{eq:two-scale-A}. We first work with a piecewise-constant curve and then smooth its downward jumps before applying Lemma~\ref{lem:certificate-realization}. Since both the two-scale inequalities and the sign of the weighted mass are invariant under positive scaling of the curve, we normalize the tail to $-1$. Take $\gamma=1/2$ and define
\[
A_0(y)=
\begin{cases}
0, & 0\le y<\rho^2,\\
-4, & \rho^2\le y<\rho,\\
2, & \rho\le y<1,\\
-1, & y\ge1.
\end{cases}
\]

The two-scale inequality~\eqref{eq:two-scale-A} is immediate. If $0\le y<\rho^2$, then both terms are zero. If $\rho^2\le y<\rho$, then $A_0(y)+\frac12 A_0(\rho y)=-4$. If $\rho\le y<1$, then $A_0(y)+\frac12 A_0(\rho y)=2+\frac12(-4)=0$. If $1\le y<1/\rho$, then $A_0(y)+\frac12 A_0(\rho y)=-1+\frac12\cdot2=0$. Finally, if $y\ge1/\rho$, then $A_0(y)+\frac12 A_0(\rho y)=-1-\frac12<0$. Hence $A_0(y)+\frac12 A_0(\rho y)\le0$ for every $y\ge0$.

It remains to verify that $\mathcal L(A_0)>0$. Direct integration gives
\[
\mathcal L(A_0) = -1 + e\left[ 2\left(e^{-\rho}-e^{-1}\right) - 4\left(e^{-\rho^2}-e^{-\rho}\right) \right].
\]
Using the elementary bounds $e^{-\rho}\ge1-\rho$ and $e^{-\rho^2}\le1$, we obtain $e^{-\rho^2}-e^{-\rho}\le\rho$ and $e^{-\rho}-e^{-1}\ge1-\rho-e^{-1}$. Consequently,
\[
\mathcal L(A_0) \ge -1 + e\left[ 2(1-\rho-e^{-1})-4\rho \right] = 2e-3-6e\rho.
\]
Since $5/2<e<3$ and $\rho\le1/10$,
\[
\mathcal L(A_0) > 5-3-\frac{3}{2} = \frac12>0.
\]

Applying Lemma~\ref{lem:one-sided-smoothing} with sufficiently small smoothing intervals, followed by Lemmas~\ref{lem:certificate-realization} and~\ref{lem:two-scale-farkas-certificate}, gives a randomized ski-rental algorithm supported on $[0,1]$ with simultaneous competitive ratio strictly smaller than $R_{\mathrm{ski}}$.
\end{proof}

\subsubsection{Simple impossibility for $\rho\geq 3/5$}
\label{sec:3/5}

We next give a simple impossibility proof for any algorithm when $\rho$ is sufficiently close to $1$. The proof introduces the dual weighting argument that will also underlie the tight impossibility results.

\begin{theorem}
Suppose $3/5\leq\rho<1$. Then no randomized ski-rental algorithm, even with arbitrary buying times, can have simultaneous competitive ratio strictly smaller than $R_{\mathrm{ski}}$.
\end{theorem}

\begin{proof}
Suppose for contradiction that an algorithm $F$ achieves some target ratio $R<R_{\mathrm{ski}}$. Let $A(y):=C_F(y)-R\OPT(y)$. By Lemma~\ref{lem:two-scale-farkas-certificate}, there is some $\gamma>0$ such that
\begin{equation}
A(y)+\gamma A(\rho y)\leq0
\qquad\text{for every }y\geq0.
\label{eq:simple-impossibility-two-scale}
\end{equation}
By Lemma~\ref{lem:positive-compensation-mass}, $\mathcal L(A)>0$, while Lemma~\ref{lem:compensation-tail} gives $A(t)\leq0$ for every $t\geq1$. We consider two cases according to the value of $\gamma$.

\begin{claim}
\label{cl:warmup-small-gamma}
If $\gamma\leq\rho$, direct charging gives a contradiction to $\mathcal L(A)>0$.
\end{claim}

\begin{proof}
Let $P:=\{y\in(0,1):A(y)>0\}$. For every $y\in P$, the two-scale inequality gives
\[
A(\rho y)\leq-\frac1\gamma A(y)<0.
\]
In particular, $P$ and $\rho P$ are disjoint. By the change of variables $t=\rho y$,
\[
\int_{\rho P}A(t)e^{-t}dt
=\rho\int_P A(\rho y)e^{-\rho y}dy
\leq-\frac{\rho}{\gamma}\int_P A(y)e^{-\rho y}dy
\leq-\int_P A(y)e^{-y}dy,
\]
where the last inequality uses $\gamma\leq\rho$ and $e^{-\rho y}\geq e^{-y}$. Thus the negative mass on $\rho P$ pays for the positive mass on $P$. Outside $P\cup\rho P$, the residual is nonpositive below time $1$, and it is nonpositive on the entire tail by Lemma~\ref{lem:compensation-tail}. Hence $\mathcal L(A)\leq0$, a contradiction.
\end{proof}

The next claim considers the other case, and derives the dual function that underpins all our impossibility results.

\begin{claim}
If $\gamma>\rho$, a nonnegative dual weighting of the two-scale inequalities gives a contradiction to $\mathcal L(A)>0$.
\end{claim}

\begin{proof}
Multiplying \eqref{eq:simple-impossibility-two-scale} by a nonnegative multiplier $\lambda(y)$ and integrating gives
\begin{equation}
\int_0^\infty \lambda(y)\bigl(A(y)+\gamma A(\rho y)\bigr)dy
\leq 0.
\label{eq:dual-multiplier}
\end{equation}
We can convert the left hand side into a weighted integral over $A$:
\begin{equation}
\begin{aligned}
\int_0^\infty \lambda(y)\bigl(A(y)+\gamma A(\rho y)\bigr)dy &= \int_0^\infty \lambda(t)A(t)dt+\frac{\gamma}{\rho}\int_0^\infty\lambda(t/\rho)A(t)dt \\
&= \int_0^\infty\left(\lambda(t)+\frac{\gamma}{\rho}\lambda(t/\rho)\right)A(t)dt .
\end{aligned}
\label{eq:simple-dual-weighting}
\end{equation}
 We now choose $\lambda\geq0$ so that
\begin{equation}
\begin{aligned}
\lambda(t)+\frac{\gamma}{\rho}\lambda(t/\rho)&=e^{-t}, &&0\leq t\leq1,\\
0\leq\lambda(t)+\frac{\gamma}{\rho}\lambda(t/\rho)&\leq e^{-t}, &&t>1.
\end{aligned}
\label{eq:simple-dual-target}
\end{equation}
By Lemma~\ref{lem:compensation-tail}, we have $A(t)\leq0$ for $t\geq1$. Therefore,  combining \eqref{eq:dual-multiplier},  \eqref{eq:simple-dual-weighting}, and \eqref{eq:simple-dual-target} would imply
\[
\frac{\mathcal{L}(A)}{e}=\int_0^\infty A(t)e^{-t}dt
\leq
\int_0^\infty\left(\lambda(t)+\frac{\gamma}{\rho}\lambda(t/\rho)\right)A(t)dt
\leq0,
\]
contradicting $\mathcal L(A)>0$.

It remains to construct such a multiplier. Set $\xi:=\rho/\gamma\in(0,1)$. \eqref{eq:simple-dual-target} requires 
\[
\lambda(t)+\frac1\xi\lambda(t/\rho)=e^{-t}, \qquad \text{for all }t\in[0,1],
\]
which, after replacing \(t\) by \(\rho t\), is equivalent to
\begin{equation}
\lambda(t)=\xi e^{-\rho t}-\xi\lambda(\rho t),
\qquad \text{for all } t\in[0,1/\rho].
\label{eq:simple-dual-match}
\end{equation}
Since \(\xi<1\), this suggests the absolutely convergent solution
$$
\lambda^\star(t)
:=\xi e^{-\rho t}-\xi^2 e^{-\rho^2 t}+\xi^3 e^{-\rho^3 t}-\cdots = \sum_{k=1}^\infty (-1)^{k-1}\xi^k e^{-\rho^k t}.
$$
This recurrence will be useful later, and so it is convenient to define the normalized series
\begin{equation}
U_{\rho,\xi}(t)
:=\sum_{k=0}^\infty(-\xi)^k e^{-t\rho^k}.
\label{eq:USeries}
\end{equation}
Now, re-indexing once more, we get:
$$
\xi U_{\rho,\xi}(\rho t)
=\sum_{k=0}^\infty (-1)^k\xi^{k+1}e^{-\rho^{k+1}t}
=\sum_{j=1}^\infty (-1)^{j+1}\xi^j e^{-\rho^j t}
=\lambda^\star(t).
$$

By Lemma~\ref{lem:geometric-crossing-representation} proved below, we have $U_{\rho,\xi}(t)>0$ for $0\leq t\leq 1/\rho$ as long as
\[
e^{(1-\rho)/\rho}<2\qquad\text{and}\qquad \rho e^{1-\rho}\leq1.
\]
The second condition always holds due to $-\log \rho \geq 1-\rho$, and the first condition holds under the assumption $\rho\geq 3/5$, where $e^{(1-\rho)/\rho}\leq e^{2/3}<2$.

Therefore, we can set 
\begin{align*}
\lambda(t)=\begin{cases}
\xi U_{\rho,\xi}(\rho t), & 0\leq t\leq 1/\rho,\\
0, & t>1/\rho.
\end{cases}
\end{align*}
We now show that $\lambda$ satisfies \eqref{eq:simple-dual-target}. By the discussion above, $\lambda$ is nonnegative for all $t\geq 0$ and satisfies \eqref{eq:simple-dual-match}. Hence it satisfies \eqref{eq:simple-dual-target} on $0\leq t\leq 1$. Further, for $1<t\leq 1/\rho$, we have $\lambda(t/\rho)=0$, hence
\[
\lambda(t)+\frac{\gamma}{\rho}\lambda(t/\rho)=\lambda(t)=\xi U_{\rho,\xi}(\rho t) = e^{-t}-U_{\rho,\xi}(t) \in[0,e^{-t}] ~.
\]
For $t>1/\rho$, we have $\lambda(t)=\lambda(t/\rho)=0$. Therefore \eqref{eq:simple-dual-target} is also satisfied for all $t>1$. This completes the proof.
\end{proof}

The two claims together imply that assuming $\rho\geq 3/5$,  $\mathcal{L}(A)>0$ is incompatible with \eqref{eq:simple-impossibility-two-scale} for all $\gamma>0$, thus no algorithm can have simultaneous competitive ratio strictly smaller than $R_{\mathrm{ski}}$, completing the proof of the theorem.
\end{proof}

\paragraph{Non-negativity of the dual weights.} The preceding proof showed an impossibility result provided the function $U_{\rho,\xi}(t) \ge 0$. We now give an alternative probabilistic representation of $U_{\rho,\xi}$ which provides general conditions under which it is non-negative. These conditions will also be specialized in the tight results later.

Let $E$ have the exponential distribution with $\Pr[E>s]=e^{-s}$. For $0<\rho<1$ and $t\geq0$, set $K_{\rho,t}:=\min\{k\geq0:E>t\rho^k\}$. For $0\leq\xi\leq1$, define
\[
U_{\rho,\xi}(t):=\frac{1}{1+\xi}\mathbb E[(-\xi)^{K_{\rho,t}}].
\]
For $\xi<1$, the first part of the following lemma shows that this agrees with the series definition used above.

\begin{lemma}[Geometric crossing representation]
\label{lem:geometric-crossing-representation}
For $0<\rho<1$, $0\leq\xi<1$, and $t\geq0$, $U_{\rho,\xi}(t)$ satisfies \eqref{eq:USeries}.  Further, for every $0\leq\xi\leq1$ and $t\geq0$,
\begin{equation}
U_{\rho,\xi}(t)+\xi U_{\rho,\xi}(\rho t)=e^{-t}.
\label{eq:U-functional-identity}
\end{equation}
If $T\geq0$ satisfies $e^{(1-\rho)T}<2$ and $\rho e^{\rho(1-\rho)T}\leq1$, then $U_{\rho,\xi}(t)>0$ for every $0\leq\xi\leq1$ and $0\leq t\leq T$.
\end{lemma}
\begin{proof}
The random variable $K_{\rho,t}$ is finite almost surely because
$t\rho^k\to0$ and $E>0$ almost surely. For $t>0$, let
$p_k:=\Pr[K_{\rho,t}=k]$. Then $p_0=e^{-t}$ and, for $k\ge1$,
$p_k=e^{-t\rho^k}-e^{-t\rho^{k-1}}$. Hence, when $\xi<1$, absolute
convergence gives
\[
\E[(-\xi)^{K_{\rho,t}}]
=e^{-t}+\sum_{k\ge1}(-\xi)^k
  \bigl(e^{-t\rho^k}-e^{-t\rho^{k-1}}\bigr)
=(1+\xi)\sum_{k\ge0}(-\xi)^k e^{-t\rho^k}.
\]
This proves the series representation. The case $t=0$ follows directly
from $K_{\rho,0}=0$.

Scaling $t$ by $\rho$ removes the first interval crossed, except when
$K_{\rho,t}=0$. More precisely,
$K_{\rho,\rho t}=(K_{\rho,t}-1)_+$. Therefore, pointwise for
$K=K_{\rho,t}$,
\[
(-\xi)^K+\xi(-\xi)^{(K-1)_+}
=(1+\xi)\mathbf 1_{\{K=0\}}.
\]
Taking expectations and using $\Pr[K_{\rho,t}=0]=e^{-t}$ proves the
recurrence~\eqref{eq:U-functional-identity}, including the endpoint $\xi=1$.
Figure~\ref{fig:U-interval-shift} shows this cancellation on each interval
of the exponential random variable $E$.

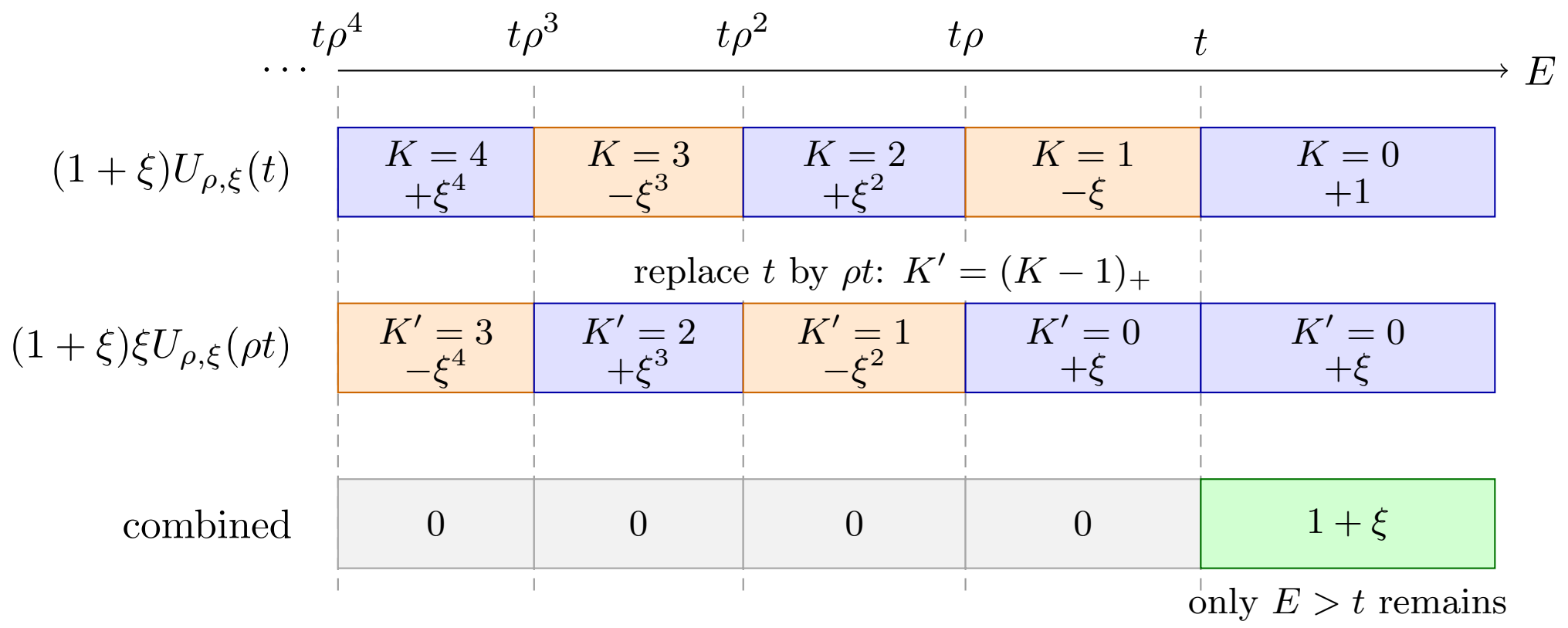
\begin{figure}[t]
\centering
\begin{tikzpicture}[
    x=1.12cm,
    y=1.06cm,
    font=\small,
    positive/.style={draw=blue!65!black,fill=blue!12},
    negative/.style={draw=orange!80!black,fill=orange!18},
    cancelled/.style={draw=black!35,fill=black!5},
    survivor/.style={draw=green!45!black,fill=green!18}
]
    \def\ya{3.00}
    \def\yb{1.58}
    \def\yc{0.16}
    \def\hh{0.72}

    \draw[->] (0.50,4.18) -- (9.45,4.18) node[right] {$E$};
    \node at (0.12,4.18) {$\cdots$};

    \foreach \x/\lab in {
        0.50/{t\rho^4},
        2.00/{t\rho^3},
        3.60/{t\rho^2},
        5.30/{t\rho},
        7.10/{t}}
    {
        \draw[densely dashed,black!40] (\x,-0.02) -- (\x,4.08);
        \node[above] at (\x,4.18) {$\lab$};
    }

    \node[anchor=east] at (0.27,\ya+0.36)
        {$(1+\xi)U_{\rho,\xi}(t)$};
    \node[anchor=east] at (0.27,\yb+0.36)
        {$(1+\xi)\xi U_{\rho,\xi}(\rho t)$};
    \node[anchor=east] at (0.27,\yc+0.36) {combined};

    \foreach \xa/\xb/\style/\k/\w in {
        0.50/2.00/positive/4/{+\xi^4},
        2.00/3.60/negative/3/{-\xi^3},
        3.60/5.30/positive/2/{+\xi^2},
        5.30/7.10/negative/1/{-\xi},
        7.10/9.35/positive/0/{+1}}
    {
        \draw[\style] (\xa,\ya) rectangle (\xb,\ya+\hh);
        \node[font=\footnotesize] at ({(\xa+\xb)/2},\ya+0.51) {$K=\k$};
        \node[font=\footnotesize] at ({(\xa+\xb)/2},\ya+0.19) {$\w$};
    }

    \node[font=\footnotesize,fill=white,inner sep=2pt] at (4.75,2.55)
        {replace $t$ by $\rho t$: $K'=(K-1)_+$};

    \foreach \xa/\xb/\style/\k/\w in {
        0.50/2.00/negative/3/{-\xi^4},
        2.00/3.60/positive/2/{+\xi^3},
        3.60/5.30/negative/1/{-\xi^2},
        5.30/7.10/positive/0/{+\xi},
        7.10/9.35/positive/0/{+\xi}}
    {
        \draw[\style] (\xa,\yb) rectangle (\xb,\yb+\hh);
        \node[font=\footnotesize] at ({(\xa+\xb)/2},\yb+0.51) {$K'=\k$};
        \node[font=\footnotesize] at ({(\xa+\xb)/2},\yb+0.19) {$\w$};
    }

    \foreach \xa/\xb in {0.50/2.00,2.00/3.60,3.60/5.30,5.30/7.10}
    {
        \draw[cancelled] (\xa,\yc) rectangle (\xb,\yc+\hh);
        \node[font=\footnotesize] at ({(\xa+\xb)/2},\yc+0.36) {$0$};
    }
    \draw[survivor] (7.10,\yc) rectangle (9.35,\yc+\hh);
    \node[font=\footnotesize] at (8.225,\yc+0.36) {$1+\xi$};
    \node[font=\footnotesize,below] at (8.225,\yc-0.03)
        {only $E>t$ remains};
\end{tikzpicture}
\caption{The interval shift behind the functional identity. The first two
rows show the contributions before division by $1+\xi$. Replacing $t$ by
$\rho t$ shifts each interval one place, and multiplication by $\xi$
reverses its sign. Every interval cancels except $E>t$. Dividing the last
row by $1+\xi$ and taking expectations gives the
recurrence~\eqref{eq:U-functional-identity}. The horizontal spacing is schematic.}
\label{fig:U-interval-shift}
\end{figure}

Since $U_{\rho,\xi}(0)=1/(1+\xi)>0$, it remains to prove positivity for
$t>0$. Since $K_{\rho,t}$ takes values in the nonnegative integers and
$p_k=\Pr[K_{\rho,t}=k]$, its expectation can be expanded as
\[
\begin{aligned}
(1+\xi)U_{\rho,\xi}(t)
&=\E\!\left[(-\xi)^{K_{\rho,t}}\right]
=\sum_{k=0}^{\infty}(-\xi)^k p_k\\
&=\sum_{j=0}^{\infty}
\left((-\xi)^{2j}p_{2j}
      +(-\xi)^{2j+1}p_{2j+1}\right)\\
&=\sum_{j=0}^{\infty}
\xi^{2j}\bigl(p_{2j}-\xi p_{2j+1}\bigr).
\end{aligned}
\]
This grouping is valid also at $\xi=1$, since
$\sum_{k\ge0}\xi^k p_k\le\sum_{k\ge0}p_k=1$.

It is therefore enough to show $p_{2j}\ge p_{2j+1}$ for every $j$, with strict
inequality for $j=0$. First, $p_0>p_1$ is equivalent to
$e^{(1-\rho)t}<2$, which follows from $t\le T$ and the first assumed
inequality.

It remains to compare $p_{2j}$ and $p_{2j+1}$ for $j\ge1$. Changing
variables $x=z/\rho$ in the integral for $p_{2j}$ gives
\[
p_{2j}=\frac1\rho
\int_{t\rho^{2j+1}}^{t\rho^{2j}}e^{-z/\rho}\,dz,
\qquad
p_{2j+1}=\int_{t\rho^{2j+1}}^{t\rho^{2j}}e^{-z}\,dz.
\]
On the interval of integration, $z\le t\rho^{2j}\le T\rho^2$.
Consequently,
\[
\frac{\rho^{-1}e^{-z/\rho}}{e^{-z}}
=\rho^{-1}e^{-z(1-\rho)/\rho}
\ge\rho^{-1}e^{-T\rho(1-\rho)}\ge1,
\]
where the last inequality is the second assumption. Thus
$p_{2j}\ge p_{2j+1}$.
Every summand in the grouped expression is nonnegative, and its first
summand is positive. This proves the positivity statement.
\end{proof}

\subsection{Why buying after the breakeven time helps}
\label{sec:strict-late-buying-separation}

The preceding bounds illustrate the main ideas behind the positive and negative arguments. We now extend these arguments to show that allowing purchases after the breakeven time strictly enlarges the set of scales $\rho$ admitting a positive result. Specifically, we show that at $\rho = 1/3$, every algorithm supported on $[0,1]$ has simultaneous competitive ratio at least $e/(e-1)$, whereas an unrestricted algorithm can beat this competitive ratio. 

The proof of this result serves to illustrate the additional analytic ideas and algorithmic constructions required for obtaining our tight results for both algorithm classes in Appendices~\ref{app:tight-buy-by-one} and~\ref{app:tight-general}. Specifically, it illustrates the functional form of the weighted tail bound for the impossibility results, and the oscillating curve that arises from our dual function that leads to the possibility results.

\subsubsection{A weighted tail bound and the function $D(\rho,\xi)$} 
We start with a general lemma relating $U_{\rho,\xi}$ to the weighted tail $\mathcal L(A)$. This result holds more generally than the settings of $\rho$ we consider, and will therefore form the basis of our tight results in Appendices~\ref{app:tight-buy-by-one} and~\ref{app:tight-general}.

\begin{lemma}[Weighted tail bound]
\label{lem:general-weighted-tail}
Let $0<\rho<1$, $\gamma>\rho$, and $\xi:=\rho/\gamma \in [0,1)$. Let $A$ be a bounded measurable function satisfying \eqref{eq:two-scale-A}. Suppose also that $U_{\rho,\xi}(t)\ge0$ for $0\le t\le1$. Then
\begin{equation}
\frac{\mathcal L(A)}e
\le
\int_1^{1/\rho}A(y)U_{\rho,\xi}(y)\,dy
+\int_{1/\rho}^{\infty}A(y)e^{-y}\,dy.
\label{eq:general-weighted-tail}
\end{equation}
\end{lemma}

\begin{proof}
For $0\le y\le1/\rho$, we have $\rho y\in[0,1]$, so the assumption gives $U_{\rho,\xi}(\rho y)\ge0$. Multiply \eqref{eq:two-scale-A} by $\xi U_{\rho,\xi}(\rho y)$ and integrate over $[0,1/\rho]$. This gives
\[
\xi\int_0^{1/\rho}U_{\rho,\xi}(\rho y)A(y)\,dy
+
\xi\gamma\int_0^{1/\rho}U_{\rho,\xi}(\rho y)A(\rho y)\,dy
\le 0.
\]
In the second integral, substitute $x=\rho y$. Since $\xi=\rho/\gamma$, we have $\xi\gamma/\rho=1$, and therefore
\[
\xi\int_0^{1/\rho}U_{\rho,\xi}(\rho y)A(y)\,dy
+
\int_0^1U_{\rho,\xi}(x)A(x)\,dx
\le 0.
\]
Reorganizing and applying \eqref{eq:U-functional-identity} gives
\begin{equation}
\begin{aligned}
& \int_0^1
\left(
U_{\rho,\xi}(y)+\xi U_{\rho,\xi}(\rho y)
\right)A(y)\,dy
+
\xi\int_1^{1/\rho}U_{\rho,\xi}(\rho y)A(y)\,dy\\
&= \int_0^1 A(y)e^{-y}dy + \xi\int_1^{1/\rho}U_{\rho,\xi}(\rho y)A(y)\,dy \leq 0.
\end{aligned}
\label{eq:general-weighted-tail-below-one}
\end{equation}

By definition,
\[
\frac{\mathcal L(A)}e
=
\int_0^1A(y)e^{-y}\,dy
+
\int_1^{1/\rho}A(y)e^{-y}\,dy
+
\int_{1/\rho}^{\infty}A(y)e^{-y}\,dy.
\]
Substituting \eqref{eq:general-weighted-tail-below-one} for the first integral gives
\[
\frac{\mathcal L(A)}e
\le
\int_1^{1/\rho}
\left(
e^{-y}-\xi U_{\rho,\xi}(\rho y)
\right)A(y)\,dy
+
\int_{1/\rho}^{\infty}A(y)e^{-y}\,dy.
\]
Applying \eqref{eq:U-functional-identity} to the first integral completes the proof of \eqref{eq:general-weighted-tail}.
\end{proof}

\paragraph{Specialization to buy-by-breakeven time algorithms.}
We now specialize the above bound to buy-by-breakeven time algorithms. Previously we only used the property that $A(t)\leq 0$ for all $t\geq 1$. For algorithms that buy by the breakeven time, $A$ is actually constant on $t\geq 1$. This is because for $t \ge 1$, the algorithm's and OPT's cost are constant, which means $A(t)$ is a constant. This allows us to derive stronger impossibility and possibility results for this case.

Suppose that $A(y)=a$ for every $y\ge1$. Then \eqref{eq:general-weighted-tail} gives
\begin{align} \label{eq:L-break}
\frac{\mathcal L(A)}e
\le
a\int_1^{1/\rho}U_{\rho,\xi}(y)\,dy
+
a\int_{1/\rho}^{\infty}e^{-y}\,dy
=
a\left(
\int_1^{1/\rho}U_{\rho,\xi}(y)\,dy
+
e^{-1/\rho}
\right).
\end{align}
To simplify this coefficient, integrate \eqref{eq:U-functional-identity} over $[1,1/\rho]$:
\[
\int_1^{1/\rho}U_{\rho,\xi}(y)\,dy
+
\xi\int_1^{1/\rho}U_{\rho,\xi}(\rho y)\,dy
=
\int_1^{1/\rho}e^{-y}\,dy.
\]
Substituting $x=\rho y$ in the second integral gives
\[
\int_1^{1/\rho}U_{\rho,\xi}(y)\,dy
+
\frac{\xi}{\rho}\int_\rho^1U_{\rho,\xi}(x)\,dx
=
e^{-1}-e^{-1/\rho}.
\]
Equivalently,
\[
e\left(
\int_1^{1/\rho}U_{\rho,\xi}(y)\,dy
+
e^{-1/\rho}
\right)
=
1-\frac{e\xi}{\rho}\int_\rho^1U_{\rho,\xi}(x)\,dx.
\]
This motivates the definition
\begin{equation}
D(\rho,\xi):=\frac{e\xi}{\rho}\int_\rho^1U_{\rho,\xi}(x)\,dx,
\qquad 0\le\xi\le1.
\label{eq:D-def}
\end{equation}
For the present argument $\xi<1$, and the preceding calculation gives
\begin{equation}
\mathcal L(A)\le a\bigl(1-D(\rho,\xi)\bigr).
\label{eq:general-constant-tail}
\end{equation}
Thus $D(\rho,\xi)$ is the coefficient produced by the dual weighting when the residual has a constant tail. Appendix~\ref{app:tight-buy-by-one} shows that the same quantity also governs the matching positive construction.

\subsubsection{Analytic bounds on functionals of $U_{1/3,\xi}$} 
We now specialize to $\rho = 1/3$ and show the possibility and impossibility results. For $0\le\xi\le1$, write $U_\xi(t):=U_{1/3,\xi}(t)$. For $0\le\xi<1$, \eqref{eq:USeries} gives
\begin{equation}
U_\xi(t)=\sum_{k=0}^{\infty}(-\xi)^k e^{-3^{-k}t}.
\label{eq:one-third-U-geometric}
\end{equation}
At $\xi=1$, $U_1$ is given by the probabilistic definition in Lemma~\ref{lem:geometric-crossing-representation}, or equivalently by the power-series extension in Lemma~\ref{lem:U-power-series}.
At $\rho=1/3$, the quantity in \eqref{eq:D-def} becomes
\[
D(1/3,\xi)=3e\xi\int_{1/3}^1U_\xi(t)\,dt.
\]
The following lemma collects the three bounds needed for showing the separation between buy by breakeven time algorithms and general algorithms at $\rho = 1/3$. The first two make the dual certificate above valid and show that a constant residual tail cannot support positive weighted mass. The third shows that a suitable nonconstant tail can support this mass. Their generalizations in Appendices~\ref{app:tight-buy-by-one} and~\ref{app:tight-general} yield the corresponding sharp thresholds for the two algorithm classes. We give machine-checked proofs in Lean for the second and third estimates, which also admit purely analytic proofs.

\begin{lemma}[One-third bounds for $U$]
\label{lem:one-third-kernel}
For every $\xi\in[0,1]$,
\begin{equation}
    U_\xi(t)\ge0
    \qquad(0\le t\le1)
    \label{eq:one-third-U-positive}
\end{equation}
and
\begin{equation}
    D_\xi
    :=3e\xi\int_{1/3}^1U_\xi(t)\,dt
    <1.
    \label{eq:one-third-D}
\end{equation}
At $\xi=1$, we also have
\begin{equation}
    \int_1^3(t-4)U_1(t)\,dt>0.
    \label{eq:one-third-late-sign}
\end{equation}
\end{lemma}
\leanverified{The complete statement of this lemma, including all endpoints;
see Appendix~\ref{app:lean-one-third-kernel}. The proof uses as an axiom the power series representation of $U$ from Lemma~\ref{lem:U-power-series}.}

\begin{proof}
Appendix~\ref{app:lean-one-third-kernel} gives a complete Lean verification\footnote{The Lean code for this paper is available at
\url{https://github.com/relyt871/BWCA-Online-Lean}.}
of all three assertions.  The first inequality also has a simple proof by verifying the two preconditions in Lemma~\ref{lem:geometric-crossing-representation} for $\rho = 1/3$.
\end{proof}

\subsubsection{Impossibility for buy-by-breakeven-time algorithms at $\rho = 1/3$} We now show that no algorithm supported on $[0,1]$ can beat $e/(e-1)$ for simultaneous competitive ratio at $\rho = 1/3$.

\begin{theorem}[Breakeven-time impossibility]
\label{cor:one-third-buy-by-breakeven-impossible}
At $\rho=1/3$, no randomized ski-rental algorithm supported on $[0,1]$ can have simultaneous distributional competitive ratio strictly below $R_{\mathrm{ski}}$.
\end{theorem}

\begin{proof}
Suppose that such an algorithm exists, and let $A$ be its residual at a target ratio below $R_{\mathrm{ski}}$. Lemma~\ref{lem:positive-compensation-mass} gives $\mathcal L(A)>0$, while Lemma~\ref{lem:two-scale-farkas-certificate} gives some $\gamma>0$ such that \eqref{eq:two-scale-A} holds with $\rho=1/3$. Since the algorithm is supported on $[0,1]$, its residual is constant on $[1,\infty)$; write $A(y)=a$ there. Lemma~\ref{lem:compensation-tail} gives $a\le0$.

If $\gamma\le1/3$, Claim~\ref{cl:warmup-small-gamma} gives $\mathcal L(A)\le0$, a contradiction. Otherwise set $\xi=1/(3\gamma)\in(0,1)$. By Lemma~\ref{lem:one-third-kernel}, $U_\xi(t)\ge0$ on $[0,1]$, so Lemma~\ref{lem:general-weighted-tail} applies. Using \eqref{eq:general-constant-tail} and \eqref{eq:one-third-D},
\[
\mathcal L(A)\le a(1-D_\xi)\le0,
\]
where the final inequality follows from $a\le0$ and $D_\xi<1$. This contradicts $\mathcal L(A)>0$.
\end{proof}

\subsubsection{Possibility for general algorithms at $\rho = 1/3$} 
We now turn to unrestricted algorithms, and show that allowing $A$ to be nonconstant above $1$ gives a positive result. The construction begins with a function on $[1,\infty)$ and extends it recursively to earlier times. We first state the general form of this extension useful for our tight results.

\begin{lemma}[Recursive extension]
\label{lem:general-recursive-extension}
Let $0<\rho<1$, $\gamma>\rho$, and $\xi:=\rho/\gamma$.  Let
$g:[1,\infty)\to\mathbb R$ be bounded and measurable, and suppose that
$g(z)\le0$ for $1\le z<1/\rho$ and
$g(y)+\gamma g(\rho y)\le0$ for $y\ge1/\rho$.
For an even integer $N\ge0$, define
\begin{equation}
    A_N(y)
    :=
    \begin{cases}
        g(y),&y\ge1,\\[1mm]
        (-1)^k\gamma^{-k}g(y/\rho^k),
            &\rho^k\le y<\rho^{k-1},\quad k=1,\ldots,N,\\[1mm]
        0,&0\le y<\rho^N.
    \end{cases}
    \label{eq:general-recursive-extension}
\end{equation}
Then $A_N$ satisfies \eqref{eq:two-scale-A} with $A$ replaced by $A_N$,
and
\begin{equation}
    \lim_{\substack{N\to\infty\\N\text{ even}}}
    \frac{\mathcal L(A_N)}e
    =
    \int_1^{1/\rho}g(z)U_{\rho,\xi}(z)\,dz
    +\int_{1/\rho}^{\infty}g(y)e^{-y}\,dy.
    \label{eq:general-recursive-limit}
\end{equation}
\end{lemma}

\begin{proof}
For $k=0,\ldots,N$, let
$J_k=[\rho^k,\rho^{k-1})$, with the convention
$\rho^{-1}=1/\rho$. On each of these intervals,
\begin{equation}
    A_N(y)=(-1)^k\gamma^{-k}g(y/\rho^k).
    \label{eq:general-recursive-scaled-copy}
\end{equation}
For $k=0$, this follows from the first case in
\eqref{eq:general-recursive-extension}; for $k\ge1$, it is the second
case in that definition.

We first prove the two-scale inequality~\eqref{eq:two-scale-A}. Suppose that $y\in J_k$ for
some $k<N$, and put $z=y/\rho^k$. Then $\rho y\in J_{k+1}$, and
\eqref{eq:general-recursive-scaled-copy} gives
\[
    A_N(y)+\gamma A_N(\rho y)
    =(-1)^k\gamma^{-k}g(z)
    +(-1)^{k+1}\gamma^{-k}g(z)=0.
\]
If $y\in J_N$, then $\rho y<\rho^N$, so $A_N(\rho y)=0$. Since $N$ is
even and $z=y/\rho^N\in[1,1/\rho)$, we have
$A_N(y)=\gamma^{-N}g(z)\le0$. If $0\le y<\rho^N$, both terms vanish.
Finally, if $y\ge1/\rho$, then $y,\rho y\ge1$, and the first case in
\eqref{eq:general-recursive-extension} and the hypothesis on $g$ give
$A_N(y)+\gamma A_N(\rho y)=g(y)+\gamma g(\rho y)\le0$. These cases
cover every $y\ge0$.

It remains to calculate the limiting weighted mass. The intervals
$J_0,\ldots,J_N$ cover $[\rho^N,1/\rho)$, while $A_N$ vanishes below
$\rho^N$ and equals $g$ on $[1/\rho,\infty)$. For each $k$, substitute
$z=y/\rho^k$ in the integral over $J_k$. Since
$dy=\rho^k\,dz$ and $\rho^k\gamma^{-k}=\xi^k$,
\eqref{eq:general-recursive-scaled-copy} yields
\begin{align*}
    \frac{\mathcal L(A_N)}e
    &=\sum_{k=0}^N\int_{J_k}A_N(y)e^{-y}\,dy
      +\int_{1/\rho}^{\infty}g(y)e^{-y}\,dy\\
    &=\int_1^{1/\rho}g(z)
      \left(\sum_{k=0}^N(-\xi)^ke^{-\rho^kz}\right)\,dz
      +\int_{1/\rho}^{\infty}g(y)e^{-y}\,dy.
\end{align*}
Because $0<\xi<1$, \eqref{eq:USeries}
shows that the finite sum converges pointwise to
$U_{\rho,\xi}(z)$. If $|g(z)|\le M$, then the absolute value of the
first integrand is at most $M\sum_{k=0}^N\xi^k\le M/(1-\xi)$ on the
finite interval $[1,1/\rho]$. Dominated convergence therefore gives
\eqref{eq:general-recursive-limit} for the even values
of $N$.
\end{proof}

For buy-by-breakeven-time algorithms, the residual $A(t)$ is constant for $t \ge 1$. Appendix~\ref{app:tight-buy-by-one} therefore applies
Lemma~\ref{lem:general-recursive-extension} with the constant function
$g(z)=-1$. In this case, 
\eqref{eq:general-recursive-limit} implies
\[
\lim_{\substack{N\to\infty\\N\text{ even}}}
    \frac{\mathcal L(A_N)}e
    = -\int_1^{1/\rho}U_{\rho,\xi}(z)\,dz-e^{-1/\rho}
=
\frac{D(\rho,\xi)-1}{e},
\]
where the second equality follows from the analysis just preceding~\eqref{eq:D-def}. This is the same function
$D(\rho,\xi)$ that appears in the weighted upper bound
\eqref{eq:general-constant-tail}, so the construction and the
impossibility argument match.

The point now is that $A$ may vary after time $1$ for general algorithms. The function $g$ specifies this tail, and \eqref{eq:general-recursive-extension} extends it to earlier times while preserving the two-scale inequality~\eqref{eq:two-scale-A}. The limit in \eqref{eq:general-recursive-limit} describes the resulting value of the weighted integral $\mathcal L(A)$. At $\rho=1/3$, a suitable non-constant choice of $g$ makes this limit positive, while every constant tail gives a limit at most zero. This is the source of the benefit from allowing later purchases.

To show the positive result, we now apply the lemma with $\rho=1/3$ and
\[
g(z)=
\begin{cases}
z-4, & 1\le z<3,\\
0, & z\ge3.
\end{cases}
\]
The intuition for this function is presented in Appendix~\ref{app:tight-general}, and relates to the deficit of buying at time $T = 1/\rho = 3$. Let $I_\xi:=\int_1^3g(z)U_\xi(z)\,dz$. Lemma~\ref{lem:one-third-kernel}
gives $I_1>0$. Joint continuity of $U_\xi(z)$ on
$[0,1]\times[1,3]$ makes $I_\xi$ continuous in $\xi$, so fix $\xi<1$
sufficiently close to $1$ that $I_\xi>0$, and put
$\gamma=1/(3\xi)>1/3$.

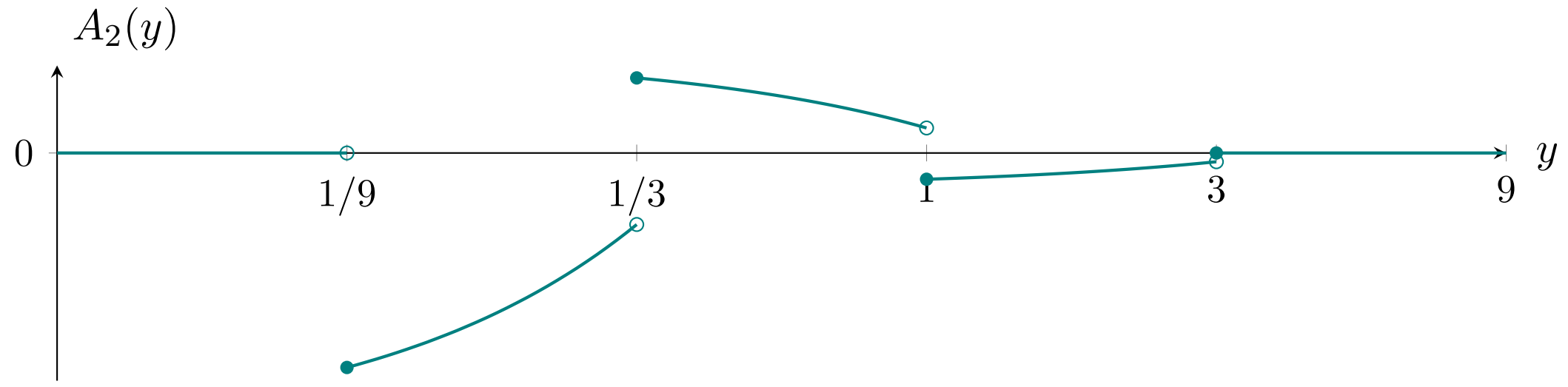
\begin{figure}[!htbp]
\centering
\begin{tikzpicture}
\begin{axis}[
    width=0.88\linewidth,
    height=4.4cm,
    xmode=log,
    log basis x=3,
    axis x line=middle,
    axis y line=left,
    xmin=0.037037,
    xmax=9,
    ymin=-26,
    ymax=10,
    xtick={0.111111,0.333333,1,3,9},
    xticklabels={$1/9$,$1/3$,$1$,$3$,$9$},
    ytick={0},
    xlabel={$y$},
    xlabel style={at={(axis description cs:1.01,0.71)},anchor=west},
    tick label style={font=\small},
    clip=false
]
\addplot[thick,teal,domain=0.037037:0.111111,samples=30] {0};
\addplot[thick,teal,domain=0.111111:0.333333,samples=30]
    {(9*x-4)/(0.35^2)};
\addplot[thick,teal,domain=0.333333:1,samples=30]
    {(4-3*x)/0.35};
\addplot[thick,teal,domain=1:3,samples=30] {x-4};
\addplot[thick,teal,domain=3:9,samples=30] {0};
\addplot[only marks,mark=*,mark size=1.5pt,teal]
    coordinates {(0.111111,-24.489796) (0.333333,8.571429)
                 (1,-3) (3,0)};
\addplot[only marks,mark=o,mark size=1.7pt,teal,fill=white]
    coordinates {(0.111111,0) (0.333333,-8.163265)
                 (1,2.857143) (3,-1)};
\node[anchor=south west] at (rel axis cs:0,1) {$A_2(y)$};
\end{axis}
\end{tikzpicture}
\caption{The recursive residual curve for $N=2$, shown with
$\gamma=0.35$ and a logarithmic horizontal axis. Each interval closer to
zero contains a compressed, reflected, and enlarged copy of $g(y)=y-4$ on
$[1,3)$. Larger even values of $N$ continue the pattern on smaller intervals.}
\label{fig:one-third-recursive-curve}
\end{figure}

For an even integer $N$, we therefore
define
\begin{equation}
    A_N(y)
    :=
    \begin{cases}
        y-4,&1\le y<3,\\
        -\gamma^{-1}A_N(3y),&3^{-N}\le y<1,\\
        0,&\text{otherwise}.
    \end{cases}
    \label{eq:one-third-recursive-construction}
\end{equation}
The recursion constructs the function from larger times to smaller ones. Each new interval is a copy of the preceding one, compressed horizontally by $3$, reflected across zero, and enlarged vertically by $1/\gamma$. Figure~\ref{fig:one-third-recursive-curve} shows the first three pieces. 

For the present choice of $g$, the general construction in \eqref{eq:general-recursive-extension} is exactly the recursion in \eqref{eq:one-third-recursive-construction}. The following corollary gives the two properties needed for the positive result.

\begin{corollary}
\label{lem:one-third-recursive-extension}
For every sufficiently large even $N$, the function $A_N$ satisfies
\eqref{eq:two-scale-A} with $A$ replaced by $A_N$ and $\rho=1/3$, and
$\mathcal L(A_N)>0$.
\end{corollary}

\begin{proof}
The function $g$ is bounded and measurable, and $g(z)=z-4<0$ for
$1\le z<3$. If $3\le y<9$, then $g(y)=0$ and
$g(y/3)=y/3-4<0$; if $y\ge9$, then both terms vanish. Hence
$g(y)+\gamma g(y/3)\le0$ for every $y\ge3$. Lemma~\ref{lem:general-recursive-extension} therefore applies. Its limiting expression becomes
\[
    \lim_{\substack{N\to\infty\\N\text{ even}}}
    \frac{\mathcal L(A_N)}e
    =\int_1^3g(z)U_\xi(z)\,dz
    =I_\xi>0.
\]
The two conclusions follow.
\end{proof}

We can now realize one finite member of the sequence of $A_N$ as an algorithm.

\begin{theorem}[Strict benefit of late buying]
\label{thm:strict-late-buying}
At $\rho=1/3$, there is an unrestricted randomized ski-rental algorithm
whose distributional competitive ratio is strictly below
$R_{\mathrm{ski}}$. Every algorithm achieving such an improvement assigns
positive probability to buying after the breakeven time.
\end{theorem}

\begin{proof}
Fix a sufficiently large even $N$ as in Corollary~\ref{lem:one-third-recursive-extension}. The function $A_N$ is right-continuous and piecewise affine, has finitely many jumps, and its only jump above time $1$ is the upward jump at time $3$. Apply Lemma~\ref{lem:one-sided-smoothing} to its downward jumps at or below time $1$. As explained there, the smoothing can be chosen to give a curve $\widetilde A_N$ with $\mathcal L(\widetilde A_N)>0$, while preserving the two-scale inequality~\eqref{eq:two-scale-A} and leaving the tail unchanged:
\[
\widetilde A_N(y)=
\begin{cases}
y-4, & 1\leq y<3,\\
0, & y\geq3.
\end{cases}
\]
Lemmas~\ref{lem:late-tail-realization-check}, \ref{lem:certificate-realization}, and~\ref{lem:two-scale-farkas-certificate} therefore give an unrestricted algorithm with simultaneous competitive ratio strictly smaller than $R_{\mathrm{ski}}$.
Theorem~\ref{cor:one-third-buy-by-breakeven-impossible} shows that no algorithm supported on $[0,1]$ can achieve such an improvement, while Lemma~\ref{lem:compensation-tail} shows that every improving algorithm buys eventually almost surely, hence buying after the breakeven time with positive probability.
\end{proof}

\subsection{Unknown $\rho$: Impossibility}
\label{sec:imposs_unknown}
We finally show that if only an upper bound on $\rho$ is known, then no single algorithm can beat the classical ratio against distributional adversaries for all sufficiently small $\rho$.

\begin{theorem}[Unknown-ratio distributional impossibility]
\label{thm:unknown-rho-distributional}
Fix $\alpha\in(0,1)$. There does not exist a single randomized ski-rental
algorithm and a constant $\eta>0$ such that, for every
$\rho\in(0,\alpha]$, the algorithm beats $R_{\mathrm{ski}}$ by $\eta$ against
every distributional adversary.
\end{theorem}

\begin{proof}
Suppose such an algorithm $F$ and such an $\eta>0$ exist. By decreasing $\eta$ if necessary, we may assume that $R:=R_{\mathrm{ski}}-\eta>1$. Let
$
A(y):=C_F(y)-R\OPT(y),
$ and write $a:=R-1>0$.

We first locate a positive residual below the breakeven time. Fix any $\rho_0\in(0,\alpha]$. Since the algorithm achieves the target ratio for $\rho_0$, Lemma~\ref{lem:compensation-tail} gives $A(t)\leq0$ for every $t\geq1$. On the other hand, Lemma~\ref{lem:positive-compensation-mass} gives $\mathcal L(A)=(e-1)\eta>0$. Consequently there is a point $y^\ast\in(0,1)$ such that $M:=A(y^\ast)>0$.

For every admissible $\rho$,
Lemma~\ref{lem:two-scale-farkas-certificate} gives some
$\gamma=\gamma(\rho)>0$ for which $A$ satisfies
\eqref{eq:two-scale-A}.
Every ski-rental algorithm has cost at least the offline optimum, so
\[
A(z)=C_F(z)-R\OPT(z)\geq-(R-1)\OPT(z)=-a\OPT(z)
\qquad\text{for every }z\geq0.
\]
We will choose $\rho$ so small that the two inequalities become incompatible. Specifically, choose an admissible ratio satisfying
\[
0<\rho<\min\left\{\alpha,y^\ast,\frac{M^2}{a^2y^\ast}\right\}.
\]

First evaluate the Farkas inequality at $y=y^\ast$. Since $\rho y^\ast<1$, we have $A(\rho y^\ast)\geq-a\rho y^\ast$, and therefore
\[
0\geq M+\gamma A(\rho y^\ast)\geq M-\gamma a\rho y^\ast.
\]
It follows that $\gamma\geq M/(a\rho y^\ast)$.

Next evaluate the same inequality at $y=y^\ast/\rho$. Now $y^\ast/\rho>1$ and the scaled point is $\rho y=y^\ast$, so $A(y^\ast/\rho)\geq-a$ and
\[
0\geq A(y^\ast/\rho)+\gamma M\geq-a+\gamma M.
\]
Hence $\gamma\leq a/M$. But our choice of $\rho$ gives $M/(a\rho y^\ast)>a/M$, a contradiction.
\end{proof}

\section{Conclusion and Discussion}
We view our results as a first step in understanding the robustness of worst-case instances for online algorithms. Several open questions remain. First, analogous to the prophet inequality result, what is the exact optimal simultaneous competitive ratio for ski rental as a function of $\rho$? Our results locate the transition between improvement and impossibility, while the full
performance curve below the threshold remains open. Second, how much information about $\rho$ does a
ski-rental algorithm need? Intermediate models include a finite set of possible ratios or a short piece of advice.  

The two problems also show that greater separation between the rates (that is, when $\rho$ is closer to $0$) can have different effects on the simultaneous competitive ratio. For prophet inequalities, a block of values can be almost certain to survive at the higher rate but appear only rarely at the lower rate. The same block then provides a reliable choice at one rate and a rare prize worth waiting for at the other. The adversary can exploit this difference to make the competitive ratio approach $2$ as the rates separate. For distributional ski rental, separation instead helps the competitive ratio. A buying strategy can accept higher costs at some horizons in exchange for lower costs at others, and sufficiently separated horizons allow this tradeoff to improve the guarantee at one of the two rates. When the horizons are closer, an adversarial distribution can prevent such an improvement. Understanding what determines these different responses to scale separation is a natural direction for further work.

More broadly, our work motivates the need to precisely identify structural conditions under which worst-case instances to online problems remain robust to subsampling and other random perturbations. Candidate problems include online bipartite matching, load balancing, and deterministic paging. Our ski rental results also show that the answer can  depend on the separation between the two rates. For other problems, it could additionally be the case that the improvement or lack thereof depends on what combinatorial object is sampled. For online matching, the classical worst-case construction~\cite{KVV} suggests robustness under sub-sampling edges, but its fragility under vertex subsampling remains unclear. We leave such extensions as future work.

\section*{Generative AI Disclosure and Proof Verifier Use} The authors conceptualized the overall two-scale problem and proved the tight prophet inequality results and the simpler arguments presented in the main paper. We also conceptualized the distributional ski rental model. The proofs of the tight results for ski-rental were obtained via extensive conversations with GPT 5.6Pro, which also formalized some analytic lemmas in Lean. These conversations involved non-trivial inputs from the authors. We used Codex to help with the initial writing of the paper. We have kept the main paper intuitive and easy to understand via presenting results for special cases that convey the main ideas. 
For ease of readability and checking, we used Lean verification for certain analytic characterizations of the $U_{\rho,\xi}(t)$ function (which otherwise involved tedious algebraic proofs).
All content was verified by the authors who take full responsibility for this work.

\clearpage
\bibliographystyle{alpha}
\bibliography{refs}

\clearpage
\appendix

\section{Proof of the Prophet Inequality Tight Result}
\label{app:prophet-tight-proof}

Throughout the appendix, we fix $0<q<p\le 1$ and set $\rho=q/p\in(0,1)$. When defining variables that depend on $p$, $q$, or $\rho$, we may therefore omit these parameters from the notation. Lower-case letters denote fixed values, while upper-case letters denote random variables.

We first establish two structural properties that compare the dynamic programs at the two sampling rates.

\begin{lemma}
\label{lem:dp_monotonicity_scaling}
For every suffix $i$, we have $v_i(q)\le v_i(p)$ and $v_i(q)/q\ge v_i(p)/p$.
\end{lemma}

\begin{proof}
The first inequality is immediate from a coupling: an item that survives $q$-subsampling can be viewed as also surviving $p$-subsampling, while the larger rate may reveal additional items.

For the scaling inequality, we use backward induction. It is trivial for $i=n+1$. Assuming it holds for $i+1$, the recurrence gives
\begin{equation}
\begin{aligned}
    \frac{v_i(q)}q &=\mathbb E[(X_i-v_{i+1}(q))^+]+\frac{v_{i+1}(q)}q,\\
    \frac{v_i(p)}p &=\mathbb E[(X_i-v_{i+1}(p))^+]+\frac{v_{i+1}(p)}p.
\end{aligned}
\label{eq:scaled-dp-recurrences}
\end{equation}
Since $v_{i+1}(q)\le v_{i+1}(p)$, the first expectation in the first line is at least the corresponding expectation in the second line; the induction hypothesis gives the same comparison for the second terms. Therefore $v_i(q)/q\ge v_i(p)/p$.
\end{proof}

\subsection{Proof of Theorem~\ref{thm:tight_two_rate_prophet}}

The proof associates each instance with a parameter $\phi\in[0,1]$ that measures the separation between its online values at rates $q$ and $p$. We first split the prophet's value into a part above the higher DP thresholds and a bounded part between the two collections of thresholds. Two lemmas control the bounded part by a product of failure probabilities. At rate $q$, the DP recurrence makes this product telescope and produces $\phi$. The survival conversion function $\Gamma_{q,p}$ then translates the same product bound to rate $p$. This yields the two bounds $CR(q)\le f(\phi)$ and $CR(p)\le g(\phi)$, whose crossing gives the result.

\subsubsection{Setup and decomposition}
Set $a=v_1(q)$ and $b=v_1(p)$. For each $i$, let $\tau_i(q)=v_{i+1}(q)$ and $\tau_i(p)=v_{i+1}(p)$. By Lemma~\ref{lem:dp_monotonicity_scaling}, we have $\tau_i(q)\le\tau_i(p)$.

We split the possible values of $X_i$ into the ranges $[0,\tau_i(q))$, $[\tau_i(q),\tau_i(p))$, and $[\tau_i(p),+\infty)$. Define
\begin{align}
    m_i &=\mathbb E\left[ (X_i-\tau_i(q))^+-(X_i-\tau_i(p))^+ \right], \notag\\
    h_i &=\mathbb E[(X_i-\tau_i(p))^+]. \label{eq:mi_hi_definitions}
\end{align}
The quantity $m_i$ is the expected contribution between the two thresholds: the random difference inside its expectation is $0$ below $\tau_i(q)$, equals $X_i-\tau_i(q)$ between the thresholds, and is capped at $\tau_i(p)-\tau_i(q)$ above the higher threshold. The quantity $h_i$ is the expected excess above $\tau_i(p)$. In particular, $m_i\ge0$. Moreover,
\[
\mathbb E[(X_i-\tau_i(q))^+]=m_i+h_i.
\]
Telescoping the two DP recurrences therefore gives
\begin{align}
    a=q\sum_{i=1}^n(m_i+h_i), \qquad b=p\sum_{i=1}^n h_i. \label{eq:ab_mh}
\end{align}
Let $m=\sum_i m_i$. Equation~\eqref{eq:ab_mh} then implies $m=a/q-b/p$.

Define the random variable
\begin{equation} Y_i=\bigl(\min\{X_i,\tau_i(p)\}-a\bigr)^+. \label{eq:Yi_definition} \end{equation}
If $\tau_i(p)\le a$, then $Y_i$ is identically zero. If $\tau_i(p)>a$, then $Y_i=0$ for $X_i\le a$, it equals $X_i-a$ for $a<X_i\le\tau_i(p)$, and it is capped at $\tau_i(p)-a$ above the higher threshold. Thus, in all cases, $0\le Y_i\le b-a$, where we used $\tau_i(p)\le v_1(p)=b$.

The point of this definition is that a prophet value above the baseline $a$ can be split into two pieces. The part above $\tau_i(p)$ is paid for by the usual DP exceedance at the higher sampling rate, while $Y_i$ is exactly the middle part that remains.

For $r\in(0,1]$, let $I_i^r$ be the indicator that item $i$ survives $r$-subsampling. The indicators are mutually independent and independent of the values $X_i$. Thus the prophet's realized value at rate $r$ is $M_r=\max_i I_i^rX_i$.

\begin{lemma}
\label{lem:opt_upper_midrange}
For every $r\in(0,1]$,
\begin{equation} \OPT(r) \le a+\frac rp\,b +\mathbb E\left[\max_{i\in[n]}I_i^rY_i\right]. \label{eq:opt_upper_midrange} \end{equation}
\end{lemma}

\begin{proof}
Fix a realization of the values and the survival indicators. If $M_r$ is attained at a surviving index $k$, then
\begin{equation}
\begin{aligned}
    (M_r-a)^+ &=(X_k-a)^+\\
    &\le (X_k-\tau_k(p))^++Y_k\\
    &\le \sum_{i=1}^n I_i^r(X_i-\tau_i(p))^+ +\max_{i\in[n]}I_i^rY_i.
\end{aligned}
\label{eq:pointwise-prophet-split}
\end{equation}
The first inequality follows directly from the definition of $Y_k$. If $\tau_k(p)>a$, it is an equality: the two terms split $(X_k-a)^+$ at $\tau_k(p)$. If $\tau_k(p)\le a$, then $Y_k=0$ and $(X_k-\tau_k(p))^+\ge(X_k-a)^+$. The final inequality in \eqref{eq:pointwise-prophet-split} replaces the contribution of index $k$ by the sum and maximum over all surviving indices. The same inequality is immediate if no item survives, because then $M_r=0$.

Taking expectations in \eqref{eq:pointwise-prophet-split} and using independence of $I_i^r$ and $X_i$ gives
\begin{equation}
\begin{aligned}
    \mathbb E[(M_r-a)^+] &\le r\sum_{i=1}^n\mathbb E[(X_i-\tau_i(p))^+] +\mathbb E\left[\max_i I_i^rY_i\right]
    =\frac rp\,b+\mathbb E\left[\max_i I_i^rY_i\right],
\end{aligned}
\label{eq:expected-prophet-split}
\end{equation}
where the last equality follows from \eqref{eq:ab_mh}. Since $M_r\le a+(M_r-a)^+$ pointwise, the  bound follows.
\end{proof}

\subsubsection{Bounding the middle contribution}
It remains to control the expected maximum of the bounded random variables $I_i^rY_i$. To obtain a convenient upper bound, we consider replacing each $I_i^r Y_i$ by a Bernoulli variable that takes value either $0$ or the maximum possible value $b-a$, so that the expected maximum does not decrease. Then for such Bernoulli variables, the expected maximum has a simple formula.

For notational convenience, define the fixed quantity $d_i=a-\tau_i(q)\ge0$; its nonnegativity follows from $\tau_i(q)=v_{i+1}(q)\le v_1(q)=a$.

\begin{lemma}
\label{lem:midrange_bernoulli_replacement}
For every $r\in(0,1]$,
\begin{equation} \mathbb E\left[\max_{i\in[n]}I_i^rY_i\right] \le (b-a)\left( 1-\prod_{i=1}^n \left(1-\frac{rm_i}{b-a+d_i}\right) \right). \label{eq:bernoulli_product_bound} \end{equation}
\end{lemma}

\begin{proof}
If $a=b$, then every $Y_i$ is identically $0$ and the claim is trivial, so assume $a<b$.

First we relate $m_i$ to the size and probability of $Y_i$. If $Y_i>0$, then
$\min\{X_i,\tau_i(p)\}>a$, while $a\ge\tau_i(q)$. If $X_i>\tau_i(p)$, then the random difference inside the expectation defining $m_i$ satisfies
\begin{align*}
    (X_i-\tau_i(q))^+-(X_i-\tau_i(p))^+ &=\tau_i(p)-\tau_i(q)\\
    &=Y_i+d_i.
\end{align*}
If instead $X_i\le\tau_i(p)$, the same difference is $X_i-\tau_i(q)=Y_i+d_i$. Outside the event $Y_i>0$, that random difference remains nonnegative. Taking expectations therefore gives
\begin{equation} m_i\ge \mathbb E[Y_i]+d_i\Pr[Y_i>0]. \label{eq:mi_controls_Yi} \end{equation}
Moreover, the random difference inside the definition of $m_i$ is pointwise at most
$\tau_i(p)-\tau_i(q)=(\tau_i(p)-a)+d_i\le b-a+d_i$.
Consequently $0\le rm_i/(b-a+d_i)\le1$.

Fix an index $i$. Let $M_i$ denote the maximum of all current coordinates other than the $i$-th one. Initially these coordinates are $I_j^rY_j$, and later in the argument some of them may already have been replaced by independent Bernoulli variables. In either case, $M_i$ is independent of $I_i^r$ and $Y_i$ and satisfies $0\le M_i\le b-a$.

We claim that
\begin{equation} (Y_i-M_i)^+ \le \mathbf 1_{\{Y_i>0\}} \frac{Y_i+d_i}{b-a+d_i} (b-a-M_i). \label{eq:Yi_excess_comparison} \end{equation}
When $Y_i\le M_i$, the left-hand side is $0$. When $Y_i>M_i$, after dividing by the positive denominators this is equivalent to \[ \frac{Y_i-M_i}{b-a-M_i} \le \frac{Y_i+d_i}{b-a+d_i}, \] and cross-multiplication leaves $((b-a)-Y_i)(M_i+d_i)\ge0$.

Since $I_i^r=1$ with probability $r$ independently of everything else, \eqref{eq:Yi_excess_comparison} and \eqref{eq:mi_controls_Yi} imply
\begin{align}
    \mathbb E[(I_i^rY_i-M_i)^+] &=r\,\mathbb E[(Y_i-M_i)^+] \notag\\
    &\le \frac{r\bigl(\mathbb E[Y_i]+d_i\Pr[Y_i>0]\bigr)}{b-a+d_i} \mathbb E[b-a-M_i] \notag\\
    &\le \frac{rm_i}{b-a+d_i}\mathbb E[b-a-M_i]. \label{eq:current_coordinate_gain}
\end{align}
The second line also uses the independence of $M_i$ and $Y_i$ to factor the expectation of the product in \eqref{eq:Yi_excess_comparison}.

Now choose $Y_i'$ independently of all variables currently present, so that it equals $b-a$ with probability $rm_i/(b-a+d_i)$ and equals $0$ otherwise. We also choose the replacement variables for different coordinates to be mutually independent. Then
\begin{align*}
    \mathbb E[(Y_i'-M_i)^+] &=\frac{rm_i}{b-a+d_i}\mathbb E[b-a-M_i]
    \ge \mathbb E[(I_i^rY_i-M_i)^+].
\end{align*}
Since $\max\{W,M_i\}=M_i+(W-M_i)^+$ for any random variable $W$, replacing $I_i^rY_i$ by $Y_i'$ cannot decrease the expected maximum.

Perform this replacement one coordinate at a time. After all coordinates have been replaced, the resulting random variables are independent, each is either $0$ or $b-a$, with $\Pr[Y_i'=b-a]=rm_i/(b-a+d_i)$. Their maximum equals $b-a$ unless every coordinate is zero. Therefore
\begin{align*}
    \mathbb E\left[\max_i I_i^rY_i\right] \le \mathbb E\left[\max_i Y_i'\right]
    =(b-a)\left( 1-\prod_{i=1}^n \left(1-\frac{rm_i}{b-a+d_i}\right) \right),
\end{align*}
which proves the lemma.
\end{proof}

\subsubsection{The lower sampling rate}
We first apply Lemmas~\ref{lem:opt_upper_midrange} and~\ref{lem:midrange_bernoulli_replacement} at $r=q$. Telescoping the $q$-rate DP gives
\begin{align}
    d_i =v_1(q)-v_{i+1}(q)
    =q\sum_{j=1}^i\mathbb E[(X_j-\tau_j(q))^+]
    \ge q\sum_{j=1}^i m_j.
    \label{eq:di-lower-middle-mass}
\end{align}
The inequality follows because the definition \eqref{eq:mi_hi_definitions} gives
$m_j\le\mathbb E[(X_j-\tau_j(q))^+]$ for every $j$. Substituting
\eqref{eq:di-lower-middle-mass} into each factor in
\eqref{eq:bernoulli_product_bound} yields
\begin{align}
    1-\frac{qm_i}{b-a+d_i} &\ge 1-\frac{qm_i}{b-a+q\sum_{j=1}^i m_j}
    =\frac{b-a+q\sum_{j=1}^{i-1}m_j} {b-a+q\sum_{j=1}^{i}m_j}.
    \label{eq:q-product-factor}
\end{align}
Multiplying these inequalities causes the right-hand side to telescope:
\begin{equation} \prod_{i=1}^n \left(1-\frac{qm_i}{b-a+d_i}\right) \ge \frac{b-a}{b-a+qm}. \label{eq:q_product_telescope} \end{equation}
Using $m=a/q-b/p$ and $\rho=q/p$, the denominator simplifies to
\begin{equation}
    b-a+qm=(b-a)+q\left(\frac aq-\frac bp\right)=(1-\rho)b.
    \label{eq:q-product-denominator}
\end{equation}
This suggests the natural parameter
\begin{equation} \phi=\frac{b-a}{(1-\rho)b}. \label{eq:phi_definition_appendix} \end{equation}
Lemma~\ref{lem:dp_monotonicity_scaling} gives $a\le b$ and
$a/q\ge b/p$, with the latter inequality equivalent to $a\ge\rho b$.
The standing nontriviality assumption gives $b>0$, so
$\rho b\le a\le b$ implies $\phi\in[0,1]$.

If $a=b$, then $\phi=0$ and every $Y_i$ is zero. Lemma~\ref{lem:opt_upper_midrange} at rates $q$ and $p$ respectively gives
$\OPT(q)\le a+\rho b=(1+\rho)a$ and $\OPT(p)\le a+b=2b$.
Equivalently, $CR(q)\le1+\rho$ and $CR(p)\le2$, which are precisely the bounds $f(0)$ and $g(0)$ derived below. We therefore assume for the intermediate calculations that $a<b$; the resulting formulas extend continuously to $\phi=0$.

Equations~\eqref{eq:q_product_telescope}--\eqref{eq:phi_definition_appendix}
show that the product in \eqref{eq:bernoulli_product_bound} is at least
$\phi$. Applying Lemma~\ref{lem:midrange_bernoulli_replacement} with $r=q$ therefore gives
\begin{align*}
    \mathbb E\left[\max_i I_i^qY_i\right] &\le (b-a)(1-\phi)
    =(1-\rho)b\phi(1-\phi).
\end{align*}
The definition \eqref{eq:phi_definition_appendix} also gives
$a=b-(b-a)=b(1-(1-\rho)\phi)$. Lemma~\ref{lem:opt_upper_midrange} now yields
\begin{align*}
    \OPT(q) &\le a+\rho b+\mathbb E\left[\max_i I_i^qY_i\right]\\
    &\le b\bigl(1-(1-\rho)\phi\bigr)+\rho b +(1-\rho)b\phi(1-\phi)\\
    &=b\bigl(1+\rho-(1-\rho)\phi^2\bigr).
\end{align*}
Therefore
\begin{equation} CR(q)=\frac{\OPT(q)}{a} \le \frac {1+\rho-(1-\rho)\phi^2}{1-(1-\rho)\phi} :=f(\phi). \label{eq:f_phi_bound} \end{equation}

\subsubsection{Transferring the product bound}
To bound $CR(p)$, define the fixed numbers
$\theta_i=m_i/(b-a+d_i)\in[0,1]$. The proof of
Lemma~\ref{lem:midrange_bernoulli_replacement} established the stated range.
Equations~\eqref{eq:q_product_telescope}--\eqref{eq:phi_definition_appendix}
give the product bound
\begin{equation} \prod_{i=1}^n(1-q\theta_i)\ge\phi. \label{eq:q_theta_product} \end{equation}
We now ask the following question: among all collections $\theta_i\in[0,1]$ satisfying \eqref{eq:q_theta_product}, how small can the corresponding product $\prod_{i=1}^n (1-p\theta_i)$ be? Define
\begin{equation} \Gamma_{q,p}(\phi) =\inf\left\{ \prod_{i=1}^k(1-p\theta_i): k\ge1,\ \theta_i\in[0,1],\ \prod_{i=1}^k(1-q\theta_i)\ge\phi \right\}, \label{eq:Gamma_definition_appendix} \end{equation}
with $\Gamma_{q,p}(0)=0$, as in Definition~\ref{def:Gamma}. Thus \eqref{eq:q_theta_product} immediately implies $\prod_i(1-p\theta_i)\ge\Gamma_{q,p}(\phi)$.

The collection giving the smallest product has a simple form: as many coordinates as possible are pushed to $1$, at most one coordinate is fractional, and the remaining coordinates are $0$. We give the exact formula because the same structure will later appear in the tight example.

\begin{lemma}
\label{lem:Gamma_explicit}
For $\phi\in(0,1]$, choose an integer $\ell\ge0$ satisfying $(1-q)^{\ell+1}\le\phi\le(1-q)^\ell$. Then
\begin{equation} \Gamma_{q,p}(\phi) =(1-p)^\ell \left( 1-\frac1\rho \left(1-\frac{\phi}{(1-q)^\ell}\right) \right). \label{eq:Gamma_explicit_appendix} \end{equation}
In particular, $\Gamma_{q,p}$ is continuous and nondecreasing on $[0,1]$.
\end{lemma}

\begin{proof}
First suppose $p<1$. Let $u_i=1-q\theta_i$, so $u_i\in[1-q,1]$. Since $p/q=1/\rho$,
\begin{align*}
    1-p\theta_i &=1-\frac1\rho(1-u_i) =\frac1\rho u_i-\left(\frac1\rho-1\right).
\end{align*}
The right-hand side is increasing in $u_i$. Starting from any permitted collection with $\prod_i u_i>\phi$, we may therefore decrease one or more coordinates until the product equals $\phi$, without increasing the product we seek to minimize. If the existing coordinates reach the lower endpoint $1-q$ before equality is reached, we append a coordinate initially equal to $1$ and decrease it; repeating this operation reaches equality after finitely many steps because $\phi>0$. It therefore suffices to minimize under the equality constraint $\prod_i u_i=\phi$.

Write $z_i=-\log u_i$, let $\lambda=-\log(1-q)$, and let $s=-\log\phi$. Then $z_i\in[0,\lambda]$ and the product constraint becomes $\sum_i z_i=s$. Taking logarithms of the product reduces the problem to minimizing
\begin{align*}
    \sum_i h(z_i), \qquad h(z)=\log\left( \frac1\rho e^{-z}-\left(\frac1\rho-1\right) \right).
\end{align*}
The argument of the logarithm is positive on $[0,\lambda]$ because its minimum is the value $1-p>0$ attained at $z=\lambda$.
A direct calculation gives
\begin{equation} h''(z) =-\frac{\frac1\rho\left(\frac1\rho-1\right)e^{-z}} {\left( \frac1\rho e^{-z}-\left(\frac1\rho-1\right) \right)^2}<0, \label{eq:h_concavity_appendix} \end{equation}
so $h$ is concave.

Suppose two coordinates $z_i,z_j$ both lie strictly inside $(0,\lambda)$. Keeping $z_i+z_j$ fixed, the function $t\mapsto h(t)+h(z_i+z_j-t)$ is concave on its permitted interval, so its minimum is attained at an endpoint. We may therefore move mass between the two coordinates until at least one reaches $0$ or $\lambda$, without increasing the sum being minimized. Repeating this operation transforms every permitted vector into one in which all but at most one coordinate belong to $\{0,\lambda\}$, without increasing that sum.

The choice of $\ell$ in the lemma is equivalent to $\ell\lambda\le s\le(\ell+1)\lambda$. Hence the resulting vector has $\ell$ coordinates equal to $\lambda$, one coordinate equal to $s-\ell\lambda$, and all remaining coordinates equal to $0$. Translating back to the $u_i$ variables, there are $\ell$ coordinates equal to $1-q$, one coordinate equal to $\phi/(1-q)^\ell$, and all remaining coordinates equal to $1$.

Substituting these values into the $p$-rate product gives
\begin{align*}
    \Gamma_{q,p}(\phi) &=(1-p)^\ell \left( \frac1\rho\frac{\phi}{(1-q)^\ell} -\left(\frac1\rho-1\right) \right)
    =(1-p)^\ell \left( 1-\frac1\rho \left(1-\frac{\phi}{(1-q)^\ell}\right) \right),
\end{align*}
which is \eqref{eq:Gamma_explicit_appendix}. The displayed choice satisfies the constraint, and the preceding transformations show that no permitted choice gives a smaller product. This proves the formula for $p<1$.

It remains to consider $p=1$. If $0<\phi\le1-q$, choosing one coordinate $\theta_1=1$ satisfies the constraint and makes the product zero. This agrees with \eqref{eq:Gamma_explicit_appendix}, where one may choose $\ell\ge1$. If $1-q<\phi\le1$, then $\ell=0$ and $s<\lambda$. No coordinate with $\theta_i=1$ can satisfy the constraint. Because the coordinates $z_i$ are nonnegative and sum to $s<\lambda$, the same concavity argument leaves one coordinate equal to $s$ and all others equal to zero. Hence $u_1=\phi$, so $\theta_1=(1-\phi)/q$ and
\[
\Gamma_{q,1}(\phi)=1-\frac{1-\phi}{q},
\]
which is again \eqref{eq:Gamma_explicit_appendix}, since $\rho=q$ when $p=1$.

Monotonicity follows directly from the definition: increasing $\phi$ leaves fewer permitted collections. For continuity away from zero, the explicit piecewise formulas agree at their common endpoints, including endpoints where either allowed choice of $\ell$ may be used. When $p<1$, the remaining fractional factor in \eqref{eq:Gamma_explicit_appendix} lies in $[1-p,1]$, so $0\le\Gamma_{q,p}(\phi)\le(1-p)^\ell$. As $\phi\downarrow0$, we have $\ell\to\infty$, and hence $\Gamma_{q,p}(\phi)\to0$. When $p=1$, the function is identically zero on $[0,1-q]$. This proves continuity at zero and completes the proof.
\end{proof}

\subsubsection{The higher sampling rate}
Applying Lemma~\ref{lem:midrange_bernoulli_replacement} with $r=p$ and using
$\prod_{i=1}^n(1-q\theta_i)\ge\phi$ gives
\begin{align*}
    \mathbb E\left[\max_i I_i^pY_i\right] &\le (b-a)\bigl(1-\Gamma_{q,p}(\phi)\bigr)
    =(1-\rho)b\phi\bigl(1-\Gamma_{q,p}(\phi)\bigr).
\end{align*}
Applying Lemma~\ref{lem:opt_upper_midrange} at rate $p$ therefore yields
\begin{align*}
    \OPT(p) &\le a+b+\mathbb E\left[\max_i I_i^pY_i\right]\\
    &\le b\bigl(1-(1-\rho)\phi\bigr)+b +(1-\rho)b\phi\bigl(1-\Gamma_{q,p}(\phi)\bigr)\\
    &=b\bigl(2-(1-\rho)\phi\Gamma_{q,p}(\phi)\bigr).
\end{align*}
Thus
\begin{equation} CR(p)=\frac{\OPT(p)}{b} \le 2-(1-\rho)\phi\Gamma_{q,p}(\phi):=g(\phi). \label{eq:g_phi_bound} \end{equation}

\subsubsection{Completing the proof}
The preceding arguments are summarized in the following lemma.

\begin{lemma}
\label{lem:joint_lower_bound}
For every nontrivial instance, define $a=v_1(q)$, $b=v_1(p)$, and $\phi=(b-a)/((1-\rho)b)$. Then $\phi\in[0,1]$ and
\begin{equation} CR(q)\le f(\phi), \qquad CR(p)\le g(\phi), \label{eq:two_rate_frontier_summary} \end{equation}
\end{lemma}

\begin{proof}
The range $\phi\in[0,1]$ was established after
\eqref{eq:phi_definition_appendix}. The two bounds are exactly
\eqref{eq:f_phi_bound} and \eqref{eq:g_phi_bound}, including the endpoint
$\phi=0$ handled there.
\end{proof}

The two bounds in Lemma~\ref{lem:joint_lower_bound} make the source of the simultaneous improvement explicit. The function $f$ is continuous and nondecreasing on $[0,1]$. Indeed,
\begin{equation}
f'(\phi)=\frac{(1-\rho)(1-\phi)\bigl(1+\rho-(1-\rho)\phi\bigr)}{\bigl(1-(1-\rho)\phi\bigr)^2}\ge0,
\label{eq:f_derivative_appendix}
\end{equation}
while $g$ is continuous and nonincreasing because $\Gamma_{q,p}$ is nonnegative and nondecreasing, and hence so is $\phi\Gamma_{q,p}(\phi)$. At the endpoints,
\begin{align*}
(f(0),g(0))=(1+\rho,2), \qquad
(f(1),g(1))=(2,1+\rho).
\end{align*}
We have $f(0)<g(0)$ and $f(1)>g(1)$, so continuity gives a crossing point $\phi^*\in(0,1)$. Moreover, \eqref{eq:f_derivative_appendix} is strictly positive for $0\le\phi<1$, while $g$ is nonincreasing. Hence the crossing point is unique. Since $\phi^*<1$ and $f(1)=2$, its common value satisfies $f(\phi^*)=g(\phi^*)<2$.

The crossing also gives the worst tradeoff. If $\phi\le\phi^*$, then
$\min\{f(\phi),g(\phi)\}\le f(\phi)\le f(\phi^*)$. If $\phi\ge\phi^*$, then
$\min\{f(\phi),g(\phi)\}\le g(\phi)\le g(\phi^*)$. Therefore
\[
\zeta(q,p)=\max_{\phi\in[0,1]}\min\{f(\phi),g(\phi)\},
\]
and the maximum is attained at $\phi^*$ and is strictly smaller than $2$.

Finally, every nontrivial instance has an associated parameter $\phi$, and
Lemma~\ref{lem:joint_lower_bound} gives
\[
\min\{CR(q),CR(p)\}
\le \min\{f(\phi),g(\phi)\}
\le \zeta(q,p)<2.
\]
This proves Theorem~\ref{thm:tight_two_rate_prophet}. The construction below shows that this upper bound is tight.

\subsection{Tight construction}

The construction follows the equality conditions suggested by the proof. The last stage creates the single-rate indifference at the higher sampling rate. The middle stage realizes the extremal product in Lemma~\ref{lem:Gamma_explicit}. The first stage then sets the low-rate baseline so that the algorithm is also indifferent there.

Fix $\phi\in(0,1]$, and let $\ell\ge0$ satisfy $(1-q)^{\ell+1}\le\phi\le(1-q)^\ell$. Define
\begin{align*}
    \theta &=\frac1q\left(1-\frac{\phi}{(1-q)^\ell}\right), & c&=1-(1-\rho)\phi.
\end{align*}
The choice of $\ell$ ensures $\theta\in[0,1]$. We construct an instance with three stages.

\begin{itemize}
\item \textbf{Stage 1.} There are $N$ deterministic variables, each equal to $c$.

\item \textbf{Stage 2.} There are $\ell$ deterministic variables, each equal to $1$, followed by a random variable $W$ with
    \[
        W=
        \begin{cases}
            1, & \text{with probability }\theta,\\
            0, & \text{otherwise}.
        \end{cases}
    \]

\item \textbf{Stage 3.} There is one rare large random variable $Z$ with
    \[
        Z=
        \begin{cases}
            \dfrac1\varepsilon, & \text{with probability }\dfrac{\varepsilon}{p},\\
            0, & \text{otherwise},
        \end{cases}
    \]
    where $0<\varepsilon<p$.
\end{itemize}

We take the limits first as $N\to\infty$ and then as $\varepsilon\to0$.

We first compute the DP values. Under $p$-subsampling, Stage 3 has continuation value $p\mathbb E[Z]=1$. Hence the algorithm is indifferent between accepting any realized value $1$ in Stage 2 and waiting for Stage 3. The value entering Stage 2 is therefore exactly $1$. Since $c\le1$, Stage 1 cannot improve this value, so $v_1(p)=1$.

Under $q$-subsampling, the Stage 3 continuation value is $q\mathbb E[Z]=q/p=\rho$. The probability that no positive value from Stage 2 survives is
\begin{align*}
    (1-q)^\ell(1-q\theta) &=(1-q)^\ell\frac{\phi}{(1-q)^\ell} =\phi.
\end{align*}
Thus the DP value upon entering Stage 2 is
\begin{align*}
    (1-\phi)\cdot1+\phi\cdot\rho &=1-(1-\rho)\phi=c.
\end{align*}
The Stage 1 values are exactly $c$, so the algorithm is indifferent throughout Stage 1 and $v_1(q)=c$.

We next compute the prophet benchmarks. Under $q$-subsampling, the event that $Z$ realizes its large value and survives has probability $q\varepsilon/p=\rho\varepsilon$. Since all earlier values are at most $1$, for $\varepsilon<1$ the rare large value dominates whenever this event occurs. As $N\to\infty$, at least one Stage 1 value survives with probability tending to $1$. Therefore, after taking the two limits in the prescribed order,
\begin{align*}
    \OPT(q) &\longrightarrow \rho+(1-\phi)\cdot1+\phi c
    =1+\rho-(1-\rho)\phi^2.
\end{align*}
Consequently
\begin{align*}
    CR(q) &\longrightarrow \frac{1+\rho-(1-\rho)\phi^2}{1-(1-\rho)\phi} =f(\phi).
\end{align*}

For $p$-subsampling, Stage 3 contributes asymptotic expected value $1$. The probability that no positive Stage 2 value survives is $(1-p)^\ell(1-p\theta)$. By Lemma~\ref{lem:Gamma_explicit} and our choice of $\theta$, this is exactly $\Gamma_{q,p}(\phi)$. Hence
\begin{align*}
    \OPT(p) &\longrightarrow 1+\bigl(1-\Gamma_{q,p}(\phi)\bigr)\cdot1 +\Gamma_{q,p}(\phi)\cdot c
    =2-(1-\rho)\phi\Gamma_{q,p}(\phi),
\end{align*}
while $v_1(p)=1$. Therefore
\begin{align*}
    CR(p) &\longrightarrow {2-(1-\rho)\phi\Gamma_{q,p}(\phi)} =g(\phi).
\end{align*}

Thus every $\phi\in(0,1]$ is asymptotically attainable on the frontier; the endpoint $\phi=0$ follows by taking $\phi\downarrow0$. Choosing a crossing point $\phi^\star$ maximizing $\min\{f(\phi),g(\phi)\}$ gives instances for which
\begin{align*}
    \min\{CR(q),CR(p)\} &\longrightarrow \zeta(q,p).
\end{align*}
This proves asymptotic tightness and completes the proof of the theorem.

\subsection{Proof of the closed-form corollary}

\begin{proof}[Proof of Corollary~\ref{cor:closed_form_prophet}]
We next derive a closed-form upper bound that depends only on $\rho=q/p$.

Set $\phi_0=1-\rho/2$. For the parameter $\phi$ associated with an arbitrary instance by Lemma~\ref{lem:joint_lower_bound}, there are two cases. If $\phi\le\phi_0$, monotonicity of $f$ gives $CR(q)\le f(\phi)\le f(\phi_0)$. If $\phi\ge\phi_0$, monotonicity of $g$ gives $CR(p)\le g(\phi)\le g(\phi_0)$. It therefore remains to compare $g(\phi_0)$ with $f(\phi_0)$.

We first lower-bound $\Gamma_{q,p}(\phi_0)$ without using its explicit piecewise formula. Couple the two subsamples by generating the $p$-subsample first and then retaining each surviving item independently with probability $\rho=q/p$. For any fixed collection of independent ``success'' events, conditional on at least one success surviving the $p$-subsample, the probability that at least one of those successes remains after the additional $\rho$-thinning is at least $\rho$. Hence
\begin{equation} \Pr[\text{success under }q] \ge \rho\,\Pr[\text{success under }p]. \label{eq:success_thinning_coupling} \end{equation}
If the no-success probability under $q$ is at least $\phi_0=1-\rho/2$, then the success probability under $q$ is at most $\rho/2$. Equation~\eqref{eq:success_thinning_coupling} implies that the success probability under $p$ is at most $1/2$, and therefore the no-success probability under $p$ is at least $1/2$. By the definition of the conversion function, $\Gamma_{q,p}(\phi_0)\ge1/2$.

Consequently,
\begin{align*}
g(\phi_0)&=2-(1-\rho)\phi_0\Gamma_{q,p}(\phi_0)
\le 2-\frac{(1-\rho)\phi_0}{2}
\le f(\phi_0),
\end{align*}
where the last inequality is a direct algebraic simplification for $0<\rho<1$. Thus in both cases $\min\{CR(q),CR(p)\}\le f(\phi_0)$.

It remains to evaluate $f(\phi_0)$. Since
\begin{align*}
1-(1-\rho)\phi_0&=\frac{\rho(3-\rho)}{2},\\
1+\rho-(1-\rho)\phi_0^2&=\frac{\rho(\rho^2-5\rho+12)}{4},
\end{align*}
we obtain
\begin{align*}
\min\{CR(q),CR(p)\}&\le f(\phi_0)=\frac{\rho^2-5\rho+12}{2(3-\rho)}
=2-\frac{\rho(1-\rho)}{2(3-\rho)}<2.
\end{align*}
This is the closed-form bound in Corollary~\ref{cor:closed_form_prophet}.
\end{proof}

\section{Realizing Curves as Ski-Rental Algorithms}
\label{sec:curve-realization-appendix}

This appendix proves the residual-curve realization lemma used in the ski-rental results. The proof is based on an elementary identity for the expected cost of a buy-time distribution. It also records two simple admissibility tools used by the positive constructions: a tail calculation for late buying and a one-sided smoothing lemma for downward jumps.

\subsection{General residual realization}
\label{sec:general-residual-realization}

For any buy-time CDF $F$ and any finite horizon $y$, integration by parts gives
\begin{equation}
C_F(y)=\int_{[0,y]}(x+1)\,dF(x)+y(1-F(y))=y+F(y)-\int_0^yF(x)dx.
\label{eq:cost-from-cdf}
\end{equation}
Thus perturbing a buying CDF by a function $\Phi$ perturbs the cost curve by $\Phi(y)-\int_0^y\Phi(x)dx$. The definition of $\Phi_A$ in the main text is exactly the inverse of this relation.

\begin{proof}[Proof of Lemma~\ref{lem:certificate-realization}]
Recall that
\[
\eta_A=\frac{\mathcal L(A)}{e-1}
\]
and
\[
\Phi_A(y)=A(y)-\eta_A\OPT(y)+e^y\int_0^y\bigl(A(x)-\eta_A\OPT(x)\bigr)e^{-x}dx.
\]
Since $\mathcal L(\OPT)=e-1$, the definition of $\eta_A$ gives
\begin{equation}
\int_0^\infty\bigl(A(x)-\eta_A\OPT(x)\bigr)e^{-x}dx=0.
\label{eq:residual-zero-mass}
\end{equation}

We first verify the inverse relation. By Fubini's theorem,
\[
\begin{aligned}
\int_0^y\Phi_A(t)dt
&=\int_0^y\bigl(A(t)-\eta_A\OPT(t)\bigr)dt+\int_0^y e^t\int_0^t\bigl(A(x)-\eta_A\OPT(x)\bigr)e^{-x}dx\,dt\\
&=\int_0^y\bigl(A(x)-\eta_A\OPT(x)\bigr)dx+\int_0^y(e^y-e^x)\bigl(A(x)-\eta_A\OPT(x)\bigr)e^{-x}dx\\
&=e^y\int_0^y\bigl(A(x)-\eta_A\OPT(x)\bigr)e^{-x}dx.
\end{aligned}
\]
Therefore
\begin{equation}
\Phi_A(y)-\int_0^y\Phi_A(x)dx=A(y)-\eta_A\OPT(y).
\label{eq:phi-inverse}
\end{equation}

Let $F^\star$ be the classical equalizer CDF,
\[
F^\star(y)=
\begin{cases}
\dfrac{e^y-1}{e-1}, & 0\leq y\leq1,\\
1, & y\geq1,
\end{cases}
\]
and define
\[
F_\varepsilon(y):=F^\star(y)+\varepsilon\Phi_A(y).
\]
Assuming for the moment that $F_\varepsilon$ is a valid CDF, Equations~\eqref{eq:cost-from-cdf} and~\eqref{eq:phi-inverse} give
\[
C_{F_\varepsilon}(y)-C_{F^\star}(y)=\varepsilon\left(\Phi_A(y)-\int_0^y\Phi_A(x)dx\right)=\varepsilon\bigl(A(y)-\eta_A\OPT(y)\bigr).
\]
Since $C_{F^\star}(y)=R_{\mathrm{ski}}\OPT(y)$,
\[
C_{F_\varepsilon}(y)=\bigl(R_{\mathrm{ski}}-\varepsilon\eta_A\bigr)\OPT(y)+\varepsilon A(y),
\]
so the residual at target ratio $R_{\mathrm{ski}}-\varepsilon\eta_A$ is exactly $\varepsilon A$.

It remains to verify that $F_\varepsilon$ is a CDF for sufficiently small $\varepsilon$. We have $\Phi_A(0)=0$. Using \eqref{eq:residual-zero-mass}, we may also write
\begin{equation}
\Phi_A(y)=A(y)-\eta_A\OPT(y)-e^y\int_y^\infty\bigl(A(x)-\eta_A\OPT(x)\bigr)e^{-x}dx.
\label{eq:phi-tail-form}
\end{equation}
Since $A$ has a finite limit and $\OPT(y)\to1$, the two terms on the right cancel in the limit, so $\Phi_A(y)\to0$. Hence $F_\varepsilon(0)=0$ and $\lim_{y\to\infty}F_\varepsilon(y)=1$.

The integral term in the definition of $\Phi_A$ is continuous, as is $\OPT$, so every jump of $\Phi_A$ is exactly the corresponding jump of $A$. In particular, all jumps at or below time $1$ are nonnegative. On a smooth piece below time $1$,
\[
\Phi_A'(y)=\Phi_A(y)+A'(y)-\eta_A.
\]
Because there are finitely many pieces and by the assumptions of the lemma statement, $\Phi_A'$ is bounded below on $(0,1)$. Hence
\[
F_\varepsilon'(y)=\frac{e^y}{e-1}+\varepsilon\Phi_A'(y)\geq0
\]
for sufficiently small $\varepsilon$. On $[1,\infty)$, $F^\star$ is constant and $\Phi_A$ is nondecreasing by hypothesis. Thus $F_\varepsilon$ is a valid CDF.

Finally, if $A$ is constant on $[1,\infty)$, then for $y\geq1$ Equation~\eqref{eq:phi-tail-form} gives $\Phi_A(y)=0$. Hence $F_\varepsilon(y)=F^\star(y)=1$ for every $y\geq1$, so the realized algorithm is supported on $[0,1]$.
\end{proof}

\subsection{Admissibility checks}
\label{sec:realization-admissibility-checks}

The tail form \eqref{eq:phi-tail-form} simplifies further after the breakeven time. Since $\OPT(x)=1$ for $x\geq1$, the $\eta_A$ terms cancel and
\begin{equation}
\Phi_A(y)=A(y)-e^y\int_y^\infty A(x)e^{-x}dx,
\qquad y\geq1.
\label{eq:phi-tail-A-only}
\end{equation}
Thus the realizability of a nonconstant tail can be checked directly from $A$, independently of the improvement $\eta_A$. In particular, a constant tail gives $\Phi_A=0$ and hence a buy-by-breakeven-time algorithm.

The unrestricted positive constructions use the following simple nonconstant tail.

\begin{lemma}[Tail check for the late-buying construction]
\label{lem:late-tail-realization-check}
Let $T>1$, and suppose a candidate residual curve $A$ has
\[
A(y)=
\begin{cases}
y-T-1, & 1\leq y<T,\\
0, & y\geq T.
\end{cases}
\]
Then $\Phi_A(y)=-1$ for $1\leq y<T$ and $\Phi_A(y)=0$ for $y\geq T$. Consequently, $\Phi_A$ is nondecreasing on $[1,\infty)$, with a single upward jump of size one at $T$.
\end{lemma}
\begin{proof}
For $1\leq y<T$, Equation~\eqref{eq:phi-tail-A-only} gives
\[
\Phi_A(y)=y-T-1-e^y\int_y^T(x-T-1)e^{-x}dx=-1.
\]
For $y\geq T$, both terms in \eqref{eq:phi-tail-A-only} are zero. Thus $\Phi_A$ is constant on $(1,T)$ and $(T,\infty)$ and has an upward jump of size one at $T$.
\end{proof}

\subsection{One-sided smoothing}

The constructions below are easiest to describe using piecewise-defined residual curves, which may initially have downward jumps. Since $\Phi_A$ has exactly the same jumps as $A$, a downward jump would require a negative buying atom. The smoothing lemma below replaces each downward jump by a short continuous transition. Since the smoothing only lowers $A$, it preserves the two-scale inequality~\eqref{eq:two-scale-A}, while weighted convergence preserves positivity of $\mathcal L(A)$ for sufficiently short smoothing intervals.

\begin{lemma}[One-sided smoothing of downward jumps]
\label{lem:one-sided-smoothing}
Let $A$ be a right-continuous, piecewise continuously differentiable
function with finitely many pieces and jumps, and suppose that $A$
satisfies \eqref{eq:two-scale-A} for some $\gamma>0$. Let
\[
\mathcal B_-:=\{b>0:A(b^-)>A(b)\}
\]
be its finite set of downward jumps, and write $\Delta_b:=A(b^-)-A(b)>0$. Then, for every sufficiently small $\delta>0$, there is a right-continuous function $A_\delta$ such that:
\begin{enumerate}
\item $A_\delta\leq A$ pointwise;
\item $A_\delta$ has no downward jumps;
\item $A_\delta=A$ outside left neighborhoods of the points in $\mathcal B_-$ whose total length tends to zero with $\delta$;
\item $A_\delta$ satisfies \eqref{eq:two-scale-A} with $A$ replaced by
$A_\delta$;
\item for every integrable nonnegative weight $w$,
\[
\int_0^\infty|A_\delta(y)-A(y)|w(y)dy\longrightarrow0
\qquad(\delta\downarrow0).
\]
\end{enumerate}
\end{lemma}
\begin{proof}
Choose a continuously differentiable nondecreasing function $\chi:[0,1]\to[0,1]$ with $\chi(0)=0$ and $\chi(1)=1$. Choose $\delta$ small enough that the intervals
\[
J_b=[b(1-\delta),b],
\qquad b\in\mathcal B_-,
\]
are pairwise disjoint and contain no jumps of $A$ except their right endpoints. Define $A_\delta=A$ outside the interiors of these intervals, and for $y\in[b(1-\delta),b)$ set
\[
A_\delta(y)=A(y)-\Delta_b\chi\left(\frac{y-b(1-\delta)}{b\delta}\right).
\]
At the lower endpoint of $J_b$ the correction is zero. At the upper endpoint,
\[
\lim_{y\uparrow b}A_\delta(y)=A(b^-)-\Delta_b=A(b),
\]
so the downward jump at $b$ is replaced by a continuous transition. Since all downward jumps are treated and no other jumps are introduced, $A_\delta$ has no downward jumps.

Every correction is nonnegative, so $A_\delta\leq A$. Therefore
\[
A_\delta(y)+\gamma A_\delta(\rho y)\leq A(y)+\gamma A(\rho y)\leq0.
\]
Finally,
\[
0\leq A(y)-A_\delta(y)\leq\sum_{b\in\mathcal B_-}\Delta_b\mathbf 1_{J_b}(y).
\]
The intervals $J_b$ shrink to finitely many points as $\delta\downarrow0$. The absolute continuity of the Lebesgue integral for the integrable weight $w$ gives the claimed convergence.
\end{proof}

\section{Deterministic Adversaries and Buy-by-Breakeven Time}
\label{app:extra}

In this section we give a formal proof for the assertion in Lemma~\ref{lem:det-wlog} that when competing against deterministic adversaries, buying by the breakeven time is without loss of generality. 

\begin{proof}[Proof of Lemma~\ref{lem:det-wlog}]
Fix $\rho\in(0,1)$, and suppose that an algorithm has simultaneous competitive ratio at most $R$, so that for every $y>0$ the target is met at one of the two horizons $y$ and $\rho y$. Let $X$ denote its random buy time and let $C(y)=\mathbb E[c_X(y)]$ be its expected cost at horizon $y$.

For any $t\geq1$, consider the pair of horizons $t/\rho$ and $t$. The offline optimum has cost $1$ at both horizons. The simultaneous guarantee therefore implies that at least one of $C(t/\rho)$ and $C(t)$ is at most $R$. Moreover, the cost of every deterministic algorithm is nondecreasing in the horizon, so $C(t)\leq C(t/\rho)$. It follows that $C(t)\leq R$ for every $t\geq1$.

We next consider the behavior of the algorithm on long horizons. For every realization of $X$, the cost $c_X(t)$ increases to $X+1$ as $t\to\infty$, where the limit is infinite when $X=\infty$. Hence, by monotone convergence, $\mathbb E[X+1]=\lim_{t\to\infty}C(t)\leq R$. In particular, the algorithm buys eventually with probability one and has finite expected buy time.

Now truncate the buy time at the breakeven point by setting $X'=\min\{X,1\}$. If $y<1$, this modification does not change the cost: realizations with $X<1$ are unchanged, while both algorithms continue renting through horizon $y$ whenever $X\geq1$. If $y\geq1$, the truncated algorithm has expected cost $1+\mathbb E[\min\{X,1\}]\leq\mathbb E[X+1]\leq R$. Since $\OPT(y)=1$ for such horizons, its competitive ratio is at most $R$.

To see that the simultaneous guarantee is preserved, consider any pair $y,\rho y$ and choose a horizon in this pair at which the original algorithm has competitive ratio at most $R$. If that horizon is smaller than $1$, its cost is unchanged by the truncation. If it is at least $1$, the preceding bound shows directly that the truncated algorithm has competitive ratio at most $R$ there. Thus truncating the buy time at $1$ preserves the simultaneous competitive ratio.

The same conclusion holds when the algorithm knows only that $\rho\leq\alpha$. Indeed, apply the preceding argument to any one admissible value of $\rho$, for example $\rho=\alpha$, to obtain $\mathbb E[X+1]\leq R$. The truncation argument itself is pointwise and does not depend on $\rho$, so the resulting algorithm preserves the guarantee for every $\rho\leq\alpha$.
\end{proof}

\section{Proofs for the Buy-by-Breakeven-Time Tight Threshold}
\label{app:tight-buy-by-one}
\label{sec:tight-buy-by-one}

Section~\ref{sec:strict-late-buying-separation} gives the weighted tail bound~\eqref{eq:general-weighted-tail},  derives $D(\rho,\xi)$ in \eqref{eq:D-def}, and shows the general form of the positive construction in Lemma~\ref{lem:general-recursive-extension}. This appendix shows that these ingredients can be put together to obtain a tight threshold for algorithms that buy by the breakeven time.

Recall that fixing a target competitive ratio below
$R_{\mathrm{ski}}$ gives a residual curve $A$ that satisfies the
two-scale inequality~\eqref{eq:two-scale-A} for some $\gamma>0$, while
Lemma~\ref{lem:positive-compensation-mass}
requires $\mathcal L(A)>0$.  Since both the cost curve and the offline
optimum are constant after time $1$, there is a number $a=A(1)$ such that
$A(y)=a$ for every $y\ge1$.  Applying \eqref{eq:two-scale-A} at any
$y\ge1/\rho$ gives
$(1+\gamma)a\le0$, and hence
\begin{equation}
    A(y)=A(1)=a\le 0
    \qquad\text{for every }y\ge 1.
    \label{eq:constant-tail}
\end{equation}
Then both the positive and negative results reduce to the behavior of $D(\rho,\xi)$. The analytic bounds below determine when $D(\rho,\xi)\le1$ for every admissible $\xi$, which yields impossibility, and when $D(\rho,\xi)>1$ for some $\xi<1$, which gives a positive construction by applying Lemma~\ref{lem:general-recursive-extension} with the constant tail $g=-1$.

\subsection{Analytic bounds}
We first extend the analytic results from Lemma~\ref{lem:one-third-kernel}. Recall in~\eqref{eq:USeries}, we define the function $U_{\rho,\xi}$ as  
\begin{equation*}
U_{\rho,\xi}(t)
:=\sum_{k=0}^\infty(-\xi)^k e^{-t\rho^k}.
\end{equation*}
We first need the following alternate power series representation for $U_{\rho,\xi}(t)$
\emph{All results in this paper which are verified in Lean concern various analytic properties of $U_{\rho,\xi}(t)$ and related functions; hence, they all use this representation as an axiom}.
\begin{lemma}[Power-series representation]
\label{lem:U-power-series}
For $0<\rho<1$, $0\le\xi\le1$, and $t\ge0$, we have
\begin{equation}
    U_{\rho,\xi}(t)
    =
    \sum_{n=0}^{\infty}
    \frac{(-t)^n}{n!(1+\xi\rho^n)}.
    \label{eq:U-power-extension}
\end{equation}
\end{lemma}

\begin{proof}
First suppose $0\le\xi<1$.  By
\eqref{eq:USeries},
$U_{\rho,\xi}(t)=\sum_{j\ge0}(-\xi)^j e^{-t\rho^j}$.
Expanding each exponential produces terms
$(-\xi)^j(-t)^n\rho^{jn}/n!$.  Before exchanging the two sums, we check
their absolute convergence.  Since $0<\rho<1$ and $t\ge0$,
\begin{align*}
    \sum_{j=0}^{\infty}\sum_{n=0}^{\infty}
    \left|
    \frac{(-\xi)^j(-t)^n\rho^{jn}}{n!}
    \right|
    &=
    \sum_{j=0}^{\infty}\xi^j
    \sum_{n=0}^{\infty}\frac{(t\rho^j)^n}{n!}
    =
    \sum_{j=0}^{\infty}\xi^j e^{t\rho^j}
    \le
    e^t\sum_{j=0}^{\infty}\xi^j
    =
    \frac{e^t}{1-\xi}<\infty.
\end{align*}
Therefore, the order of summation may
be exchanged.  We obtain
\begin{align*}
    U_{\rho,\xi}(t)
    =
    \sum_{j=0}^{\infty}(-\xi)^j
    \sum_{n=0}^{\infty}\frac{(-t)^n\rho^{jn}}{n!}
    =
    \sum_{n=0}^{\infty}\frac{(-t)^n}{n!}
    \sum_{j=0}^{\infty}(-\xi\rho^n)^j
    =
    \sum_{n=0}^{\infty}
    \frac{(-t)^n}{n!(1+\xi\rho^n)}.
\end{align*}
The last equality uses $|\xi\rho^n|<1$, so the inner sum is a geometric
series with value $1/(1+\xi\rho^n)$.

It remains to pass to $\xi=1$.  From the probabilistic definition,
$
    U_{\rho,\xi}(t)
    =
    \E\!\left[
    \frac{(-\xi)^{K_{\rho,t}}}{1+\xi}
    \right].
$
The random variable $K_{\rho,t}$ is finite almost surely, and the integrand
converges pointwise to $(-1)^{K_{\rho,t}}/2$ as $\xi\uparrow1$.  Its absolute
value is at most $1$ for every $0\le\xi\le1$.  Dominated convergence
therefore gives
$\lim_{\xi\uparrow1}U_{\rho,\xi}(t)=U_{\rho,1}(t)$.

On the series side, for $0\le\xi\le1$ the absolute value of the $n$th term
is at most $t^n/n!$, independently of $\xi$, and
$\sum_{n\ge0}t^n/n!=e^t$.  Hence the series converges uniformly in
$\xi\in[0,1]$, and its limit as $\xi\uparrow1$ is obtained by setting
$\xi=1$ in every term.  Taking limits in the identity already proved for
$\xi<1$ establishes \eqref{eq:U-power-extension} at $\xi=1$ as well.
\end{proof}

Our main analytic lemma is now the following, and it extends the first and  second inequalities in Lemma~\ref{lem:one-third-kernel} to all $\rho$. This lemma is not specific to buy-by-breakeven time algorithms, and will also be extended and used for the impossibility proofs for general algorithms in the next section.

\begin{lemma}[Buy-by-breakeven threshold bound]
\label{lem:buy-by-breakeven-threshold-bound}
For $0.31\le\rho<1$, the function $D(\rho,\xi)$ is continuous in
$\xi\in[0,1]$.  For every $\xi,t\in[0,1]$, we have
$U_{\rho,\xi}(t)\ge 0$ and $D(\rho,\xi)\le D(\rho,1)$.  The equation
$D(\rho,1)=1$ has a unique
solution in $[0.31,0.32]$, with
\begin{equation}
    \rho_\star
    \approx
    0.311,
    \label{eq:rho-star}
\end{equation}
and, for $0.31\le\rho<1$,
\[
    D(\rho,1)
    \begin{cases}
        >1,&0.31\le\rho<\rho_\star,\\
        =1,&\rho=\rho_\star,\\
        <1,&\rho>\rho_\star.
    \end{cases}
\]
\end{lemma}
\leanverified{The complete statement, including continuity, the sign of
$U_{\rho,\xi}$, the comparison $D(\rho,\xi)\le D(\rho,1)$, strict
monotonicity in $\rho$, and the unique threshold; see
Appendix~\ref{app:lean-buy-by-breakeven-threshold-bound}.}

\begin{proof}
Appendix~\ref{app:lean-buy-by-breakeven-threshold-bound} gives a complete
Lean proof.  
\end{proof}

A technical issue is that
Lemma~\ref{lem:buy-by-breakeven-threshold-bound} begins at $\rho=0.31$, while the
positive construction also needs the sign of $D(\rho,1)-1$ for smaller
$\rho$.  The following auxiliary bound supplies that range.

\begin{lemma}[The lower-$\rho$ bridge]
\label{lem:lower-rho-bridge}
For every $0<\rho\le31/100$, we have $D(\rho,1)>1$.
\end{lemma}
\leanverified{The complete statement of this lemma; see
Appendix~\ref{app:lean-lower-rho-bridge}.}

\begin{proof}
Appendix~\ref{app:lean-lower-rho-bridge} certifies this lemma directly from
the factorial series for $D(\rho,1)$. 
\end{proof}

We finally need the following continuity result.

\begin{lemma}[Continuity of $D$ in $\xi$]
\label{lem:D-continuity-all-rho}
For every fixed $0<\rho<1$, the function
$\xi\mapsto D(\rho,\xi)$ is continuous on $[0,1]$.
\end{lemma}

\begin{proof}
Fix $0<\rho<1$.  For $x,\xi\in[0,1]$, the terms in
\eqref{eq:U-power-extension} satisfy
\[
    \left|
        \frac{(-x)^n}{n!(1+\xi\rho^n)}
    \right|
    \le \frac{1}{n!}.
\]
Since $\sum_{n=0}^{\infty}1/n!$ converges, the Weierstrass
$M$-test shows that the series defining $U_{\rho,\xi}(x)$
converges uniformly on $[0,1]\times[0,1]$.  Each term is
continuous in $(x,\xi)$, so $U_{\rho,\xi}(x)$ is jointly
continuous there.  It follows from \eqref{eq:D-def} that
$\xi\mapsto D(\rho,\xi)$ is continuous on $[0,1]$.
\end{proof}

\subsection{Proof of Theorem~\ref{thm:tight-buy-by-one}}
With the above analytic results in place, the proof is now simple.
\begin{proof}
\noindent {\bf Negative direction.} Suppose $\rho\ge\rho_\star$, and suppose that an improving algorithm exists.
As above, there is a function $A$ with $\mathcal L(A)>0$ satisfying
\eqref{eq:two-scale-A} and \eqref{eq:constant-tail} for some $\gamma>0$. If $\gamma\le\rho$, the direct charging argument in
Claim~\ref{cl:warmup-small-gamma} gives $\mathcal L(A)\le0$, a
contradiction. Hence assume
$\gamma>\rho$ and let $\xi=\rho/\gamma$. Since $\rho\ge\rho_\star>0.31$,
Lemma~\ref{lem:buy-by-breakeven-threshold-bound} gives both
$U_{\rho,\xi}(t)\ge0$ on $[0,1]$ and
$D(\rho,\xi)\le D(\rho,1)\le1$. The first conclusion verifies the
remaining sign hypothesis of Lemma~\ref{lem:general-weighted-tail}. The tail is
constant with value $a=A(1)$, so that
\begin{equation}
    \mathcal L(A)
    \le
    a\bigl(1-D(\rho,\xi)\bigr).
    \label{eq:L-dual-final}
\end{equation}

The second conclusion gives $1-D(\rho,\xi)\ge0$. Since $A(1)\le0$,
\eqref{eq:L-dual-final} gives $\mathcal L(A)\le0$, contradicting the
necessary condition $\mathcal L(A)>0$.

\medskip
\noindent {\bf Positive direction.}
Now suppose $\rho<\rho_\star$. If $\rho\le31/100$, then
Lemma~\ref{lem:lower-rho-bridge} gives $D(\rho,1)>1$; otherwise the same
conclusion follows from Lemma~\ref{lem:buy-by-breakeven-threshold-bound}. By Lemma~\ref{lem:D-continuity-all-rho}, choose $0<\xi<1$ sufficiently close to $1$ that $D(\rho,\xi)>1$, and put $\gamma:=\rho/\xi>\rho$.

Apply Lemma~\ref{lem:general-recursive-extension} with the constant
function $g(y)=-1$. Its hypotheses hold because
$g(y)+\gamma g(\rho y)=-(1+\gamma)$. Set $T:=1/\rho$.
For the resulting functions $A_N$,
\eqref{eq:general-recursive-limit} gives
\[
    \lim_{\substack{N\to\infty\\N\text{ even}}}
    \frac{\mathcal L(A_N)}{e}
    =
    -\int_1^T U_{\rho,\xi}(y)\,dy-e^{-T}.
\]

To evaluate the right-hand side, integrate the
recurrence~\eqref{eq:U-functional-identity} over $[1,T]$. This gives
\[
    \int_1^T U_{\rho,\xi}(y)\,dy
    +
    \xi\int_1^T U_{\rho,\xi}(\rho y)\,dy
    =
    e^{-1}-e^{-T}.
\]
In the second integral, the substitution $x=\rho y$ gives
\[
    \xi\int_1^T U_{\rho,\xi}(\rho y)\,dy
    =
    \frac{\xi}{\rho}\int_\rho^1 U_{\rho,\xi}(x)\,dx
    =
    \frac{D(\rho,\xi)}{e},
\]
where the last equality follows from \eqref{eq:D-def}. Therefore
\[
    -\int_1^T U_{\rho,\xi}(y)\,dy-e^{-T}
    =
    \frac{D(\rho,\xi)-1}{e}.
\]
Combining the preceding identities yields
\[
    \lim_{\substack{N\to\infty\\N\text{ even}}}
    \mathcal L(A_N)
    =
    D(\rho,\xi)-1
    >0.
\]

Thus $\mathcal L(A_N)>0$ for some even $N$. This is the construction in
Lemma~\ref{lem:general-recursive-extension}, started from the constant
tail required by a buy-by-breakeven-time algorithm.

Applying Lemma~\ref{lem:one-sided-smoothing} with sufficiently small smoothing intervals, followed by Lemmas~\ref{lem:certificate-realization} and~\ref{lem:two-scale-farkas-certificate}, gives the required buy-by-breakeven-time algorithm with a simultaneous improvement over $R_{\mathrm{ski}}$.
\end{proof}

\section{Proofs for the Arbitrary-Algorithm Tight Threshold}
\label{app:tight-general}
\label{sec:tight-general}

Appendix~\ref{app:tight-buy-by-one} gives the tight analysis for algorithms that buy by the breakeven time.  We now allow arbitrary buying times.  The new work concerns the possible shapes of the residual $A$ after time $1$, since the algorithm can now buy after the breakeven time. This renders the proof much more intricate than the previous section, though we reuse several ideas from there.

We proceed in three steps. First, we express the tail of an arbitrary
algorithm in terms of its buying time; this is a piecewise linear function leading to the shape of the residual curve in Figure~\ref{fig:one-third-recursive-curve}. Second, we evaluate its contribution
to the weighted integral $\mathcal L(A)$. This introduces
$h_{\rho,\xi}(\tau)$, which measures the contribution of buying at time
$\tau$. We then show that the sign of this function is controlled by its
value at $T=1/\rho$. The scaled value at this point is the function
$K(\rho,\xi)$ that determines the threshold $\rho$ value between possibility and impossibility. Finally, we determine when $K$ changes sign. The positive direction then uses the recursive extension from Lemma~\ref{lem:general-recursive-extension} and realization lemmas from Appendix~\ref{sec:curve-realization-appendix} to complete the proof.  

For a target ratio $R<R_{\mathrm{ski}}$, the residual curve is $A=C_F-R\OPT=A_{F,\eta}$. The quantities below depend instead on $\rho$ and $\xi$ assuming $\eta = R_{\mathrm{ski}} - R$ is given, and further assume $\eta$ is fixed.  We retain the functions $U_{\rho,\xi}$ and $D(\rho,\xi)$ from Appendix~\ref{app:tight-buy-by-one}.  
Recall that $\mathcal L(A)=e\int_0^\infty A(y)e^{-y}\,dy$. By Lemma~\ref{lem:positive-compensation-mass}, an improvement over $R_{\mathrm{ski}}$ requires $\mathcal L(A)>0$.

\subsection{Reducing $\mathcal L(A)$ via the weighted cost $h_{\rho,\xi}$}

This subsection rewrites the bound on $\mathcal{L}(A)$ from Lemma~\ref{lem:general-weighted-tail} in terms of the buying time $X$. The function $s_\tau(y)$ is the part of the eventual cost of buying at time $\tau$ that has yet to be paid at horizon $y$. The weight $W_{\rho,\xi}$ combines the two integrals in the bound, and $h_{\rho,\xi}(\tau)$ is the weighted contribution of $s_\tau$. These quantities give the bound on $\mathcal L(A)$ in \eqref{eq:general-tail-cone-dual}. This bound combines a constant term from the previous section (buying by breakeven time) and a second term encoded by $h_{\rho,\xi}(\tau)$ that captures buying at later time points.

We first describe the part of a residual curve that comes from buying after time $1$.  For $\tau\ge0$ and $y\ge1$, set
\[
    s_\tau(y)
    :=
    \begin{cases}
        \tau+1-y,&y<\tau,\\
        0,&y\ge\tau.
    \end{cases}
\]
As we show below,  $s_\tau(y)$ is the difference between the eventual cost $\tau+1$ of buying at time $\tau$
and the cost at horizon $y$. 

\begin{lemma}[Residual after time $1$]
\label{lem:general-residual-tail}
Let $X$ be the buying time of an algorithm with residual curve
$A=C_F-R\OPT$, where $R<R_{\mathrm{ski}}$, and suppose that $A$ satisfies
\eqref{eq:two-scale-A}. Then $\mathbb E[X] < \infty$. Set $A_\infty:=\mathbb E[X+1]-R$.  Then
$A_\infty\le0$, and, for every $y\ge1$,
\begin{equation}
    A(y)=A_\infty-
    \mathbb E\!\left[\mathbf 1_{\{X>1\}}s_X(y)\right].
    \label{eq:general-tail-decomposition}
\end{equation}
\end{lemma}

\begin{proof}
Set $\eta:=R_{\mathrm{ski}}-R$, so that
$A=C_F-R\OPT=A_{F,\eta}$. Since $A$ satisfies
\eqref{eq:two-scale-A}, Lemma~\ref{lem:two-scale-farkas-certificate}
shows that $F$ has simultaneous distributional competitive ratio at most
$R$. Lemma~\ref{lem:compensation-tail} therefore shows that the algorithm
buys eventually almost
surely,
\[
    \lim_{y\to\infty} C_F(y)=\mathbb E[X+1]<\infty,
\]
and $A(y)\le0$ for every $y\ge1$. Since $\OPT(y)=1$ for $y\ge1$,
\[
    \lim_{y\to\infty}A(y)
    =
    \lim_{y\to\infty}\bigl(C_F(y)-R\bigr)
    =
    \mathbb E[X+1]-R
    =
    A_\infty.
\]
Taking the limit in $A(y)\le0$ therefore gives $A_\infty\le0$.

It remains to describe how $A(y)$ differs from this limiting value. Fix a
buying time $\tau$. Then
\[
    c_\tau(y)
    =
    \begin{cases}
        y,&y<\tau,\\
        \tau+1,&y\ge\tau.
    \end{cases}
\]
Subtracting  $\tau+1$ gives
\[
    c_\tau(y)-(\tau+1)
    =
    \begin{cases}
        y-\tau-1,&y<\tau,\\
        0,&y\ge\tau,
    \end{cases}
    =
    -s_\tau(y).
\]
Now fix $y\ge1$. If $\tau\le1$, then $y\ge\tau$, and hence
$s_\tau(y)=0$. Consequently,
\[
    s_\tau(y)=\mathbf 1_{\{\tau>1\}}s_\tau(y)
    \qquad (y\ge1).
\]

Since $C_F(y)=\mathbb E[c_X(y)]$, we may subtract and add
$\mathbb E[X+1]$ to obtain
\begin{align*}
    A(y)
    &=
    C_F(y)-R\\
    &=
    \mathbb E[c_X(y)]-\mathbb E[X+1]
      +\mathbb E[X+1]-R\\
    &=
    A_\infty+\mathbb E\!\left[c_X(y)-(X+1)\right]\\
    &=
    A_\infty-
    \mathbb E\!\left[\mathbf 1_{\{X>1\}}s_X(y)\right].
\end{align*}
The expectations are finite because Lemma~\ref{lem:compensation-tail}
gives $\mathbb E[X+1]<\infty$, while
$0\le s_X(y)\le X+1$. This proves
\eqref{eq:general-tail-decomposition}.
\end{proof}

Lemma~\ref{lem:general-weighted-tail} bounds $\mathcal L(A)$ by two
integrals over $[1,1/\rho]$ and $[1/\rho,\infty)$.  We combine them by
defining
\begin{equation}
    W_{\rho,\xi}(y)
    :=
    \begin{cases}
        U_{\rho,\xi}(y),&1\le y<1/\rho,\\[1mm]
        e^{-y},&y\ge1/\rho.
    \end{cases}
    \label{eq:general-W-piecewise}
\end{equation}

\begin{lemma}[Total weight of $W$]
\label{lem:general-B-tail}
For $0<\rho<1$ and $0\le\xi\le1$,
\[
    \int_1^\infty W_{\rho,\xi}(y)\,dy
    =\frac{1-D(\rho,\xi)}e.
\]
In particular, $D(\rho,\xi)\le1$ implies that this integral is
nonnegative.
\end{lemma}

\begin{proof}
By the recurrence~\eqref{eq:U-functional-identity}, when $1\le y<1/\rho$,
\[
    W_{\rho,\xi}(y)
    =
    U_{\rho,\xi}(y)
    =
    e^{-y}-\xi U_{\rho,\xi}(\rho y).
\]
When $y\ge1/\rho$, the definition gives
$W_{\rho,\xi}(y)=e^{-y}$. Combining the two cases yields
\begin{equation}
    W_{\rho,\xi}(y)
    =
    e^{-y}
    -
    \xi U_{\rho,\xi}(\rho y)
    \mathbf 1_{\{1\le y<1/\rho\}}.
    \label{eq:general-W}
\end{equation}

Integrating \eqref{eq:general-W} gives
\begin{align*}
    \int_1^\infty W_{\rho,\xi}(y)\,dy
    &=
    \int_1^\infty e^{-y}\,dy
    -
    \xi\int_1^{1/\rho}
    U_{\rho,\xi}(\rho y)\,dy
    =
    \frac1e
    -
    \xi\int_1^{1/\rho}
    U_{\rho,\xi}(\rho y)\,dy
    = 
    \frac1e
    -
    \frac{\xi}{\rho}
    \int_\rho^1 U_{\rho,\xi}(x)\,dx.
\end{align*}
Finally, by definition
\[
    D(\rho,\xi)
    =
    \frac{e\xi}{\rho}
    \int_\rho^1 U_{\rho,\xi}(x)\,dx \qquad \Rightarrow \qquad 
    \int_1^\infty W_{\rho,\xi}(y)\,dy
    =
    \frac{1-D(\rho,\xi)}e.
\]
Therefore
$\int_1^\infty W_{\rho,\xi}(y)\,dy\ge0$ if and only if
$D(\rho,\xi)\le1$.
\end{proof}

For $\tau\ge1$, define
$h_{\rho,\xi}(\tau):=\int_1^\tau(\tau+1-y)W_{\rho,\xi}(y)\,dy$.
Since $s_\tau(y)=\tau+1-y$ for $1\le y<\tau$ and vanishes for
$y\ge\tau$, this can also be written as
\[
    h_{\rho,\xi}(\tau)
    =\int_1^\infty s_\tau(y)W_{\rho,\xi}(y)\,dy.
\]
Thus $h_{\rho,\xi}(\tau)$ measures the weighted cost that remains unpaid
at horizons before the purchase at time $\tau$.

\begin{lemma}[Weighted tail decomposition]
\label{lem:general-weighted-decomposition}
Under the assumptions of Lemma~\ref{lem:general-residual-tail}, suppose
also that $0<\rho<1$, $\gamma>\rho$, $\xi=\rho/\gamma$, and
$U_{\rho,\xi}(t)\ge0$ for $0\le t\le1$.  Then
\begin{equation}
    \frac{\mathcal L(A)}e
    \le
    \frac{1-D(\rho,\xi)}e A_\infty
    -\mathbb E\!\left[\mathbf 1_{\{X>1\}}h_{\rho,\xi}(X)\right].
    \label{eq:general-tail-cone-dual}
\end{equation}
\end{lemma}

\begin{proof}
The cost curve is bounded on $[0,1]$, and
Lemma~\ref{lem:compensation-tail} bounds it on $[1,\infty)$.  Hence $A$
is bounded.  Lemma~\ref{lem:general-weighted-tail} and
\eqref{eq:general-W-piecewise} give
\[
    \frac{\mathcal L(A)}e
    \le
    \int_1^{1/\rho}A(y)U_{\rho,\xi}(y)\,dy
    +\int_{1/\rho}^\infty A(y)e^{-y}\,dy
    =\int_1^\infty A(y)W_{\rho,\xi}(y)\,dy.
\]
Substituting \eqref{eq:general-tail-decomposition} into the last integral
gives
\begin{align*}
    \int_1^\infty A(y)W_{\rho,\xi}(y)\,dy
    &=
    A_\infty\int_1^\infty W_{\rho,\xi}(y)\,dy\\
    &\quad-
    \int_1^\infty
    \mathbb E\!\left[\mathbf 1_{\{X>1\}}s_X(y)\right]
    W_{\rho,\xi}(y)\,dy.
\end{align*}
By Lemma~\ref{lem:general-B-tail}, the first term is
$A_\infty(1-D(\rho,\xi))/e$. For the second term, Fubini's theorem and
the definition of $s_X$ give
\begin{align*}
    -\int_1^\infty
    \mathbb E\!\left[\mathbf 1_{\{X>1\}}s_X(y)\right]
    W_{\rho,\xi}(y)\,dy
    &=-\mathbb E\!\left[
        \mathbf 1_{\{X>1\}}
        \int_1^X(X+1-y)W_{\rho,\xi}(y)\,dy
      \right]
    \\
    &=-\mathbb E\!\left[\mathbf 1_{\{X>1\}}h_{\rho,\xi}(X)\right].
\end{align*}
Indeed, $s_X(y)=X+1-y$ for $1\le y<X$ and vanishes for $y\ge X$.
The exchange is valid because
\[
    \mathbb E\!\left[
      \int_1^\infty
      \mathbf 1_{\{X>1\}}|s_X(y)W_{\rho,\xi}(y)|\,dy
    \right]
    \le
    \mathbb E[X+1]
    \int_1^\infty|W_{\rho,\xi}(y)|\,dy
    <\infty.
\]
Combining these identities proves \eqref{eq:general-tail-cone-dual}.
\end{proof}

\subsection{Non-negativity of the weighted cost $h_{\rho,\xi}$}

In order to show an impossibility result, we need to show
$\mathcal{L}(A)\le0$. Equation~\eqref{eq:general-tail-cone-dual} shows that this happens when $h_{\rho,\xi}(\tau) \ge 0$ for every possible
buying time $\tau\ge1$. We begin with a property of $U_{\rho,\xi}$ that will imply concavity of $h_{\rho,\xi}$. We then calculate $h_{\rho,\xi}$ at $1$ and at $T:=1/\rho$, the point where the definition of $W_{\rho,\xi}$ changes, and study its shape on either side of $T$. Finally, we introduce $K(\rho,\xi)$ as a series expression for the scaled value of $h_{\rho,\xi}(T)$, and reduce the sign condition to a condition on the sign of $K(\rho,\xi)$. These facts together determine the sign of $h_{\rho,\xi}(\tau)$ for every $\tau\ge1$.

\begin{lemma}[Derivative estimate for $U$]
\label{lem:general-U-derivative-estimate}
Let $1/3\le\rho\le1$, $0\le\xi\le1$, and $0\le t\le1$.  Then
\[
    U_{\rho,\xi}(t)
    +\rho\frac{d}{dt}U_{\rho,\xi}(t)
    \ge0.
\]
\end{lemma}
\leanverified{The complete statement of this lemma; see
Appendix~\ref{app:lean-general-concavity}.}

\begin{proof}
Appendix~\ref{app:lean-general-concavity} gives a Lean proof.
\end{proof}

\begin{lemma}[Endpoint values of $h_{\rho,\xi}$]
\label{lem:general-h-values}
We have $h_{\rho,\xi}(1)=0$ and
\begin{equation}
    h_{\rho,\xi}(T)
    =\frac{1-\rho}{e\rho}
    -\frac{\xi}{\rho^2}
      \int_\rho^1(1+\rho-x)U_{\rho,\xi}(x)\,dx.
    \label{eq:general-h-endpoint-integral}
\end{equation}
\end{lemma}

\begin{proof}
The first identity follows directly from the definition, since the
integral defining $h_{\rho,\xi}(1)$ has equal endpoints. For the second,
use \eqref{eq:general-W} in the definition of $h_{\rho,\xi}(T)$:
\begin{align*}
    h_{\rho,\xi}(T)
    &=\int_1^T(T+1-y)e^{-y}\,dy-
      \xi\int_1^T(T+1-y)U_{\rho,\xi}(\rho y)\,dy.
\end{align*}
To evaluate the first integral, observe that
$\frac{d}{dy}((y-T)e^{-y})=(T+1-y)e^{-y}$. Hence
\[
    \int_1^T(T+1-y)e^{-y}\,dy
    =\left[(y-T)e^{-y}\right]_1^T
    =\frac{T-1}{e}
    =\frac{1-\rho}{e\rho}.
\]
In the second integral, set $x=\rho y$, so that $dy=dx/\rho$. The
endpoints become $x=\rho$ and $x=1$, and
$T+1-y=(1+\rho-x)/\rho$. Therefore
\[
    \xi\int_1^T(T+1-y)U_{\rho,\xi}(\rho y)\,dy
    =\frac{\xi}{\rho^2}
      \int_\rho^1(1+\rho-x)U_{\rho,\xi}(x)\,dx.
\]
Combining the two calculations proves
\eqref{eq:general-h-endpoint-integral}.
\end{proof}

\begin{lemma}[Shape of $h_{\rho,\xi}$]
\label{lem:general-h-shape}
Let $1/3\le\rho<1$ and $0\le\xi\le1$.  The function $h_{\rho,\xi}$ is concave on
$[1,T]$.  If $D(\rho,\xi)\le1$, then $h_{\rho,\xi}$ is nondecreasing on
$[T,\infty)$.
\end{lemma}

\begin{proof}
On $[1,T)$, \eqref{eq:general-W-piecewise} gives
$W_{\rho,\xi}=U_{\rho,\xi}$.  Differentiating $h_{\rho,\xi}$ twice gives
\begin{align}
    h_{\rho,\xi}'(\tau)
    &=W_{\rho,\xi}(\tau)+
      \int_1^\tau W_{\rho,\xi}(y)\,dy,
    \label{eq:general-h-prime}\\
    h_{\rho,\xi}''(\tau)
    &=W_{\rho,\xi}'(\tau)+W_{\rho,\xi}(\tau).
    \label{eq:general-h-second}
\end{align}
Differentiating the recurrence~\eqref{eq:U-functional-identity} gives
\begin{equation}
    h_{\rho,\xi}''(\tau)
    =-\xi\bigl(
        U_{\rho,\xi}(\rho\tau)
        +\rho U_{\rho,\xi}'(\rho\tau)
    \bigr),
    \qquad 1<\tau<T.
    \label{eq:general-h-concavity}
\end{equation}
Since $0\le\rho\tau\le1$, the expression in parentheses is nonnegative
by Lemma~\ref{lem:general-U-derivative-estimate}.  Thus $h_{\rho,\xi}$ is concave
on $[1,T]$.

For $\tau>T$, \eqref{eq:general-W-piecewise} gives
$W_{\rho,\xi}(\tau)=e^{-\tau}$.  Hence
\begin{align*}
    h_{\rho,\xi}'(\tau)
    &=e^{-\tau}+\int_1^\tau W_{\rho,\xi}(y)\,dy
    =\int_\tau^\infty W_{\rho,\xi}(y)\,dy
      +\int_1^\tau W_{\rho,\xi}(y)\,dy
    =\int_1^\infty W_{\rho,\xi}(y)\,dy
     =\frac{1-D(\rho,\xi)}e.
\end{align*}
The second equality uses $W_{\rho,\xi}(y)=e^{-y}$ for $y\ge T$ and
$\tau>T$, which gives
$\int_\tau^\infty W_{\rho,\xi}(y)\,dy=e^{-\tau}$. The last equality is
Lemma~\ref{lem:general-B-tail}. Thus the derivative is nonnegative when
$D(\rho,\xi)\le1$.
\end{proof}

\begin{figure}[t]
\centering
\begin{tikzpicture}
\begin{axis}[
    width=0.82\textwidth,
    height=0.43\textwidth,
    xmin=1,
    xmax=3.6,
    ymin=-0.013,
    ymax=0.095,
    xlabel={buying time $\tau$},
    ylabel={$h_{\rho,1}(\tau)$},
    xtick={1,1.5,2,2.5,3,3.5},
    ytick={-0.01,0,0.02,0.04,0.06,0.08},
    tick label style={font=\small},
    label style={font=\small},
    legend style={
        font=\small,
        at={(0.03,0.97)},
        anchor=north west,
        draw=none,
        fill=white,
        cells={anchor=west}
    },
    axis line style={black!55},
    tick style={black!55},
    clip=false
]
    \addplot[black!55,densely dashed,forget plot]
        coordinates {(1,0) (3.6,0)};

    \addplot[blue!70!black,very thick,mark=none] coordinates {
        (1.00,0.00000000) (1.05,0.00350912) (1.10,0.00670876)
        (1.15,0.00961055) (1.20,0.01222582) (1.25,0.01456568)
        (1.30,0.01664095) (1.35,0.01846221) (1.40,0.02003980)
        (1.45,0.02138380) (1.50,0.02250407) (1.55,0.02341021)
        (1.60,0.02411163) (1.65,0.02461746) (1.70,0.02493666)
        (1.75,0.02507795) (1.80,0.02504984) (1.85,0.02486063)
        (1.90,0.02451841) (1.95,0.02403108) (2.00,0.02340634)
        (2.05,0.02265169) (2.10,0.02177444) (2.15,0.02078174)
        (2.20,0.01968051) (2.25,0.01847754) (2.30,0.01717941)
        (2.35,0.01579254) (2.40,0.01432319) (2.45,0.01277744)
        (2.50,0.01116121) (2.55,0.00948028) (2.60,0.00774024)
        (2.65,0.00594655) (2.70,0.00410453) (2.75,0.00221932)
        (2.80,0.00029595) (2.85,-0.00166071) (2.90,-0.00364594)
        (2.95,-0.00565512) (3.00,-0.00768379) (3.05,-0.00605294)
        (3.10,-0.00442210) (3.15,-0.00279126) (3.20,-0.00116042)
        (3.25,0.00047042) (3.30,0.00210126) (3.35,0.00373210)
        (3.40,0.00536295) (3.45,0.00699379) (3.50,0.00862463)
        (3.55,0.01025547) (3.60,0.01188631)
    };
    \addlegendentry{$\rho=1/3$}

    \addplot[orange!85!black,very thick,densely dashed,mark=none] coordinates {
        (1.00,0.00000000) (1.05,0.00395716) (1.10,0.00763536)
        (1.15,0.01104571) (1.20,0.01419901) (1.25,0.01710582)
        (1.30,0.01977641) (1.35,0.02222080) (1.40,0.02444873)
        (1.45,0.02646972) (1.50,0.02829300) (1.55,0.02992759)
        (1.60,0.03138227) (1.65,0.03266558) (1.70,0.03378582)
        (1.75,0.03475109) (1.80,0.03556927) (1.85,0.03624800)
        (1.90,0.03679474) (1.95,0.03721673) (2.00,0.03752101)
        (2.05,0.03771443) (2.10,0.03780365) (2.15,0.03779514)
        (2.20,0.03769516) (2.25,0.03750984) (2.30,0.03724509)
        (2.35,0.03690666) (2.40,0.03650015) (2.45,0.03603096)
        (2.50,0.03550437) (2.55,0.03492546) (2.60,0.03429918)
        (2.65,0.03363033) (2.70,0.03292355) (2.75,0.03218334)
        (2.80,0.03323904) (2.85,0.03656850) (2.90,0.03989796)
        (2.95,0.04322743) (3.00,0.04655689) (3.05,0.04988636)
        (3.10,0.05321582) (3.15,0.05654528) (3.20,0.05987475)
        (3.25,0.06320421) (3.30,0.06653367) (3.35,0.06986314)
        (3.40,0.07319260) (3.45,0.07652207) (3.50,0.07985153)
        (3.55,0.08318099) (3.60,0.08651046)
    };
    \addlegendentry{$\rho=0.36$}

    \addplot[only marks,mark=*,mark size=2.2pt,blue!70!black]
        coordinates {(3,-0.00768379)};
    \node[anchor=south east,font=\small,fill=white,inner sep=1.5pt]
        (negativepoint) at (axis cs:2.91,-0.011)
        {$h_{1/3,1}(3)<0$};
    \draw[->,black!65] (negativepoint.east) -- (axis cs:3,-0.00768379);

    \addplot[only marks,mark=*,mark size=2.2pt,orange!85!black]
        coordinates {(2.77777778,0.03175928)};
    \node[anchor=south west,font=\small,fill=white,inner sep=1.5pt]
        (positivepoint) at (axis cs:2.84,0.047)
        {$h_{0.36,1}(1/0.36)>0$};
    \draw[->,black!65] (positivepoint.south west) --
        (axis cs:2.77777778,0.03175928);
\end{axis}
\end{tikzpicture}
\caption{The weighted contribution $h_{\rho,1}(\tau)$ for two values of
$\rho$. Each marked point is $T=1/\rho$ for its curve. For $\rho=1/3$, the curve is
negative at $T$; for $\rho=0.36$, it is positive there. Lemma~\ref{lem:general-h-shape}
shows that the curve is concave on $[1,T]$ and, when $D(\rho,1)\le1$,
nondecreasing after $T$. Thus the sign at $T$ controls the sign condition
used in the proof of possibility versus impossibility.}
\label{fig:general-weighted-cost-h}
\end{figure}
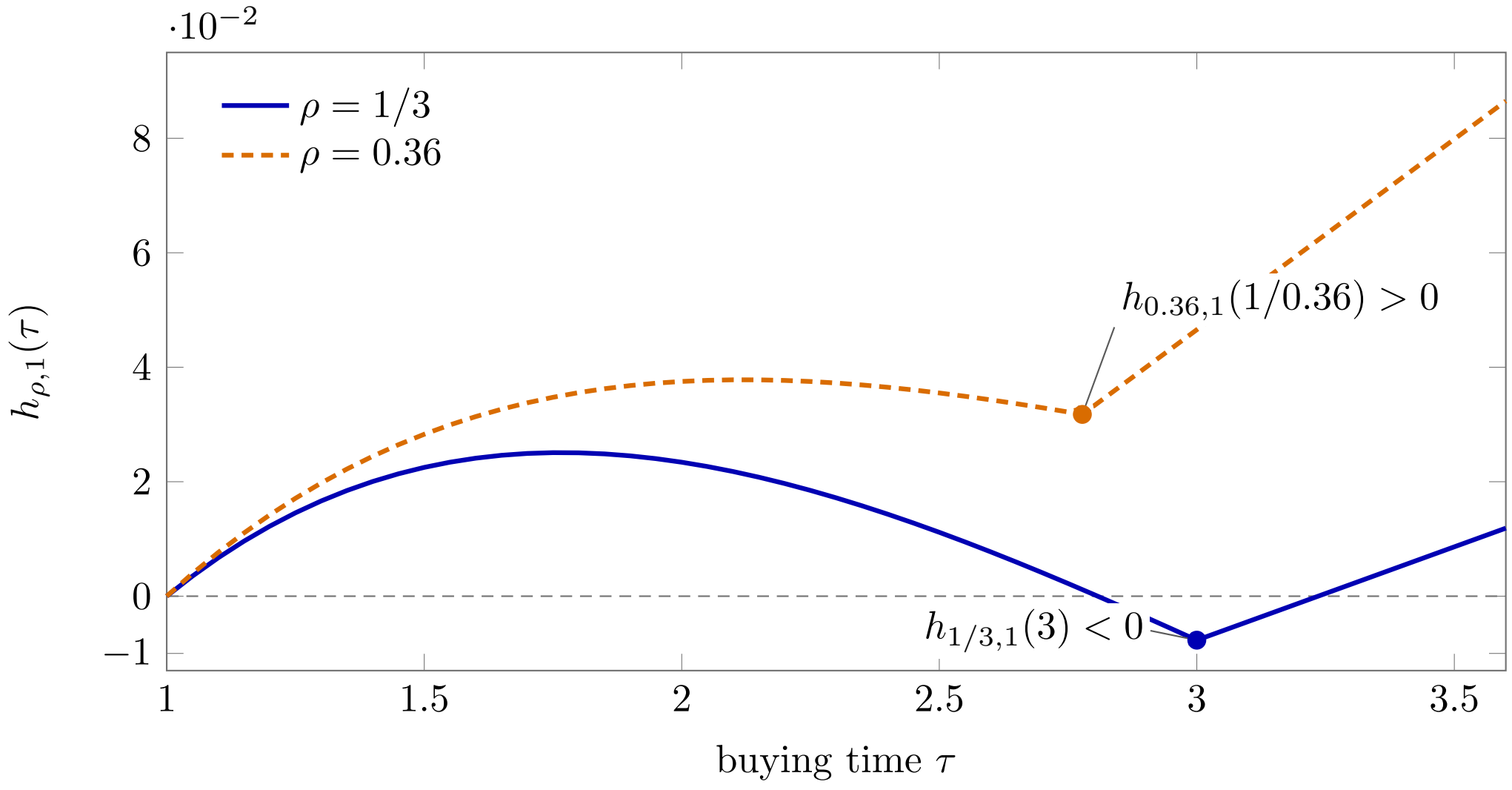

Figure~\ref{fig:general-weighted-cost-h} illustrates the two parts of
Lemma~\ref{lem:general-h-shape}.

\paragraph{Defining $K(\rho,\xi)$.} In order to bound $h_{\rho,\xi}(T)$ and to analyze the positive result, we construct an auxiliary function $K(\rho,\xi)$ whose sign we analyze. Towards this end, define
\begin{equation}
    d_n(\rho)
    :=
    \frac1{1-\rho}
    \int_\rho^1(1+\rho-x)x^n\,dx,
    \label{eq:general-dn}
\end{equation}
and
\begin{equation}
    K(\rho,\xi)
    :=
    \frac{\rho}{e}
    -\xi\sum_{n=0}^{\infty}
    \frac{(-1)^n d_n(\rho)}{n!(1+\xi\rho^n)}.
    \label{eq:general-K}
\end{equation}
The power series for $U_{\rho,\xi}$ gives the following relation, which
will be needed for both the possibility and impossibility results. Indeed, this equality is the crux of the proof.
\begin{equation}
    \frac{1-\rho}{e\rho}
    -\frac{\xi}{\rho^2}
    \int_\rho^1(1+\rho-x)U_{\rho,\xi}(x)\,dx
    =\frac{1-\rho}{\rho^2}K(\rho,\xi).
    \label{eq:general-K-scaled}
\end{equation}
\leanverified{The identity in \eqref{eq:general-K-scaled}, including the
identification of the integral coefficients with $d_n(\rho)$ and the
sum-integral exchange; see Appendix~\ref{app:lean-general-K}.}
Combining Lemma~\ref{lem:general-h-values} with
\eqref{eq:general-K-scaled} gives
\begin{equation}
    h_{\rho,\xi}(T)
    =\frac{1-\rho}{\rho^2}K(\rho,\xi).
    \label{eq:general-h-endpoint-K}
\end{equation}
Thus $K$ is the scaled contribution of buying at time $T$. This yields a condition for non-negativity of $h$ in terms of non-negativity of $K$.

\begin{corollary}
\label{cor:general-h-nonnegative}
If $1/3\le\rho<1$, $0\le\xi\le1$, $D(\rho,\xi)\le1$, and
$K(\rho,\xi)\ge0$, then $h_{\rho,\xi}(\tau)\ge0 \ \forall \ \tau\ge1$.
\end{corollary}

\begin{proof}
Lemma~\ref{lem:general-h-values} gives $h_{\rho,\xi}(1)=0$, while
\eqref{eq:general-h-endpoint-K} gives $h_{\rho,\xi}(T)\ge0$.
By Lemma~\ref{lem:general-h-shape}, concavity gives
$h_{\rho,\xi}(\tau)\ge(\tau-1)h_{\rho,\xi}(T)/(T-1)\ge0$ on $[1,T]$.  The same lemma shows
that $h_{\rho,\xi}$ is nondecreasing after $T$, completing  the proof.
\end{proof}

\subsection{Bounding $K(\rho,\xi)$ and $\mathcal L(A)$}

We now combine the preceding lemmas into the tail bound used for the
impossibility result. We then determine when its condition on
$K(\rho,\xi)$ holds.

\begin{lemma}[Tail bound for arbitrary buying times]
\label{lem:general-tail-reduction}
Let $X$ be the buying time of an algorithm with residual curve
$A=C_F-R\OPT$, where $R<R_{\mathrm{ski}}$.  Suppose that $A$ satisfies
\eqref{eq:two-scale-A}, that $1/3\le\rho<1$, that $\gamma>\rho$, and that
$U_{\rho,\xi}\ge0$ on $[0,1]$ for $\xi=\rho/\gamma$.  If
$D(\rho,\xi)\le1$ and $K(\rho,\xi)\ge0$, then
$\mathcal L(A)\le0$.
\end{lemma}
\begin{proof}
Lemma~\ref{lem:general-residual-tail} gives $A_\infty\le0$.
Lemma~\ref{lem:general-B-tail} gives
$(1-D(\rho,\xi))/e\ge0$, and
Corollary~\ref{cor:general-h-nonnegative} gives $h_{\rho,\xi}(X)\ge0$ whenever
$X>1$.  Both terms on the right side of
\eqref{eq:general-tail-cone-dual} are therefore nonpositive.  The result
follows from Lemma~\ref{lem:general-weighted-decomposition}.
\end{proof}

The two lemmas below now determine when the non-negativity condition on $K$ holds. Together with the definition in  \eqref{eq:general-K-scaled}, these facts will also give the positive result.

\begin{lemma}[Dependence on $\xi$]
\label{lem:general-K-xi}
For every $0<\rho<1$, the function $K(\rho,\xi)$ is continuous for
$\xi\in[0,1]$.  For $1/3\le\rho<1$, it also satisfies
$K(\rho,\xi)\ge K(\rho,1)$.
\end{lemma}
\leanverified{The complete statement of this lemma; see
Appendix~\ref{app:lean-general-K}.}

\begin{proof}
Appendix~\ref{app:lean-general-K} gives a complete Lean verification.  
\end{proof}

\begin{lemma}[Dependence on $\rho$]
\label{lem:general-K-rho}
\label{lem:general-K}
The function $K(\rho):=K(\rho,1)$ is strictly increasing on $(0,1)$ and
has a unique zero $\rho_{\mathrm{all}}\in(1/3,1/2)$.
\end{lemma}
\leanverified{The complete statement of this lemma; see
Appendix~\ref{app:lean-general-K}.}

\begin{proof}
Appendix~\ref{app:lean-general-K} gives a complete Lean verification.  
\end{proof}

\subsection{Proof of Theorem~\ref{thm:tight-general}}

The proof now follows the sign of $K(\rho,\xi)$, which by
\eqref{eq:general-h-endpoint-K} is the sign of the contribution from
buying at $T=1/\rho$. In the range used for the impossibility result,
$D(\rho,\xi)\le1$. If $K(\rho,\xi)\ge0$, then
Corollary~\ref{cor:general-h-nonnegative} shows that every buying time
$\tau\ge1$ has $h_{\rho,\xi}(\tau)\ge0$. The weighted decomposition in
Lemma~\ref{lem:general-weighted-decomposition} then shows that no
distribution over late buying times can make $\mathcal L(A)$ positive.

If $K(\rho,\xi)<0$, the buying time $T$ itself has
$h_{\rho,\xi}(T)<0$. The construction below starts from the tail
$g=-s_T$ and extends it recursively to earlier times. For the resulting
curves, the weighted integral $\mathcal L(A_N)$ approaches a positive
limit. The realization lemmas then turn one of these curves into an algorithm.
Thus the sign of the contribution from buying at the transition time
$T$ both rules out an improvement on one side of the threshold and
provides the construction on the other. We now formalize this.

\begin{proof}
We again prove the negative direction first. Suppose $\rho\ge\rho_{\mathrm{all}}$ and an improving algorithm exists.
Its residual curve satisfies \eqref{eq:two-scale-A} for some $\gamma>0$,
together with $\mathcal L(A)>0$.
Lemma~\ref{lem:compensation-tail} gives $A(y)\le0$ for $y\ge1$.

If $\gamma\le\rho$, the direct charging argument in
Claim~\ref{cl:warmup-small-gamma} gives $\mathcal L(A)\le0$.  That argument
uses only the sign of the tail.  Hence assume $\gamma>\rho$ and put
$\xi=\rho/\gamma$.
Because $\rho\ge\rho_{\mathrm{all}}>1/3>\rho_\star>31/100$,
Lemma~\ref{lem:buy-by-breakeven-threshold-bound} gives both
$U_{\rho,\xi}(t)\ge0$ on $[0,1]$ and
$D(\rho,\xi)\le D(\rho,1)\le1$. Lemmas~\ref{lem:general-K-xi}
and~\ref{lem:general-K-rho} also give
\[
    K(\rho,\xi)
    \ge
    K(\rho,1)
    \ge
    K(\rho_{\mathrm{all}},1)
    =
    0.
\]
Lemma~\ref{lem:general-tail-reduction} now gives
$\mathcal L(A)\le0$, contradicting $\mathcal L(A)>0$.

\paragraph{Positive direction.}
Now suppose $\rho<\rho_{\mathrm{all}}$.  Lemma~\ref{lem:general-K-rho}
gives $K(\rho,1)<0$.  By the continuity in
Lemma~\ref{lem:general-K-xi}, choose $\xi<1$ sufficiently close to $1$ that
$K(\rho,\xi)<0$, and put $\gamma:=\rho/\xi>\rho$.

Let $T:=1/\rho$ and take the function
\[
    g(y)
    :=
    \begin{cases}
        y-T-1,&1\le y<T,\\
        0,&y\ge T.
    \end{cases}
\]
This function is nonpositive on $[1,T)$.  If $T\le y<T/\rho$, then
$g(y)=0$ and $g(\rho y)\le0$; if $y\ge T/\rho$, both terms vanish.  Thus
$g$ satisfies the hypotheses of
Lemma~\ref{lem:general-recursive-extension}.  For the resulting functions
$A_N$, that lemma gives the two-scale inequality~\eqref{eq:two-scale-A} and
\[
    \lim_{\substack{N\to\infty\\N\text{ even}}}
    \frac{\mathcal L(A_N)}e
    =
    \int_1^Tg(z)U_{\rho,\xi}(z)\,dz
    =
    -\frac{1-\rho}{\rho^2}K(\rho,\xi)
    >0.
\]
Indeed, $g(z)=-(T+1-z)$ on $[1,T]$, while
$W_{\rho,\xi}(z)=U_{\rho,\xi}(z)$ there.  Hence, by the definition of
$h_{\rho,\xi}$ and \eqref{eq:general-h-endpoint-K},
\[
    \int_1^T g(z)U_{\rho,\xi}(z)\,dz
    =
    -h_{\rho,\xi}(T)
    =
    -\frac{1-\rho}{\rho^2}K(\rho,\xi).
\]

Hence $\mathcal L(A_N)>0$ for some even $N$. This is exactly the
construction in Lemma~\ref{lem:general-recursive-extension}, with
$T=1/\rho$.

Applying Lemma~\ref{lem:one-sided-smoothing}, followed by Lemmas~\ref{lem:late-tail-realization-check}, \ref{lem:certificate-realization}, and~\ref{lem:two-scale-farkas-certificate}, gives an algorithm with a strict simultaneous improvement over $R_{\mathrm{ski}}$.
\end{proof}

Finally, to solve  $K(\rho,1)=0$, numerical bisection gives $\rho_{\mathrm{all}}\approx0.338$. 

\section{Lean Proofs}
\label{app:lean-proofs}
The Lean code is available at
\url{https://github.com/relyt871/BWCA-Online-Lean}.
The \verb|lean/| directory of this repository contains a standalone Lean project.
Build instructions are given in the repository's root \verb|README.md|.
All file paths below are relative to the repository root.

Some finite rational computations are checked using Lean's
\verb|native_decide| tactic.  Every such application concerns an explicit
rational equality or inequality, a finite partial sum, or positivity of a
finite list or table of coefficients.  The surrounding analytic reductions
are expressed as ordinary Lean proofs.  These computations use exact
arithmetic and rely on Lean's native compiler in addition to its kernel; their
generated axiom dependencies can be inspected with \verb|#print axioms|.

All results in this paper which are verified in Lean concern analytic properties of the power series representation (Eq.\eqref{eq:U-power-extension}) of $U_{\rho,\xi}(t)$ given in Lemma~\ref{lem:U-power-series}: for $\rho\in (0,1),\xi\in[0,1]$, and $t\ge0$:
\begin{equation*}
    U_{\rho,\xi}(t)
    =
    \sum_{n=0}^{\infty}
    \frac{(-t)^n}{n!(1+\xi\rho^n)}.
\end{equation*}

\subsection{Lean certificates for Lemma~\ref{lem:buy-by-breakeven-threshold-bound}}
\label{app:lean-buy-by-breakeven-threshold-bound}
This appendix explains the Lean certificates used in
Lemma~\ref{lem:buy-by-breakeven-threshold-bound}.  The entry point is:
\begin{verbatim}
lean/buy_by_breakeven_threshold/BuyByBreakevenRho.lean
\end{verbatim}

\paragraph{1.  The endpoint-safe definition of \(U\).} The expression from Lemma~\ref{lem:U-power-series}
is encoded as \verb|Upower|
\begin{verbatim}
theorem buy_by_breakeven_U_power_series (rho xi t : Real) :
    U rho xi t = sum' n,
      (-t)^n / ((Nat.factorial n : Real) * (1 + xi * rho^n))
\end{verbatim}

\paragraph{2.  The functional identity.}
Lean proves the identity
\[
    U_{\rho,\xi}(t)+\xi U_{\rho,\xi}(\rho t)=e^{-t}
\]
directly from the power series.  The proof is coefficientwise: for each
\(n\),
\[
    \frac{(-t)^n}{n!(1+\xi\rho^n)}
    +
    \xi
    \frac{(-\rho t)^n}{n!(1+\xi\rho^n)}
    =
    \frac{(-t)^n}{n!}.
\]
Summing this identity gives:
\begin{verbatim}
theorem buy_by_breakeven_U_functional_identity {rho xi t : Real}
    (hrho0 : 0 <= rho) (hxi0 : 0 <= xi) :
    U rho xi t + xi * U rho xi (rho * t) = Real.exp (-t)
\end{verbatim}

\paragraph{3.  The dual multiplier recurrence.}
Lean defines the orbit weight and multiplier by
\begin{verbatim}
buyByBreakevenOrbitWeight rho x k
buyByBreakevenMultiplier rho gamma x k
\end{verbatim}
corresponding to
\[
    b_k(x)=\rho^k e^{-\rho^k x},
    \qquad
    u_k(x)=
    \frac{\rho^k}{\gamma}
    U_{\rho,\rho/\gamma}(\rho^k x).
\]
Using the functional identity and elementary algebra, Lean proves:
\begin{verbatim}
theorem buy_by_breakeven_multiplier_recurrence {rho gamma x : Real}
    (hrho0 : 0 <= rho) (hgamma : 0 < gamma) (k : Nat) :
    gamma * buyByBreakevenMultiplier rho gamma x k
      + buyByBreakevenMultiplier rho gamma x (k + 1)
      = buyByBreakevenOrbitWeight rho x k
\end{verbatim}
This is the recurrence \(\gamma u_k+u_{k+1}=b_k\) used in the dual
certificate.

\paragraph{4.  Nonnegative multipliers.}
The paper proves the sign of \(U\) from the geometric crossing
representation in Lemma~\ref{lem:geometric-crossing-representation}.  Lean
also supplies an independent algebraic verification.  It proves
\begin{verbatim}
theorem buy_by_breakeven_U_nonneg {rho xi t : Real}
    (hrho0 : (31 / 100 : Real) <= rho) (hrho1 : rho <= 1)
    (hxi0 : 0 <= xi) (hxi1 : xi <= 1)
    (ht0 : 0 <= t) (ht1 : t <= 1) :
    0 <= U rho xi t
\end{verbatim}
The Lean proof keeps the first four power-series terms, proves the alternating
tail is nonnegative, clears denominators in the resulting cubic lower bound,
and checks exact rational Bernstein coefficients on the box
\(\rho\in[0.31,1]\), \(\xi,t\in[0,1]\).

Consequently, Lean also proves nonnegativity of the actual multipliers:
\begin{verbatim}
theorem buy_by_breakeven_multiplier_nonneg {rho gamma x : Real}
    (hrho0 : (31 / 100 : Real) <= rho) (hrho1 : rho <= 1)
    (hgamma : 0 < gamma) (hxi1 : rho / gamma <= 1)
    (hx0 : 0 <= x) (hx1 : x <= 1) (k : Nat) :
    0 <= buyByBreakevenMultiplier rho gamma x k
\end{verbatim}

\paragraph{5.  The endpoint comparison in \(\xi\).}
Lean first proves continuity on the full closed interval:
\begin{verbatim}
theorem buy_by_breakeven_D_continuousOn_xi {rho : Real}
    (hrho0 : 0 < rho) (hrho1 : rho <= 1) :
    ContinuousOn (fun xi => D rho xi) (Set.Icc 0 1)
\end{verbatim}
The proof bounds every term uniformly by $1/n!$.  Lean then proves the full
comparison \(D(\rho,\xi)\le D(\rho,1)\).  The proof first
identifies the integral definition of \(D\) with the termwise integrated power
series.  It differentiates this series in \(\xi\), with each derivative term
bounded by a summable factorial series.  The first four terms are the finite
expression
\[
    \frac{1-\rho}{(1+\xi)^2}
    -
    \frac{1-\rho^2}{2(1+\xi\rho)^2}
    +
    \frac{1-\rho^3}{6(1+\xi\rho^2)^2}
    -
    \frac{1-\rho^4}{24(1+\xi\rho^3)^2}.
\]
Lean certifies this finite expression:
\begin{verbatim}
theorem buy_by_breakeven_Dxi_finite_nonneg {rho xi : Real}
    (hrho0 : (31 / 100 : Real) <= rho) (hrho1 : rho <= 1)
    (hxi0 : 0 <= xi) (hxi1 : xi <= 1) :
    0 <= buyByBreakevenDxiFinite rho xi
\end{verbatim}
The proof clears positive denominators and checks exact rational Bernstein
coefficients on the square \(\rho\in[0.31,1]\), \(\xi\in[0,1]\).
It also proves that the magnitudes of the remaining terms decrease, so the
alternating tail beginning with its positive term is nonnegative.  These two
facts make the full derivative nonnegative.  The derivative test then gives
the endpoint comparison:
\begin{verbatim}
theorem buy_by_breakeven_D_le_endpoint {rho xi : Real}
    (hrho0 : (31 / 100 : Real) <= rho) (hrho1 : rho <= 1)
    (hxi0 : 0 <= xi) (hxi1 : xi <= 1) :
    D rho xi <= D rho 1
\end{verbatim}
The case \(\xi=0\) follows directly from the integral definition and the
certified nonnegativity of \(U_{\rho,1}\).  For \(\xi>0\), Lean applies the
derivative argument on the interval \([\xi,1]\).

\paragraph{6.  Strict decrease in \(\rho\).}
The entry file also formalizes the one-variable argument for
$D(\rho,1)$.  It identifies $D$ with its factorially convergent series and
differentiates that series term by term on every neighborhood contained in
$(0,1)$.  Consecutive derivative terms are then paired.  The first pair is
zero, while every later pair is strictly negative.  Summability makes this
pairing valid for the infinite series.  Lean concludes:
\begin{verbatim}
theorem buy_by_breakeven_D_strictAntiOn_rho :
    StrictAntiOn (fun rho : Real => D rho 1) (Set.Ioo 0 1)
\end{verbatim}

\paragraph{7.  Existence and uniqueness of \(\rho_\star\).}
Lean proves that the magnitudes in the alternating series for $D(\rho,1)$
decrease throughout $0\le\rho\le1$.  Even and odd finite sums therefore give
rigorous lower and upper bounds.  It combines these bounds with Mathlib's
enclosure of $e$ and exact rational arithmetic to prove opposite signs at two
rational points whose interval rounds to $0.311$.  Continuity gives a root, and the
strict decrease above makes it unique.  The paper's three-decimal value is
captured by:
\begin{verbatim}
theorem buy_by_breakeven_rho_star_eq :
    D buyByBreakevenRhoStar 1 = 1

theorem buy_by_breakeven_rho_star_rounds_to_0311 :
    |buyByBreakevenRhoStar - (311 / 1000 : Real)| < 1 / 2000
\end{verbatim}
Finally, the same monotonicity theorem proves $D(\rho,1)>1$ below this root
and $D(\rho,1)<1$ above it.  Thus Lean verifies every assertion in
Lemma~\ref{lem:buy-by-breakeven-threshold-bound}.

\subsection{Lean certificates for Lemma~\ref{lem:lower-rho-bridge}}
\label{app:lean-lower-rho-bridge}

This appendix explains the Lean certificate for
Lemma~\ref{lem:lower-rho-bridge}.  The entry point is:
\begin{verbatim}
lean/lower_rho_bridge/LowerRhoBridge.lean
\end{verbatim}

Lean formalizes the series expression
\[
    \frac e\rho
    \sum_{n=1}^{\infty}
    (-1)^{n+1}
    \frac{1-\rho^n}{n!(1+\rho^{n-1})}
\]
as \verb|lowerRhoBridgeDSeries rho|.  The entry file imports the checked
identity between this series and $D(\rho,1)$ and exports the complete paper
statement as one theorem:
\begin{verbatim}
theorem lower_rho_bridge_D_gt_one {rho : Real}
    (hrho0 : 0 < rho) (hrho1 : rho <= 31 / 100) :
    1 < D rho 1
\end{verbatim}
The proof first rewrites $D(\rho,1)$ as
\verb|lowerRhoBridgeDSeries rho| and then applies the series inequality. Lean
derives the elementary numerical bound \(e>19/7\) from Mathlib's decimal
lower bound on $e$.

The proof has three exact pieces.  First, Lean proves that the term magnitudes
in the alternating series are decreasing for \(0\le\rho\le31/100\).  Therefore
the infinite alternating sum is at least the six-term partial sum \(S_6\).

Second, Lean proves the finite lower bound
\[
    S_6(\rho)>\frac7{19}\rho
    \qquad(0<\rho\le31/100).
\]
It does this by reducing \(S_6\) to the two rational functions \(p_3+p_5\),
clearing denominators, and checking positivity of a degree-\(13\) Bernstein
polynomial on the unit interval.

Third, Lean combines the alternating-series lower bound, the finite bound on
\(S_6\), and \(e>19/7\) to conclude
\[
    \frac e\rho S_6(\rho)>1.
\]
The file also checks the exact rational endpoint arithmetic displayed in the
paper:
\begin{verbatim}
theorem lower_rho_bridge_S6_margin_certificate :
    lowerRhoBridgeS6Margin
      =
      q 1538190334986700888894969
        20237843235225380172556254400
\end{verbatim}

\subsection{Lean certificates for
Lemma~\ref{lem:general-U-derivative-estimate}}
\label{app:lean-general-concavity}

This subsection explains the Lean check for the sign estimate in
Lemma~\ref{lem:general-U-derivative-estimate}.

The formal statement proves directly that
$U_{\rho,\xi}(t)+\rho U'_{\rho,\xi}(t)\ge0$ when
$\rho\in[1/3,1]$ and $\xi,t\in[0,1]$.  The Lean proof uses the power-series
representation
\[
    U_{\rho,\xi}(t)
    =
    \sum_{n=0}^{\infty}
    \frac{(-t)^n}{n!(1+\xi\rho^n)}.
\]
This is called \verb|Upower| in the formalization.  The paper-facing theorem
is part of the general-$K$ analytic file:
\begin{verbatim}
lean/general_K/GeneralKAnalytic.lean
\end{verbatim}
\begin{verbatim}
theorem general_endpoint_concavity_factor_nonneg {rho xi t : Real}
    (hrho0 : (1 / 3 : Real) <= rho) (hrho1 : rho <= 1)
    (hxi0 : 0 <= xi) (hxi1 : xi <= 1)
    (ht0 : 0 <= t) (ht1 : t <= 1) :
    0 <= U rho xi t +
      rho * deriv (fun z => U rho xi z) t
\end{verbatim}
Its proof has three parts.

\paragraph{1.  The expression checked by Lean.}
Lean first defines the individual terms in the power series:
\begin{verbatim}
noncomputable def Upower (rho xi t : Real) : Real :=
  sum' n, (-t)^n / ((Nat.factorial n : Real) *
    (1 + xi * rho^n))
\end{verbatim}
The displayed Lean statements use ASCII notation for mathematical symbols.

\paragraph{2.  Rewriting $U+\rho U'$ as the series in the paper.}
Lean proves that the power series can be differentiated term by term on the
interval needed in the paper.  The main theorem for this step is
\verb|hasDerivAt_Upower|.  It uses a uniform bound by a factorial series on
the open interval $(-1,2)$, which contains all $t\in[0,1]$.  After
differentiating, Lean proves that the derivative series is the shifted series:
\begin{verbatim}
theorem UpowerDerivSeries_eq_neg_shifted <omitted parameters>:
    UpowerDerivSeries rho xi t
      = - sum' n, UpowerShiftTerm rho xi n t
\end{verbatim}
Then it checks the algebraic identity for every coefficient:
\[
    \frac{(-t)^n}{n!(1+\xi\rho^n)}
    -
    \rho
    \frac{(-t)^n}{n!(1+\xi\rho^{n+1})}
    =
    (1-\rho)
    \frac{(-t)^n}
    {n!(1+\xi\rho^n)(1+\xi\rho^{n+1})}.
\]
In Lean this per-term algebra step is named
\verb|Upower_term_sub_rho_shift_eq|.  Together with the derivative theorem,
it identifies the series with the expression $U+\rho U'$ used in the paper.

\paragraph{3.  Proving the series is at least zero.}
The remaining task is the series sign check on the box.
Lean writes
\[
    \frac{U_{\rho,\xi}(t)+\rho U'_{\rho,\xi}(t)}{1-\rho}
    =
    \sum_{n=0}^{\infty}
    \frac{(-t)^n}
    {n!(1+\xi\rho^n)(1+\xi\rho^{n+1})}.
\]
It keeps the first six terms and proves that their sum is already positive.
After clearing denominators, this six-term expression is a polynomial in
$\rho,\xi,t$.  Lean substitutes
\[
    \rho=\frac13+\frac23\alpha,\qquad \xi=\beta,\qquad t=\tau,
\]
so that the domain becomes the unit cube
$(\alpha,\beta,\tau)\in[0,1]^3$.  The cleared polynomial is then expressed in
the tensor-product Bernstein basis of degrees $(20,5,5)$.  All Bernstein
coefficients are exact rational numbers, stored in
\verb|lean/SkiRental/LemmaCertificates.lean|, and Lean checks that every one
is positive:
\begin{verbatim}
theorem generalVPositiveP5TBernsteinCoeffs_positive :
    allBlocksPositive generalVPositiveP5TBernsteinCoeffs = true := by
  native_decide
\end{verbatim}
The reusable lemma in \verb|lean/SkiRental/Polynomial.lean| says that a
Bernstein polynomial with positive coefficients is positive on the unit cube.
This yields positivity of the six-term lower polynomial.

Finally, Lean verifies that the alternating tail starting at term $6$ is at
least zero.  This uses the elementary fact that, on the same box, the
magnitudes of the remaining terms are decreasing.  Therefore
\[
    \sum_{n=0}^{5}
    \frac{(-t)^n}
    {n!(1+\xi\rho^n)(1+\xi\rho^{n+1})}
    \le
    \sum_{n=0}^{\infty}
    \frac{(-t)^n}
    {n!(1+\xi\rho^n)(1+\xi\rho^{n+1})}.
\]
The exact internal theorem names are
\verb|generalVPositiveSixTerm_pos_on_box| for positivity of the six-term
polynomial, \verb|generalVPositiveSixTerm_le_rawSeries| for the
alternating-tail comparison, and
\verb|generalVPositiveSeries_nonneg_on_box| for the nonnegativity of the
series.  The paper-facing theorem above combines these facts with the
termwise derivative identity.  The numerical coefficients are rational
numbers, and the positivity checks are exact.

\subsection{Lean certificates for Lemmas~\ref{lem:general-K-xi}
and~\ref{lem:general-K-rho}}
\label{app:lean-general-K}

This appendix explains the Lean checks used in
Lemmas~\ref{lem:general-K-xi} and~\ref{lem:general-K-rho}.  The entry points
are:
\begin{verbatim}
lean/general_K/GeneralK.lean
lean/general_K/GeneralKAnalytic.lean
lean/general_K/GeneralKRho.lean
\end{verbatim}

The lemma studies
\[
    K(\rho,\xi)
    =
    \frac{\rho}{e}
    -
    \xi
    \sum_{n=0}^{\infty}
    \frac{(-1)^n d_n(\rho)}
    {n!(1+\xi\rho^n)}.
\]
Lean uses the same definition:
\begin{verbatim}
noncomputable def K (rho xi : Real) : Real :=
  rho / Real.exp 1
  - xi * (sum' n,
      (-1)^n * dCoeff rho n
      / ((Nat.factorial n : Real) * (1 + xi * rho^n)))
\end{verbatim}
Here \verb|dCoeff rho n| is the closed form for \(d_n(\rho)\) appearing in
the paper.

\paragraph{1.  Monotonicity of the coefficients.}
Both alternating-tail arguments use that the magnitudes decrease.  The
coefficient input is
\[
    d_{n+1}(\rho)\le d_n(\rho),
    \qquad 0\le \rho\le1.
\]
Lean proves this from the closed form for \(d_n\):
\begin{verbatim}
theorem general_K_dCoeff_succ_le {rho : Real}
    (hrho0 : 0 <= rho) (hrho1 : rho <= 1) (n : Nat) :
    dCoeff rho (n + 1) <= dCoeff rho n
\end{verbatim}
The proof also records the elementary bounds \(0\le d_n(\rho)\le1\), which
give summability by comparison with \(\sum 1/n!\).

\paragraph{2.  The \(\xi\)-derivative alternating tail.}
After differentiating the displayed series with respect to \(\xi\), the sign
check contains the alternating series
\[
    \sum_{n=0}^{\infty}
    (-1)^n
    \frac{d_n(\rho)}
    {n!(1+\xi\rho^n)^2}.
\]
Lean names the full series and its four-term truncation as follows:
\begin{verbatim}
generalKDxiSeries rho xi
generalKDxiFourTerm rho xi
\end{verbatim}
It proves:
\begin{verbatim}
theorem general_K_Dxi_fourTerm_le_series {rho xi : Real}
    (hrho0 : 0 <= rho) (hrho1 : rho <= 1)
    (hxi0 : 0 <= xi) (hxi1 : xi <= 1) :
    generalKDxiFourTerm rho xi <= generalKDxiSeries rho xi
\end{verbatim}
This is the alternating-series estimate used in
Lemma~\ref{lem:general-K-xi} to prove
\(K(\rho,\xi)\ge K(\rho,1)\).  The finite expression consists of the first
four terms of the derivative series.

Lean also checks that this finite expression is nonnegative on the needed
box.  The first two pairs are reduced to denominator-cleared polynomial
inequalities:
\begin{verbatim}
theorem general_K_first_pair_nonneg {rho : Real}
    (hrho : 0 <= rho) :
    0 <= generalKFirstPairClear rho

theorem general_K_second_pair_nonneg {rho : Real}
    (hrho : 0 <= rho) :
    0 <= generalKSecondPairClear rho
\end{verbatim}
At \(\xi=1\), Lean clears the positive denominator, shifts
\(\rho=1/3+s\), and checks that every coefficient of the resulting polynomial
is positive:
\begin{verbatim}
theorem general_K_F_at_one_pos {rho : Real}
    (hrho : (1 / 3 : Real) <= rho) :
    0 < generalKFAtOne rho
\end{verbatim}

\paragraph{3.  The complete comparison in $\xi$.}
The file \verb|GeneralKAnalytic.lean| applies these finite inequalities to
the series that defines $K$.  It proves a uniform summable bound for the
derivative terms, differentiates the infinite series, and identifies its
derivative with the series in the preceding paragraph.  The resulting
paper-facing theorems are:
\begin{verbatim}
theorem general_K_continuousOn_xi {rho : Real}
    (hrho0 : 0 <= rho) (hrho1 : rho <= 1) :
    ContinuousOn (fun xi => K rho xi) (Set.Icc (0 : Real) 1)

theorem general_K_minimized_at_one {rho xi : Real}
    (hrho : (1 / 3 : Real) <= rho) (hrho1 : rho <= 1)
    (hxi0 : 0 <= xi) (hxi1 : xi <= 1) :
    K rho 1 <= K rho xi
\end{verbatim}
Thus Lean checks the continuity and the full inequality in
Lemma~\ref{lem:general-K-xi}, including both endpoints of the $\xi$ interval.

\paragraph{4.  The scaled weighted identity.}
Lean also formalizes the direct identity between the weighted integral and
$K$.  It first proves the closed form for $d_n(\rho)$, integrates each
weighted power, and justifies the exchange of the sum and integral by a
summable uniform bound.  The final theorem is:
\begin{verbatim}
theorem general_K_scaled_endpoint_identity {rho xi : Real}
    (hrho0 : 0 < rho) (hrho1 : rho <= 1) (hxi0 : 0 <= xi) :
    (1 - rho) / (Real.exp 1 * rho)
        - xi / rho ^ 2 *
          intervalIntegral (fun x =>
            (1 + rho - x) * U rho xi x) rho 1 MeasureTheory.volume
      = (1 - rho) / rho ^ 2 * K rho xi
\end{verbatim}
This is exactly \eqref{eq:general-K-scaled}.

\paragraph{5.  The endpoint alternating tails.}
For \(K(\rho)=K(\rho,1)\), Lean defines the endpoint series
\[
    S(\rho)
    =
    \sum_{n=0}^{\infty}
    (-1)^n
    \frac{d_n(\rho)}{n!(1+\rho^n)}.
\]
The odd truncation at \(n=5\) is a lower bound:
\begin{verbatim}
theorem general_K_partial5_le_series {rho : Real}
    (hrho0 : 0 <= rho) (hrho1 : rho <= 1) :
    generalKPartial rho 5 <= generalKSeries rho
\end{verbatim}
The even truncation at \(n=2\) is an upper bound:
\begin{verbatim}
theorem general_K_series_le_partial2 {rho : Real}
    (hrho0 : 0 <= rho) (hrho1 : rho <= 1) :
    generalKSeries rho <= generalKPartial rho 2
\end{verbatim}
These are the two alternating-series estimates used for the endpoint signs
in Lemma~\ref{lem:general-K-rho} at \(\rho=1/3\) and \(\rho=1/2\).

\paragraph{6.  Endpoint signs.}
Lean assembles the alternating-tail bounds with the exact rational arithmetic.
At the left endpoint it proves
\[
    K(1/3,1)<0
\]
from the elementary exponential bound \(125/46<e\):
\begin{verbatim}
theorem general_K_left_endpoint_negative
    (he : (125 / 46 : Real) < Real.exp 1) :
    K (1 / 3) 1 < 0
\end{verbatim}
The proof uses the certified inequality
\[
    \frac{46}{375}
    <
    \sum_{n=0}^{5}
    (-1)^n
    \frac{d_n(1/3)}{n!(1+(1/3)^n)}
    \le
    S(1/3),
\]
and \(e>125/46\), which gives \(1/(3e)<46/375\).

At the right endpoint it proves
\[
    K(1/2,1)>0
\]
using Mathlib's upper bound on $e$:
\begin{verbatim}
theorem general_K_right_endpoint_positive_unconditional :
    0 < K (1 / 2) 1
\end{verbatim}
The proof uses
\[
    S(1/2)
    \le
    \sum_{n=0}^{2}
    (-1)^n
    \frac{d_n(1/2)}{n!(1+(1/2)^n)}
    =
    \frac{127}{720},
\]
and \(e<30/11\), which gives \(1/(2e)>11/60\).  The margin is
\[
    \frac{11}{60}-\frac{127}{720}
    =
    \frac1{144}.
\]

\paragraph{7.  Dependence on $\rho$ and the unique zero.}
Lean proves the complete statement of Lemma~\ref{lem:general-K-rho}.  The
file \verb|GeneralKRho.lean| proves the coefficient derivative bounds,
justifies differentiation of the infinite series, and proves that the
derivative of $K(\rho,1)$ is positive on $(0,1)$.  Its final two theorems are:
\begin{verbatim}
theorem general_K_strictMonoOn_rho :
    StrictMonoOn (fun rho : Real => K rho 1) (Set.Ioo 0 1)

theorem general_K_existsUnique_zero :
    ExistsUnique (fun rho : Real =>
      And (Set.Ioo (1 / 3 : Real) (1 / 2 : Real) rho) (K rho 1 = 0))
\end{verbatim}
The second theorem applies the intermediate value theorem to the certified
endpoint signs and then uses strict increase for uniqueness.  Thus Lean gives
the complete proofs of Lemmas~\ref{lem:general-K-xi} and
\ref{lem:general-K-rho}, together with the exact weighted identity
\eqref{eq:general-K-scaled}.

\subsection{Lean verification of Lemma~\ref{lem:one-third-kernel}}
\label{app:lean-one-third-kernel}

Lemma~\ref{lem:one-third-kernel} contains three estimates for the function $U$ at
\(\rho=1/3\).  The paper obtains the first estimate from
Lemma~\ref{lem:geometric-crossing-representation}, and Lean supplies an
independent check.  The second estimate is a direct Lean theorem.  Lean also
checks the complete conversion of the third estimate from its integral to the
alternating series.  We describe each part in the order in which it appears
in the lemma.  The entry points are
\begin{verbatim}
lean/one_third_kernel/OneThirdKernel.lean
lean/general_K/GeneralKAnalytic.lean
\end{verbatim}

\paragraph{1.  Positivity of $U$.}
The shared Lean definition of \(U_{\rho,\xi}(t)\) is the endpoint-safe power
series
\[
    U_{\rho,\xi}(t)
    =\sum_{n=0}^{\infty}
      \frac{(-t)^n}{n!(1+\xi\rho^n)}.
\]
The following theorem is also used for
Lemma~\ref{lem:buy-by-breakeven-threshold-bound}:
\begin{verbatim}
theorem buy_by_breakeven_U_nonneg {rho xi t : Real}
    (hrho0 : (31 / 100 : Real) <= rho) (hrho1 : rho <= 1)
    (hxi0 : 0 <= xi) (hxi1 : xi <= 1)
    (ht0 : 0 <= t) (ht1 : t <= 1) :
    0 <= U rho xi t
\end{verbatim}
Since \(31/100\le 1/3\le1\), specializing this theorem to \(\rho=1/3\)
proves \(U_\xi(t)\ge0\) for every \(\xi,t\in[0,1]\), including all four
endpoints.

\paragraph{2.  The bound on \(D_\xi\).}
In Lean, \verb|D (1 / 3) xi| is exactly
\(3e\xi\int_{1/3}^{1}U_{1/3,\xi}(t)\,dt\).  The final theorem has the same
hypotheses and conclusion as the second assertion of the lemma:
\begin{verbatim}
theorem one_third_kernel_D_lt_one {xi : Real}
    (hxi0 : 0 <= xi) (hxi1 : xi <= 1) :
    D (1 / 3) xi < 1
\end{verbatim}

The proof first keeps the terms of degrees zero through four in the power
series.  The omitted tail starts with a negative term.  Lean verifies from the
alternating-series estimate that this tail is nonpositive and hence proves the
pointwise bound
\begin{verbatim}
theorem one_third_kernel_U_le_upper {xi t : Real}
    (hxi0 : 0 <= xi) (hxi1 : xi <= 1)
    (ht0 : 0 <= t) (ht1 : t <= 1) :
    U (1 / 3) xi t <= oneThirdUUpper xi t
\end{verbatim}
It then integrates this five-term polynomial exactly over \([1/3,1]\).  After
multiplication by \(3\xi\), the result is
\[
 S(\xi)=
 \frac{2\xi(1-\xi)}{(1+\xi)(3+\xi)}
 +\frac{\xi(\xi+87)}{(\xi+9)(\xi+27)}
 +\frac{121\xi}{60(\xi+81)}.
\]
The formal proof identifies the integrated expression with \(S(\xi)\).  It
then checks the uniform rational bound:
\begin{verbatim}
theorem one_third_kernel_s_le {xi : Real}
    (hxi0 : 0 <= xi) (hxi1 : xi <= 1) :
    oneThirdS xi <= 11671 / 34440
\end{verbatim}
For the latter inequality, Lean clears denominators after deriving their
positivity from \(0\le\xi\).  The resulting identity has a positive factor on
the left and \((1-\xi)P(\xi)\) on the right, where
\[
 P(\xi)=229720293-138614976\xi+13485654\xi^2
        -225696\xi^3+7781\xi^4.
\]
The assumptions \(0\le\xi\le1\) imply \(P(\xi)>0\), so the identity gives the
stated bound on \(S\).  Finally, Mathlib's rigorous estimate for \(e\) gives
\(e<11/4\), and exact rational arithmetic gives
\((11/4)(11671/34440)<1\).  Combining these inequalities proves
\(D_\xi<1\) on the full closed interval \(0\le\xi\le1\).

\paragraph{3.  The late-tail integral.}
For \(\xi=1\), define
\[
 S_{\mathrm{late}}
 :=\sum_{n=0}^{\infty}
 \frac{(-1)^n d_n(1/3)}{n!(1+3^{-n})},
 \qquad
 I:=\int_1^3(t-4)U_1(t)\,dt
   =6\left(S_{\mathrm{late}}-\frac{1}{3e}\right).
\]
The coefficient $d_n(1/3)$ from \eqref{eq:general-dn} is exactly
\verb|dCoeff (1 / 3) n| in Lean.  The paper-facing Lean definition
\verb|oneThirdLateSeries| is the series $S_{\mathrm{late}}$ in this display.  Lean proves both
the identity and the required strict inequality:
\begin{verbatim}
theorem one_third_late_integral_eq_series :
    oneThirdLateIntegral =
      6 * (oneThirdLateSeries - 1 / (3 * Real.exp 1))

theorem one_third_kernel_late_series_positive :
    (1 / 3 : Real) / Real.exp 1 < oneThirdLateSeries

theorem one_third_late_integral_positive :
    0 < oneThirdLateIntegral
\end{verbatim}
The first theorem includes the weighted power integral, the sum-integral
exchange, the functional identity for $U$, the change of variables, and the
elementary exponential integral.  The positivity theorem combines this
identity with the odd alternating truncation at degree five, exact rational
endpoint arithmetic, and Mathlib's lower bound on \(e\).

Thus the positivity of \(U_\xi\) and the strict bound on \(D_\xi\) are fully
checked in Lean, including \(\xi=0,1\).  The late-tail integral identity and
its strict positivity are also complete Lean theorems.

\end{document}